\documentclass{aa}  
\let\linenumbers\relax 
\usepackage{graphicx}
\usepackage{txfonts}
\usepackage{lipsum}
\usepackage{subcaption}         
\usepackage{lscape}             
\usepackage{placeins}           

\usepackage{xcolor}

\usepackage{txfonts}
\usepackage{pifont}
\usepackage{keyval}
\usepackage[colorlinks=true,linkcolor=blue,filecolor=blue,citecolor = blue,urlcolor = blue]{hyperref}
\usepackage{hyperref}
\usepackage{amsmath}
\usepackage{makecell}
\usepackage{subcaption}
\usepackage{wasysym}

\newcommand{\ha}{H$\alpha$}
\newcommand{\niir}{$[\ion{N}{ii}]\,\lambda6584$}
\newcommand{\niib}{$[\ion{N}{ii}]\,\lambda6549$}
\newcommand{\niiha}{$[\ion{N}{ii}]\,\lambda6584/ \rm H\alpha$}
\newcommand{\siiha}{$[\ion{S}{ii}]\,\lambda\lambda6716,6731/ \rm H\alpha$}
\newcommand{\siir}{$[\ion{S}{ii}]\,\lambda6731$}
\newcommand{\siib}{$[\ion{S}{ii}]\,\lambda6716$}

\defcitealias{Naidu+2024_alt}{Naidu \& Matthee et al. 2024}  

\begin{document}

   \title{The Gas-Phase Metallicity and Nitrogen abundances in typical galaxies at $z\sim5$}

   \subtitle{}

\titlerunning{Nitrogen based MZR in NIRCam Grism}
\authorrunning{Di Cesare et al.}


\author{Claudia Di Cesare\inst{1}\corrauth{claudia.dicesare@ista.ac.at}
   \and Gauri Kotiwale\inst{1}
   \and Jorryt Matthee\inst{1}
   \and Alice Shapley\inst{2}
   \and Alba Covelo-Paz\inst{3}
   \and Kasper E. Heintz\inst{4,5,6}
   \and Pascal A. Oesch\inst{5,6,7}
   \and Edoardo Iani\inst{1}
   \and Anna Frebel\inst{8}
   \and Daichi Kashino\inst{9}
   \and Ivan G. Kramarenko\inst{1}
   \and Sara Mascia\inst{1}
   \and Rohan P. Naidu\inst{10}
   \and Irene Shivaei\inst{11}
   \and Alberto Torralba\inst{1}
   \and Rongmon Bordoloi\inst{12}
   \and Emma Giovinazzo\inst{7}
   \and Jasleen Matharu\inst{13}
}

\institute{Institute of Science and Technology Austria (ISTA), Am Campus 1, 3400 Klosterneuburg, Austria
   \and Department of Physics \& Astronomy, University of California, Los Angeles, 430 Portola Plaza, Los Angeles, CA 90095, USA
   \and Université de Strasbourg, CNRS, Observatoire astronomique de Strasbourg, UMR 7550, 67000 Strasbourg, France
   \and DTU Space, Technical University of Denmark, Elektrovej 327, DK2800 Kgs. Lyngby, Denmark
   \and Cosmic Dawn Center (DAWN), Copenhagen, Denmark
   \and Niels Bohr Institute, University of Copenhagen, Jagtvej 128, 2200 Copenhagen N, Denmark
    \and Department of Astronomy, University of Geneva, Chemin Pegasi 51, 1290 Versoix, Switzerland
   \and Department of Physics and Kavli Institute for Astrophysics and Space Research, Massachusetts Institute of Technology, Cambridge, MA 02139, USA
   \and National Astronomical Observatory of Japan (NAOJ), 2-21-1 Osawa, Mitaka, Tokyo 181-8588, Japan
   \and Institute for Astronomy, University of Hawai`i, 2680 Woodlawn Drive, Honolulu, HI 96822, USA
   \and Centro de Astrobiolog\'ia (CAB), CSIC-INTA, Carretera de Ajalvir km 4, Torrej\'on de Ardoz 28850, Madrid, Spain
   \and Department of Physics, North Carolina State University, Raleigh, NC 27695, USA
   \and Max-Planck-Institut f\"ur Astronomie, K\"onigstuhl 17, D-69117 Heidelberg, Germany
}



   \date{}

  
  \abstract{The relation between stellar mass and gas-phase metallicity (MZR) is a key tracer of the baryon cycle in galaxies. JWST now gives access to rest-optical emission lines at $z > 3$, enabling unprecedented constraints on the chemical enrichment and the conditions of the interstellar medium of galaxies in the early Universe. In this work, we present the MZR at $z=3.8-6.6$ from stacked NIRCam/grism spectra of 1748 H$\alpha$-selected star-forming galaxies compiled over an area of 320 arcmin$^2$. The simple H$\alpha$ flux-limited selection allows us to measure representative $[\ion{N}{ii}]\,\lambda6584/\rm H\alpha$ and $[\ion{S}{ii}]\,\lambda\lambda6716,6731/\rm H\alpha$ ratios in stacks spanning $\rm log(M_\star/M_\odot)=8-10.2$, extending the upper mass range of previous high-redshift MZR measurements by $\sim1$~dex. Gas-phase metallicities are derived from these ratios using strong-line calibrations recently derived from high-redshift galaxies. The [\ion{N}{ii}]-based MZR has a slope $\gamma = 0.42^{+0.10}_{-0.07}$ and $12+\rm log(O/H) = 7.94^{+0.06}_{-0.09}$ at $\rm log(M_\star/M_\odot) = 9$, in broad agreement with recent empirical studies and hydrodynamical simulations. The [\ion{S}{ii}]-based relation yields a consistent slope and a $\sim0.3$~dex higher normalization, within the systematic uncertainties of strong-line calibrations. From the resolved $\rm [\ion{S}{ii}]$ doublet we infer a  median electron density of $n_e = 184^{+226}_{-169}\,\rm cm^{-3}$. Combining $\rm [\ion{N}{ii}]$ and $\rm [\ion{S}{ii}]$, we indirectly estimate the N/O ratio and find that it follows the trend expected from primary and secondary nitrogen production, with no evidence for nitrogen enhancement in the H\textsc{ii} region emission that dominates our galaxies. Our results provide a benchmark for the chemical enrichment and ISM conditions of typical star-forming galaxies at $z\sim5$ across a wide mass range.
  }



   \maketitle

\section{Introduction}

Metals play a central role in shaping the interstellar medium (ISM) of galaxies \citep[e.g.,][]{Kewley+2019, MaiolinoMannucci2019}. In H\,\textsc{ii} regions, metals dominate gas cooling through collisionally excited emission lines and control the efficiency of dust growth in the ISM, which is thought to play a key role in the rapid dust enrichment observed in early galaxies \citep{Asano+2013_dust, SchneiderMaiolino2024}. At the same time, the gas-phase metallicity is set by the balance between metal production through star formation, the accretion of metal-poor gas and the removal of metal-rich gas through feedback-driven outflows, making it a sensitive tracer of the baryon cycle in galaxies \citep[e.g.][]{Lilly+2013, PerouxHowk2020}. JWST now provides access to the full set of rest-frame optical emission lines, including $\rm H\alpha$ and $\rm [\ion{N}{ii}]$, beyond $z \sim 3$, enabling detailed studies of the ISM conditions and chemical enrichment of high-redshift galaxies.

Several observational campaigns have characterized the ISM metallicity of large samples of galaxies across a broad range of redshifts, establishing the existence of a positive correlation between the gas-phase metallicity and stellar mass (M$_\star$), known as the mass-metallicity relation \citep[MZR,][]{Tremonti+2004, AndrewsMartini2013, MaiolinoMannucci2019}, which has now been traced out to $z > 6$ \citep[e.g.,][]{nakajima+2023_mzr, Heintz+2023,Curti+2024_mzr,Chemerynska+2024_mzr,Kotiwale+2026, Lam+2026, Isobe+2026}. The MZR is commonly interpreted as a consequence of more massive galaxies, with deeper potential wells, retaining their metals more efficiently and losing a smaller fraction of them through outflows \citep[e.g.,][]{PeeplesShankar2011,Lilly+2013, Ma+2016_mzr, Torrey+2018}.  

At $z\sim 0$, the MZR follows a power law below $\rm M_\star \sim 3\cdot 10^{10}\, M_\odot$ and flattens into a plateau at higher masses. Its normalization decreases from $z\sim0$ to $z \sim 3$ \citep[e.g.,][]{Sanders+2021_mosdef, He+2024}, with a slower evolution at higher redshifts \citep{nakajima+2023_mzr,Curti+2024_mzr}. Lower metallicities at fixed stellar mass at earlier epochs are generally attributed to higher gas fractions and stronger feedback-driven outflows of metal-rich gas \citep{Sanders+2021_mosdef,Heintz+2023, Pollock+2026a}. The redshift evolution of the slope remains uncertain. The slope is roughly constant up to $z\sim 3$ \citep{Sanders+2021_mosdef}, while at higher redshifts some studies find a flatter relation \citep{Curti+2024_mzr, Kotiwale+2026} and others a steeper one \citep{Chemerynska+2024_mzr}. A flattening has also been reported at low masses \citep{He+2024}. These discrepancies may partly reflect differences in sample selection, the stellar mass range probed and the adopted metallicity diagnostics. Establishing the shape of the MZR at high redshift therefore requires large samples with a well-understood selection function, spanning a wide range in stellar mass and SFR.

In the local Universe, the scatter of the MZR is reduced when the SFR is included as a third parameter, giving rise to the Fundamental Metallicity Relation \citep[FMR,][]{Lara-Lopez+2010,Mannucci+2010}. The FMR is thought to reflect a quasi-equilibrium state of star-forming galaxies, regulated by the interplay between gas inflows, star formation and outflows \citep[e.g.,][]{Dave+2012}. \cite{Sanders+2021_mosdef} find that galaxies remain broadly consistent with the local FMR out to $z\sim 3$. However, recent JWST observations indicate that high-redshift galaxies have lower metallicities at fixed stellar mass and SFR than predicted by the local relation \citep[e.g.,][]{nakajima+2023_mzr, Heintz+2023,Curti+2024_mzr, Stanton+2026_excels, Lam+2026}. \cite{Lam+2026} find an offset of $\sim -0.15$ dex already at $z\gtrsim2$, while \cite{Curti+2024_mzr} find that the metal deficiency becomes more pronounced with increasing redshift. They interpret this as evidence that metals are removed more efficiently by supernova-driven outflows, or that the ISM is diluted by stochastic accretion of metal-poor gas on timescales shorter than those of subsequent enrichment by star formation. However, these results are largely based on JWST/NIRSpec samples, whose target selection and the requirement of detecting emission lines can bias the inferred metallicity distribution \citep[e.g.,][]{Lewis+2026_mzr}.

The gas-phase metallicity is typically inferred either with the direct-$T_e$ method or with strong-line calibrations \citep{Curti2025, Sanders+2024, sanders+2026_aurora_calib}. The direct method relies on faint auroral lines to measure the electron temperature of the ionized gas. These lines are more easily detected in metal-poor, highly star-forming galaxies \citep{Izotov+2006, Scholte+2025_excels}, but still require deep spectroscopy of individual sources \citep[e.g.,][]{Strom+2023_cecilia, Pollock+2026a} or stacked spectra \citep[e.g.,][]{Curti+2017, Kotiwale+2026, Isobe+2026}. For large galaxy samples, metallicities are instead derived from ratios of bright emission lines, calibrated against $T_e$-based measurements \citep[e.g.,][]{Marino+2013, curti+2020_calib, nakajima+2022_calib} or photoionization models \citep[e.g.,][]{Dopita+2016,Kewley+2019}. Since the ionized ISM at $z > 2$ differs from that of local galaxies \citep[e.g.,][]{Shapley+2015,Strom+2018,Topping+2020_ion,Isobe+2023, topping+2025_aurora}, calibrations based on $z\sim 0$ samples may be unreliable at high redshift, motivating the recent JWST-based calibrations built on auroral-line detections at $z>1$ \citep[e.g.,][]{Laseter+2024_calib, Sanders+2024, Cataldi+2025_calib, Scholte+2025_excels, Chakraborty+2025, sanders+2026_aurora_calib, Isobe+2026}. An additional caveat for the direct method concerns the electron density, which increases with redshift \citep[e.g.,][]{Isobe+2023, topping+2025_aurora}. Densities as high as $n_e \sim 10^5\,\rm cm^{-3}$ have been inferred in some $z>6$ galaxies \citep[e.g.,][]{Arellano-Cordova+2026, Torralba+2026_vms}. Adopting lower densities, from low-ionization diagnostics or from the low-density limit, leads to overestimated electron temperatures and hence underestimated metallicities, making these galaxies appear more metal-poor than they are \citep[e.g.,][]{Martinez+2025, Arellano-Cordova+2026, Hsiao+2026_ne}.

Beyond the overall metal content, the relative abundances of individual elements encode information on timescales and sources of chemical enrichment \citep{Matthee+2018, Kobayashi+2020}. The nitrogen-to-oxygen (N/O) ratio is particularly informative \citep[e.g.][]{Vincenzo+2016, MaiolinoMannucci2019}. Oxygen is an $\alpha$-element mainly produced by massive stars and returned to the ISM by core-collapse supernovae within $\lesssim40$ Myr \citep{Kobayashi+2020}. Nitrogen is instead synthesized through the CNO cycle and released via stellar winds \citep{Vincenzo+2016, MaynetMaeder2002}. Its production is primary, with a yield independent of the stellar metallicity, when the carbon and oxygen are freshly synthesized within the star itself, as in rotating massive stars \citep[e.g.,][]{MaynetMaeder2002, Chiappini+2006} and in intermediate-mass stars on the asymptotic giant branch \citep{Ventura+2013}. It is secondary when carbon and oxygen were already present in the gas from which the star formed, so that the yield increases with metallicity \citep{Vincenzo+2016, MaiolinoMannucci2019}. Depending on stellar mass, nitrogen is released over timescales ranging from a few tens of Myr to $\gtrsim 10^8$ yr \citep{Henry+2000,Vincenzo+2016}. Together, these processes produce the observed N/O-O/H relation: N/O is roughly constant at low metallicity, where primary production dominates, and rises at higher metallicity as secondary production takes over \citep[e.g.,][]{Henry+2000,Vincenzo+2016, VincenzoKobayashi2018}.

JWST has revealed a population of high-redshift galaxies with strong rest-frame UV \ion{N}{iv}]$\lambda1486$ and \ion{N}{iii}]$\lambda1750$ emission, implying super-solar N/O ratios at sub-solar O/H \citep[e.g.,][]{Cameron+2023_NO-enhancement,Isobe+2023_Nemitter, Topping+2024_Nemitter}. However, studies combining direct metallicities from rest-frame optical spectra with N/O abundances from \niir{} find a broad range of N/O values, with only a few galaxies showing super-solar ratios \citep[e.g.,][]{Arellano-Cordova+2025_NO, Cameron+2026_NO, Cataldi+2026_NOcalib, Umeda+2026_NO}. More recently, \cite{Schaerer+2026_no} found that the N/O ratios and their scatter in typical star-forming galaxies at $z\sim3$ are indistinguishable from those of low-redshift galaxies and H\,\textsc{ii} regions at the same metallicity, suggesting no redshift evolution of the N/O-O/H relation for this population. Whether this also holds at earlier epochs remains to be tested with large samples.

In this work, we present new measurements of the MZR and the N/O abundance ratio at $z=3.8-6.6$ ($\langle z \rangle = 4.6$), based on 1748 galaxies identified through their \ha{} emission in NIRCam grism spectra from the EIGER, FRESCO, CONGRESS, ALT and JWST Cycle 1 PID 1933 surveys. This provides a large sample with a simple selection function, set primarily by the \ha{} flux limit of the grism surveys rather than by photometric preselection. Although we detect metal lines in various individual sources, we focus on stacked spectra that are not biased by such detections. By stacking the spectra in bins of stellar mass, we detect the faint \niir{} and \siib{} lines, which the grism resolution ($R\sim 1600$) allows us to separate from \ha{} and \siir{}, respectively. These lines allow us to measure the gas-phase metallicity, the N/O ratio and the electron density of the stacked samples. With these measurements, we constrain the shape and normalization of the MZR at $z\sim4-6$, test whether our sample deviates from the local FMR, as reported by previous JWST studies, and assess whether the nitrogen enhancement reported in individual high-redshift galaxies is also present in the stacked \ha{}-selected population.

We derive metallicities from the \niiha{} (N2$=\rm log$(\niiha)) and \siiha{} (S2$=\rm log$(\siiha)) strong-line diagnostics, which are monotonic over the metallicity range of our sample\footnote{Based on the photoionization models of \cite{Kewley+2019}, the S2 diagnostic is expected to become double-valued at high metallicity.} and thus avoid the degeneracies affecting many oxygen-based calibrations \citep[e.g.][]{Sanders+2024}. Since sulfur, like oxygen, is an $\alpha$-element, S2 is expected to be less sensitive to abundance-ratio variations than N2, and provides an independent check on our N2-based MZR. We trace N/O with the \niir/$[\ion{S}{ii}]\,\lambda\lambda6716,6731$ ratio \citep[e.g.][]{Perez-Montero_Contini2009, Cataldi+2026_NOcalib}, which assumes that S/O varies little with O/H \citep{Berg+2020}. However, sub-solar S/O ratios have been reported in galaxies at cosmic noon \citep{Rogers+2026} and S/O may vary at low metallicity \citep{Cataldi+2026_chemical_abundances}, which would affect both our S2-based metallicities and the inferred N/O.

The paper is organized as follows. Section~\ref{sec:data} describes our observational sample. In Sect.~\ref{sec:methods}, we outline our methods, including the SED fitting (Sect.~\ref{sec:sed}), the stacking procedure (Sect.~\ref{sec:stacking}), and the derivation of physical properties (Sects.~\ref{sec:el_density} and \ref{sec:derived_quantities}). We present the [\ion{N}{ii}]-based mass-metallicity relation in Sect.~\ref{sec:mzr} and the N/O abundance ratio in Sect.~\ref{sec:NO_ratio}, while the [\ion{S}{ii}]-based relation is presented in App.~\ref{app:S2_met}. We discuss our results, including comparison with the local FMR, in Sect.~\ref{sec:discussion} and summarize our conclusions in Sect.~\ref{sec:summary}.

Throughout this work, we adopt a flat $\Lambda$CDM cosmology with $\Omega_\Lambda = 0.69$, $\Omega_M = 0.31$ and $H_0 = 67.7 \rm  \; km/s \; Mpc^{-1} $ \citep{Planck+2020} and a solar oxygen abundance of $\rm 12 + log(O/H) = 8.69$ \citep{Asplund+2009_sun}. All magnitudes are in the AB system \citep{Oke+Gunn1983}.

\section{Data}
\label{sec:data}
\subsection{NIRCam imaging and grism surveys}
\label{sec:phot}

We used imaging and wide field slitless spectroscopy (WFSS) from deep NIRCam grism surveys covering \ha{} and $[\ion{N}{ii}]$ emission lines in the redshift range $z = 3.8 - 6.6$. We collected data from the “All the Little Things" (ALT; PID 3516, \citealt{Naidu+2024_alt}), the JWST program ID 1933, hereafter “COLA1" (PIs Matthee \& Naidu), “Complete NIRCam Grism Redshift Survey" (CONGRESS; PID 3577, \citealt{Egami+2023jwst.prop, Sun+2026}), “Emission-line galaxies and Intergalactic Gas in the Epoch of Reionization" (EIGER; PID 1243, \citealt{Kashino+2023}) and  “First Reionization Epoch Spectroscopically Complete Observations" (FRESCO; PID 1985, \citealt{Oesch+2023_fresco}) surveys.

\renewcommand{\labelitemi}{$\bullet$} 
\begin{itemize}

     \item The ALT survey covers $\approx 30$ arcmin$^2$ around the lensing cluster Abell 2744 \citep[][$z=0.308$]{Abell+1989,Merten+2011_a2744} using deep NIRCam grism in the F356W ($\lambda = 3.15-3.96 \; \mu m$) filter. The ALT photometric catalog \citepalias{Naidu+2024_alt} was built combining all 20 NIRCam medium and broad bands \citep{Bezanson+2024_uncover, Suess+2024_megascience}, including ultra deep imaging in F070W and F090W taken simultaneously with the grism observations. Our observed fluxes and photometry are corrected for lensing magnification using the model from \cite{Furtak+2023}, last updated as described in \cite{Price+2025}.
     
    \item The COLA1 survey is centered on the bright double-peak Ly$\alpha$ emitter, COLA1, at $z\sim 6.6$ \citep[][]{Matthee+2018_cola1,Torralba+2024}. The area covered in the COLA1 field is 21 arcmin$^2$. It consists of NIRCam WFSS in the F356W filter, together with imaging in F115W, F150W, F200W and F356W filters in the same part of the sky.

    \item The CONGRESS survey covers 62 arcmin$^2$ in GOODS-North (GOODS-N) employing NIRCam WFSS in the F356W filter. The CONGRESS photometric catalog \citep[][]{Covelo-Paz+2025} \citep[see also][]{Robertson+2026} was built using all available HST \citep[][]{Giavalisco+2004, Grogin+2011, Koekemoer+2011} and JWST NIRCam \citep[][]{Williams+2023_jems, Eisenstein+2025_jof,Eisenstein+2026_jades} imaging from the MAST, covering a range of 0.4-5 $\mu$m in about 20 filters.

    \item The EIGER survey consists of six NIRCam WFSS and imaging mosaics targeting luminous quasars at $z>6$ \citep{Kashino+2023, Kashino2026}. Each quasar field has a survey area of $6.5 \times 4.3$ arcmin$^2$, yielding a total of 132 arcmin$^2$. Imaging data is uniformly available in the F115W, F200W and F356W filters. Archival HST/ACS imaging in F606W, F775W, F814W and F850LP filters is incorporated when available. 

    \item The FRESCO survey covers a total of 124 arcmin$^2$ in GOODS-N and GOODS-South (GOODS-S) using NIRCam/WFSS in the F444W ($\lambda=3.8-5 \; \mu m$) filter. As with the CONGRESS data in GOODS-N, the FRESCO fields are covered by HST and JWST/NIRCam photometry in about 20 filters \citep[see e.g.,][for details]{Covelo-Paz+2025}.
\end{itemize}

Together, these surveys span more than 200 hours of observation time with JWST and cover a volume of $\approx 8.5\times10^5$ cMpc$^{3}$ over the redshift range considered. The combination of relatively wide data from FRESCO with deep data on the Abell 2744 lensing cluster from ALT enables us to cover a large dynamic range in galaxies' star formation rates and stellar masses.

The ALT, CONGRESS and FRESCO photometric catalogs have been generated following the methods outlined in \cite{Weibel+2024}. Briefly, all images are matched to the point spread function (PSF) resolution in the F444W filter prior to the flux extraction, which has been performed with a circular aperture of radius $0.16"$. We apply an aperture correction based on fluxes scaled to the flux measured through Kron apertures on a PSF-matched version of the detection image. Then, it is scaled to a total flux by dividing by the encircled energy of the Kron ellipse on the F444W PSF. See \cite{Covelo-Paz+2025} and \citetalias{Naidu+2024_alt} for details on FRESCO and CONGRESS, and ALT surveys, respectively. As described in \cite{Kashino+2023, Kashino2026} and \cite{Torralba+2024}, the photometric catalogs for EIGER and COLA1 were generated by first convolving available SW images to match the resolution of the F356W data. Total photometry was measured using Kron apertures following \cite{Finkelstein+2022}, with the Kron scaling factor optimized to maximize the signal-to-noise ratio, with aperture corrections derived based on point-sources.

\subsection{NIRcam grism reduction}
\label{sec:grism_reduction}
The details of the basic reduction steps are described in \citetalias{Naidu+2024_alt} for ALT, \cite{Torralba+2024} for COLA1, \cite{Kashino+2023} for EIGER and \cite{Oesch+2023_fresco, Covelo-Paz+2025} for FRESCO and CONGRESS. Briefly, the WFSS data have been reduced with bad pixel identification, standard flat fielding and background subtraction using the {\texttt jwst} pipeline\footnote{\url{https://github.com/spacetelescope/jwst}}. We then separated the data in “continuum” and “emission-line spectra” by iteratively applying a median filter along the rows of individual grism exposures to estimate the continuum. Note that the continuum and emission-lines are both of galaxies themselves as well as from contamination (although most contamination is continuum emission). Spectra were then extracted for sources detected in the NIRCam imaging and the redshifts were identified based on the photometric redshift as well as the grism spectrum itself. 

Once we know the redshifts of the galaxies, we optimize the reduction to obtain better emission-line (EMLINE) spectra that minimize the effect of over-subtraction around emission lines, based on the method presented in \cite{Kotiwale+2026} \citep[see also][]{Matthee+2024_lrd}. The standard methodology leads to slight over-subtraction of faint wings of emission-lines that are not identified in individual spectra, but which may otherwise dominate stacked spectra and impact the line ratios that we aim to measure here. We re-extract spectra from the grism images before separating continuum and emission-line data and then remove the continuum using a similar running median filter, except we now explicitly mask the larger regions around the known positions of the emission-lines. We note that at this stage we also register the FRESCO and CONGRESS spectra (initially extracted per grism image using \texttt{grizli}; \citealt{Brammer+2019_grizli}) on the same wavelength grid as the 2D spectra extracted for the other surveys.

\subsection{Sample selection}
\begin{figure*}
   \centering
   \includegraphics[scale = 0.43]{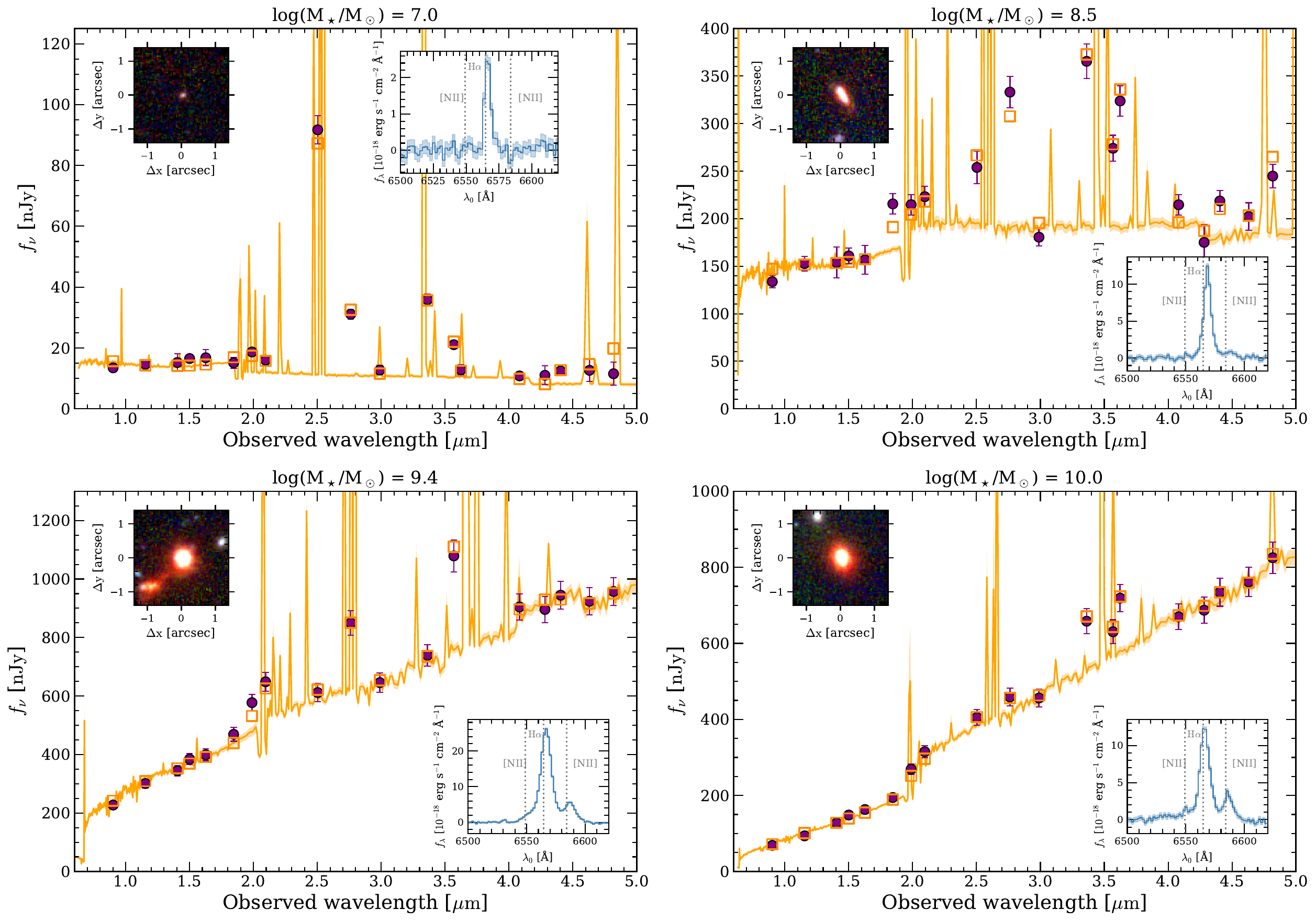}
   \caption{The observed properties of galaxies typical for stellar masses $\approx10^{7 - 10}$ M$_{\odot}$. The galaxies are selected from the ALT survey that includes the full suite of broad and medium-band NIRCam photometry. We show the \textsc{Bagpipes} SEDs for galaxies in different mass bins. The best-fit SED is in orange, and the corresponding flux-density in various filters are shown in orange squares. Photometry measurements in these filters are shown in purple. The inset panels show the RGB (F115W+F200W+F356W) image of $3\farcs0 \times3 \farcs0$ size that highlights the increasing size with mass, and the \ha{} and $[\rm \ion{N}{II}]$ emission line profiles that show the relative increase in the strength of $[\rm \ion{N}{II}]$.}
   \label{fig:spectraSED}
\end{figure*}

In this work, we focus on \ha{}-selected galaxies that are observed using NIRCam WFSS in the F356W and F444W filters, covering the $z \sim 3.7 - 6.6$ range. The parent sample ($N_{tot} = 2071$) is a composite of \ha{} emitters (HAE) from the CONGRESS, COLA1, EIGER, FRESCO and ALT surveys, which we homogenize applying the following quality checks. Since we are interested in star-forming galaxies, we removed galaxies with broad \ha{} components ($v_{\rm FWHM} > 1000$ km/s), $N_{\rm BLH\alpha} = 21$ \citep[see][]{Matthee+2024_lrd,Matthee+2024_env, Covelo-Paz+2025, Kapoor+2026_lrds}. For the remaining $N = 2050$ galaxies, we obtained the 2D grism spectra extracted at the source positions and collapsed them into 1D spectra using optimal extractions based on the source's morphology. We fit the \ha{} and $[\ion{N}{ii}]$ emission lines using a single-component Gaussian model. The three lines - \ha, \niib{} and \niir{} - share the same redshift and velocity dispersion, assuming that all three arise from the same kinematic component. The amplitude of \niir{} is parameterized as $N2H\alpha \cdot F(H\alpha)$, where $N2H\alpha$ is the \niiha{} flux ratio, and the amplitude of \niib{} is fixed to $1/2.94$ of \niir{} \citep{OsterbrockFerland2006}. The fit was performed using non-linear least squares from the \texttt{lmfit} python package. 

Based on the results of the 1D Gaussian fits to individual galaxies, we selected sources with \ha{} SNR $> 5$ and reduced $\chi_{red} ^2 < 5$. We then visually inspected the spectra to exclude galaxies whose \ha{} emission was contaminated by foreground or close-by sources. This process removed 302 galaxies, yielding a final sample of $N=1748$ galaxies spanning $z = 3.8 - 6.6$, with a median redshift of $ <z> = 4.6$. For $\sim6$\% of the sample (97 sources), we detect \niir{} with SNR > 5. These galaxies have stellar masses and star formation rates in the range $\rm log(M_\star/M_\odot) = 8.2-10.8$ and $\rm log(SFR/M_\odot \; yr^{-1}) = 0.7 - 2$, i.e. they populate the high-mass, high-SFR end of our sample. We estimate that their $\rm [\ion{N}{II}]/H\alpha$ ratio is in the range $\rm log([\ion{N}{II}]/H\alpha)= [-1.2 , -0.3]$. Additionally, we detected \siib{} or \siir{} with SNR > 5 in 16 galaxies ($\sim 1\%$ of the full sample), three of which have both lines above the SNR > 5 threshold. These galaxies span the range $\rm log(M_\star/M_\odot) = 8.6-10.4$ and $\rm log(SFR/M_\odot \; yr^{-1}) = 1.1 - 2.1$, with $\rm log([\ion{S}{II}]/H\alpha)$ in the range $[-1.0, -0.5]$. Notably, the fraction of [\ion{N}{II}]- and [\ion{S}{II}]-detected sources with $\rm SNR > 5$ is strongly mass-dependent, increasing sharply toward the high-mass end of the sample (see Sect.~\ref{sec:SFMS}).

\section{Methods}
\label{sec:methods}
\subsection{SED fitting}
\label{sec:sed}
The physical properties for all galaxies in our sample are estimated by fitting all available photometry (see Sect. \ref{sec:phot}) with \textsc{Bagpipes} \citep{Carnall+2018_bagpipes} SED fitting code. While fitting, the redshift values ($z_{spec}$) are fixed to those estimated from grism observations. We do not include emission line fluxes from grism spectra in the fit, and bands including and blueward of Ly$\alpha$ are masked. In this work, we use the default \textsc{Bagpipes} stellar population models, i.e. the \cite{BruzualCharlot2003} models, assuming a \cite{Kroupa2001} initial mass function and nebular emission is modeled with \texttt{Cloudy} \citep{Ferland+2017} grids following \cite{Byler+2017}. We adopt a non-parametric SFH \citep{Leja+2019} with time bins that are logarithmically spaced up to the age of the Universe at the highest redshift in our sample. The first two bins ($0-5$ Myr and $5-10$ Myr) are fixed, while the subsequent bins follow logarithmic spacing. This allows us to capture short-timescale bursts that power strong emission lines \citep[e.g.,][]{Whitler+2023}. Following \cite{Tacchella+2022}, we use a “bursty continuity” prior with Student's t-distribution parameters $\sigma =1.0$ and $\nu = 2.0$, which allows for a more variable SFH. Specifically, this choice relaxes the continuity conditions, present in the “continuity” prior SFH, enabling more dramatic changes in star formation rate between mass bins as predicted by high-resolution simulations \citep[e.g.,][]{Sun+2023_bursty} and shown by recent observations \citep[e.g.,][]{Endsley+2024, Perry+2025_burstySF, Clarke+2025}. Recently, \citet{Cochrane+2025_SEDfitting} tested the \textsc{Bagpipes} SED fitting code using synthetic JWST/NIRCam photometry generated from galaxies from a cosmological radiation-hydrodynamics simulation. They fit this photometry assuming six different SFH parameterizations and found that all models reproduce the observed photometry comparably well (see their Fig.~1), but that the recovered stellar masses show systematic, mass-dependent bias. Furthermore, they show that these biases can depend on the adopted SFH parameterization: in cases with strong nebular emission, more flexible SFHs better reproduce the photometry and recover more accurate stellar masses, while simpler SFHs can overestimate the stellar mass by fitting an older stellar population (see their Section~3.2). We adopt the \cite{Calzetti+2000} dust attenuation law, allowing the V-band attenuation ($A_V$) to vary between 0 and 5 mag, the stellar metallicity in the range $Z = 0.1 - 1\,Z_\odot$ and the ionization parameter in the range $\log(U) = -4$ to $-1$. Uniform priors are assumed for all parameters. We then derive the attenuation at \ha{} wavelength from the inferred $A_V$ using the same attenuation law, assuming equal attenuation of the nebular emission and the stellar continuum \citep[see also][]{Covelo-Paz+2025}. 

These assumptions introduce two potential sources of systematic uncertainty in our \ha{} dust corrections: the adopted attenuation curve in the SED fitting and the assumption of equal nebular and stellar attenuation. Recent JWST studies suggest that low-mass galaxies at $z\sim3-7$ have UV-optical attenuation curves that differ from the Calzetti law. In particular, \citep{Shivaei+2026}, using a sample of $\sim 3800$ galaxies from three JWST NIRCam/grism survey, find evidence for steeper attenuation curves, while \citet{Karthikeyan+2026}, based on NIRSpec observations from the JADES survey, find that an SMC-like attenuation curve is preferred for most galaxies at these redshifts. Adopting a different curve in the SED fitting would change the inferred $A_V$, as well as the stellar masses \citep{Reddy+2015}. Furthermore, \citet{Karthikeyan+2026} showed that nebular emission is, on average, more reddened than the stellar continuum at $z\sim4-5$ ($\Delta E(B-V)\simeq0.1$), with a smaller offset at $z>5$, while individual high-sSFR galaxies can show much larger differences \citep{Woodrum+2025}. 

To assess the robustness of our \ha{} dust corrections against these effects, we compared our median $A_{\rm H\alpha}$ per mass bin with estimates derived from the Balmer decrements of \citet{Karthikeyan+2026} for galaxies at comparable masses and redshifts \footnote{However, we note that \citet{Karthikeyan+2026} assume a \citet{Cardelli+1989} nebular extinction curve and Case B recombination.}. For $\rm log(M_\star/M_\odot)\geq8.5$, the two estimates agree within $\sim$0.1-0.2~mag, corresponding to $\lesssim0.1$~dex in (dust-corrected) SFR. 

Figure~\ref{fig:spectraSED} showcases \textsc{Bagpipes} SED fitting, together with RGB images and \ha{} and $\rm [\ion{N}{II}]$ emission lines for four galaxies at stellar masses spanning the range of our sample ($\rm log(M_\star/M_\odot) = 6 - 10.8$). These panels illustrate the mass-dependent variety among the galaxies in our sample. Low mass galaxies are typically compact young starbursts with little to none [\ion{N}{ii}] emission, a blue UV continuum and a Balmer jump. Increasingly massive galaxies have increasingly larger sizes \citep{Miller+2025}, stronger [\ion{N}{ii}] emission, stronger Balmer breaks and reddened colors indicative of more evolved stellar populations, higher metallicity and more dust attenuation.

\subsection{The star-forming main sequence}
\label{sec:SFMS}
In this section, we present the star-forming main sequence (SFMS) at $z \sim 5$ based on the galaxies in our sample. Stellar masses are obtained as output from the \textsc{Bagpipes} SED fitting code. We computed the SFR of galaxies based on the intrinsic \ha{} luminosity. The observed \ha{} fluxes were corrected for dust attenuation, estimated based on modeling the SEDs. Star formation on short timescales ($\sim 10$~Myr) is computed using the calibration

\begin{equation}
\label{eq:SFR(HA)_sphinx}
    \rm log \left( \frac{SFR}{M_\odot \; yr^{-1}} \right) = 1.06 \cdot log \left( \frac{L_{H\alpha}}{erg \; s^{-1}} \right) - 43.96, 
\end{equation}

\noindent that was recently proposed by \cite{Kramarenko+2025_sphinx}. The authors analyzed SPHINX$^{20}$ \citep{Rosdahl+2018_sphinx,Rosdahl+2022_sphix} simulations in the range $4.6 \leq z \leq 10$, selecting a sample of star-forming galaxies representative of the H$\alpha$ population observed with JWST at high-redshift to improve the SFR calibration. To assess the impact of the adopted calibration, we also computed SFRs using conversion factors commonly adopted in the literature for sub-solar metallicities \citep[see][]{Theios+2019}, as well as the metallicity-dependent calibration of \citet{sanders+2026_aurora_calib}. These alternative calibrations lower the normalization of the SFMS by $\sim 0.2$~dex, while leaving its slope unchanged. This shift is well within the uncertainties on the SFR measurements derived from the stacked samples and therefore does not affect our main conclusions, including those concerning the FMR (see Sect.~\ref{sec:fmr}).

\begin{figure}
   \centering
   \includegraphics[scale = 0.45]{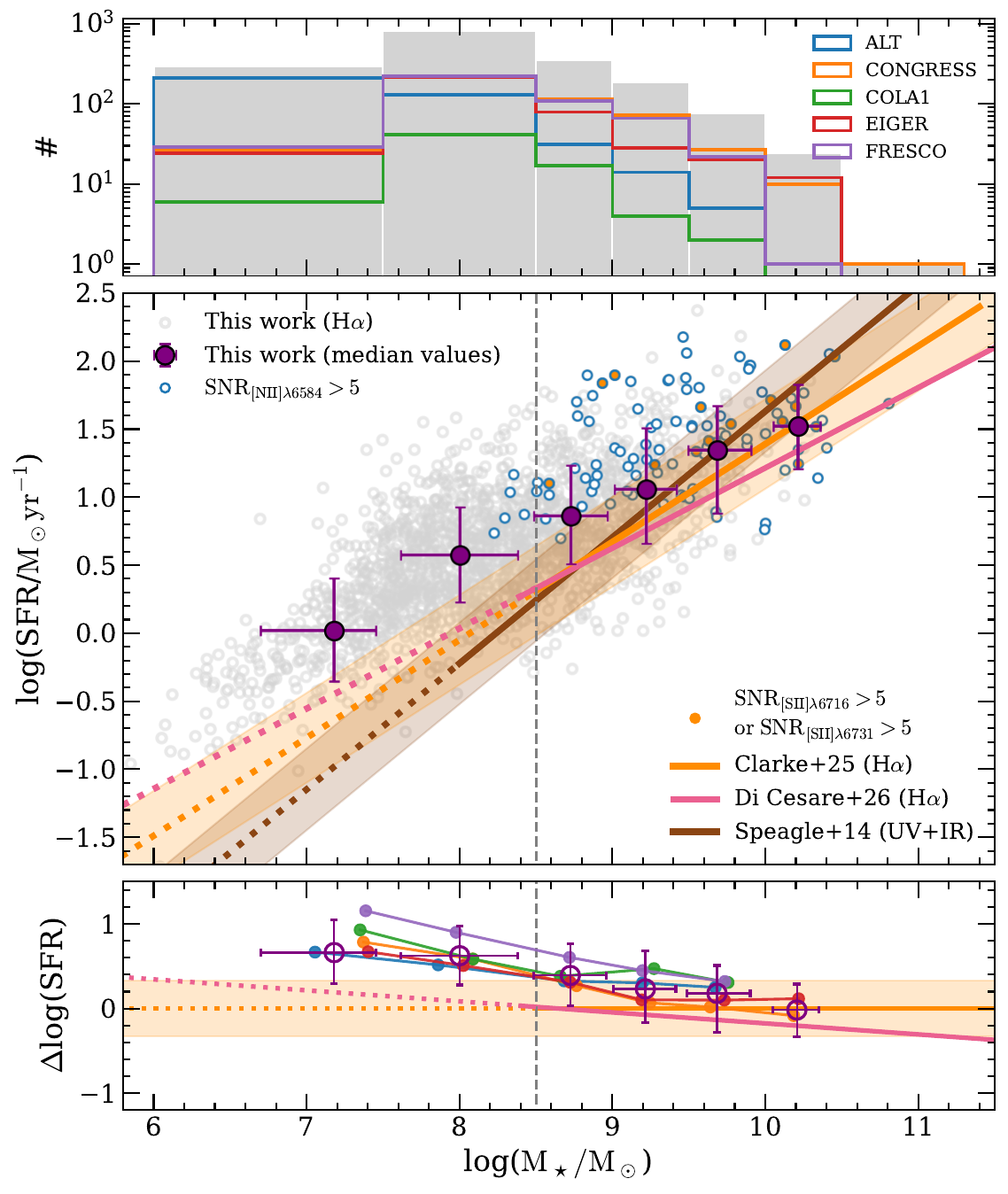}
   \caption{The star-forming main sequence at $z\sim5$. We show our sample (gray), together with median values and literature relations \citep[][]{Speagle+2014, Clarke+2025,DiCesare+2026}. Dotted lines indicate extrapolations of the literature relations below their respective lower stellar-mass limit. The vertical dashed line marks the stellar-mass completeness limit at $\rm log(M_\star/M_\odot) > 8.5$. Individual galaxies with \niir{} detected at SNR $>5$ are highlighted in blue, while those with \siib{} or \siir{} detected at SNR $> 5$ are highlighted in orange. The histogram shows the number of galaxies in each stellar-mass bin used for stacking (see Sect.~\ref{sec:mass_bins}). The bottom panel shows the offset in SFR from the SFMS of \cite{Clarke+2025}; the dotted line indicates where the relation is extrapolated and shaded band shows the intrinsic scatter. Filled circles show the median offset of each survey in the same stellar-mass bins as our stacks and open circles show our stacked measurements. The pink line shows the relation of \cite{DiCesare+2026} relative to that of \cite{Clarke+2025}.} 
   \label{fig:sfms}
\end{figure}

Figure~\ref{fig:sfms} shows the SFR-M$_\star$ plane probed by the galaxies analyzed in this work together with literature results. For individual galaxies, our sample covers stellar masses in the range $\rm log(M_\star/M_\odot) = 6 - 10.8$ and SFRs in $\rm -1.7 < log(SFR/M_\odot yr^{-1}) < 2.4$, at $z = 3.8 - 6.6$. We compare our SFMS with literature relations from \cite{Speagle+2014}, based on a compilation of UV+IR observations at $z\sim 0-5$, and with those of \cite{Clarke+2025} and \cite{DiCesare+2026} at $4 < z < 5$. The latter two are based on \ha{} emission line observations from NIRSpec and NIRCam grism data, respectively, thus probing SFR on $\sim 10$ Myr timescales. While \cite{Speagle+2014} report a slope $\approx 1$ for the SFMS, both \cite{Clarke+2025} and \cite{DiCesare+2026} find a shallower slope \citep[see also][]{Clarke+2024}. \cite{Clarke+2025} note that their $4 < z < 5$ bin lacks UV-bright objects at $\rm \sim 10^{10} \; M_\odot$ biasing the SFMS slope to lower values. Apart from the lack of high-mass galaxies, the emission-line flux limit of grism surveys, which favors galaxies with high SFRs at fixed stellar mass, may bias the observed sample and contribute to the shallower slopes found when studying scaling relations \citep[see e.g.,][]{DiCesare+2026, Kotiwale+2026}. \cite{Leja+2015} show that such a shallow SFMS slope would result in a steep low-mass end of the SMF, which is not observed \citep[e.g.,][]{Weibel+2024}. As discussed in \cite{DiCesare+2026}, their full sample is biased toward UV-bright, highly SF galaxies below $\sim 10^{8.5} \; \rm M_\odot$, although representative of the underlying population above this limit. Moreover, the offset with respect to the SFMS presented in \cite{DiCesare+2026} could be related to differences in the SED-fitting methodology\footnote{Note that they perform SED-fitting using the \texttt{Prospector} \citep{Johnson+2021} code.} and to the different calibration used to convert \ha{} luminosity into star formation rate - they adopted the \ha{}-SFR conversion from \cite{Theios+2019}. We also find a flattening in our sample (gray points in Fig.~\ref{fig:sfms}) for lower stellar masses driven primarily by the \ha{} flux limit of our flux-limited survey. Because low-mass galaxies span a wide range of luminosities and are intrinsically faint, obtaining mass-complete samples at these masses is difficult; selecting galaxies near the flux limit therefore biases the sample toward objects in a burst phase (Malmquist bias; see \citealt{Reddy+2012}). 

Following the mass completeness estimate of \cite{DiCesare+2026}, obtained by forward-modeling the stellar mass function of \cite{Weibel+2024}, we adopt a mass-completeness limit of $\rm M_\star \gtrsim 10^{8.5}\; M_\odot$ for star-forming galaxies. Since our sample shows a similar distribution in the SFMS (Fig.~\ref{fig:sfms}), we expect a comparable mass completeness, with our sample being representative of the underlying galaxy population at $\rm M_\star \gtrsim 10^{8.5}\;M_\odot$. The bottom panel of Fig.~\ref{fig:sfms} shows the offset in SFR from the SFMS of \cite{Clarke+2025} for each survey and our stacks. Below the mass completeness limit, all surveys lie above the SFMS, with our stacks offset by $\sim 0.6$~dex at $\rm log(M_\star/M_\odot) \lesssim 8$. The offset decreases with increasing stellar mass and our stacks become consistent with the SFMS within the intrinsic scatter at $\rm log(M_\star/M_\odot) \gtrsim 8.5$. This trend is common to all surveys therefore our stacks are not dominated by any individual dataset.

\subsection{Stacking}
\label{sec:stacking}

Although our sample contains galaxies with individual detections of the $\rm [\ion{N}{ii}]$ or $\rm [\ion{S}{ii}]$ lines, this is a biased sub-set of our full population and limited to the most massive and highly star-forming systems (see Fig.~\ref{fig:sfms}). In order to obtain sample-representative measurements, we obtain stacked spectra, which come with increased sensitivity but loss of information on the variation within the sub-sets that are stacked. We create median stacks of the 2D extracted emission-line spectra (Sect.~\ref{sec:grism_reduction}), after normalizing the spectrum of each galaxy with its \ha{} flux. We stack 2D spectra rather than 1D extracted spectra as this enables a better noise characterization and is less error-prone to the optimal extraction apertures defined for individual sources.

\begin{figure*}
   \centering
   \includegraphics[scale = 0.4]{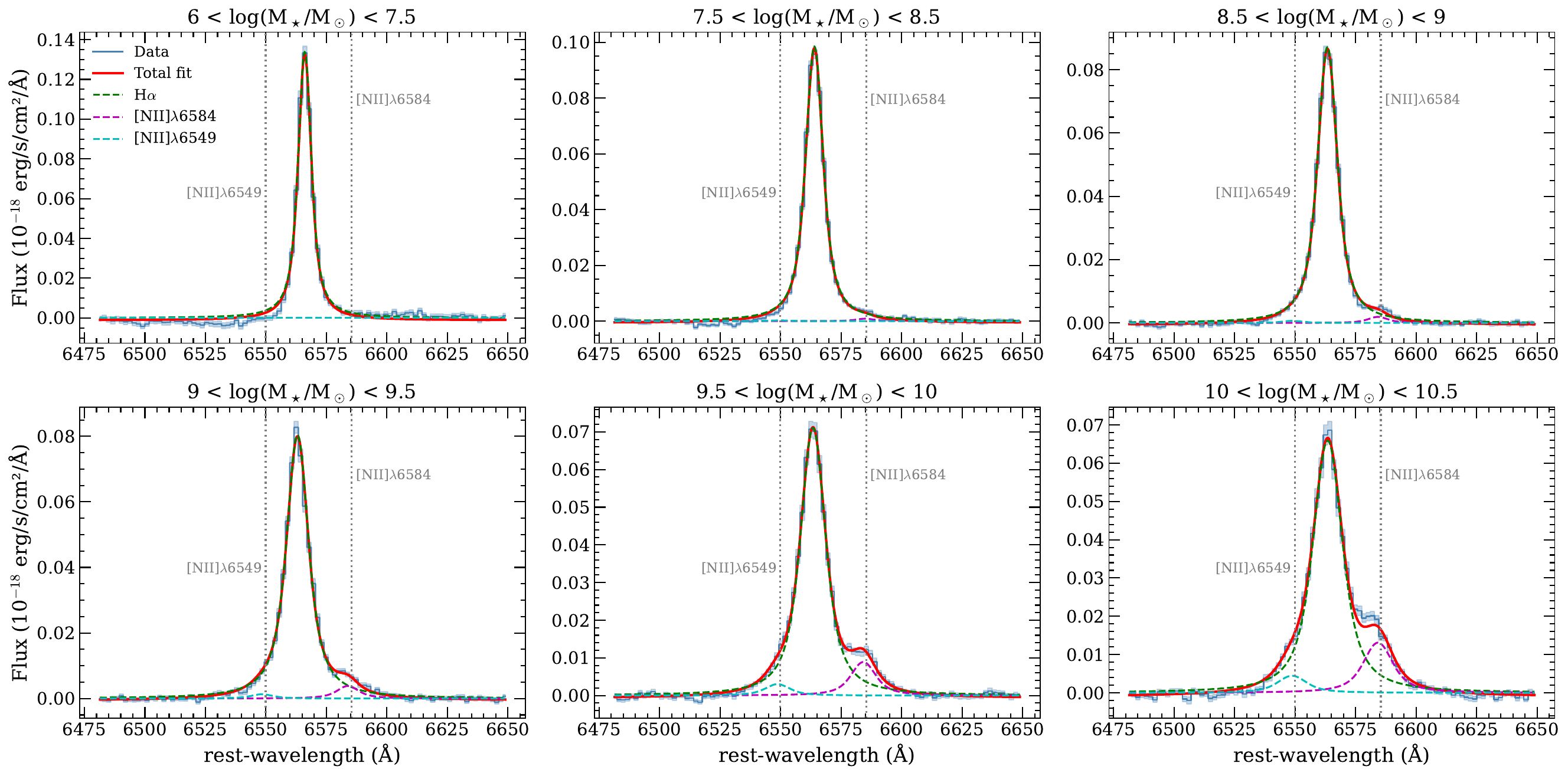}
   \caption{The \ha{} and $\rm [\ion{N}{ii}]$ emission lines in our stacked spectra in bins of stellar mass. Each panel shows the spectrum and the best-fit model in the annotated stellar mass bin. The fit is performed using a Voigt profile where the Lorentzian parameter $\gamma$ is fixed by the \ha{} fit in the spatial direction (see Sect. \ref{sec:stacking}). We show the observational data together with the associated errors in blue, the fitting components with colored dashed lines and the resulting fit in red solid lines. Dotted vertical lines highlight the position of \niib{} and \niir.}
   \label{fig:NII_stacks}
\end{figure*}

In grism spectra, the effective line-spread function is convolved with the spatial extent of the sources along the dispersion direction. For stacked spectra of large samples, it is reasonable to assume that any spatial extent of sources is randomly aligned with respect to the dispersion direction. This means that we can measure the spatial extent orthogonal to the dispersion direction and use this to estimate the spatial convolution kernel. In each of our stacks, we find that the spatial profile along this orthogonal direction is well described by a Lorentzian profile. Motivated by the fact that a Voigt profile is the convolution of a Gaussian and a Lorentzian, we therefore employ Voigt profiles when fitting the H$\alpha$, $\rm [\ion{N}{ii}]$ and $\rm [\ion{S}{ii}]$ line profiles of our stacked spectra, fixing the Lorentzian parameters to those from the spatial direction. The resulting fits accurately reproduce the observed line shapes (see Fig.~\ref{fig:NII_stacks} for the H$\alpha$ and  $\rm [\ion{N}{ii}]$ fits, and Appendix~\ref{app:SII_linefit} for the  $\rm [\ion{S}{ii}]$ fits). We use the package \texttt{lmfit} \citep{Newville+2014_lmfit} to find the best-fits for each stack, with a least-squares minimization. 

\subsection{Electron densities}
\label{sec:el_density}

Electron densities are usually derived using emission line doublets whose flux ratio depends on the occurrence of collisional excitations and de-excitations, which is a function of electron density \citep[e.g.][]{OsterbrockFerland2006}. Owing to their ionization potential ($\rm \sim 10 \; eV$), $\rm [\ion{O}{II}]\lambda\lambda 3726, 3729$ and $\rm [\ion{S}{II}]\lambda\lambda 6716, 6731$ trace the low-ionization gas, while lines such as $\rm \ion{C}{III}]\lambda\lambda 1907, 1909$  trace higher-ionization regions of the ISM, typically with higher electron densities \citep[e.g.][]{Berg+2021, Mingozzi+2022, Mainali+2023, topping+2025_aurora}. 

Thanks to the $R\sim1600$ point-source resolution of the NIRCam grism data, we are able to resolve the $[\ion{S}{II}]$ doublet, which is separated by $\sim 14 \; \AA$ (App.~\ref{app:SII_linefit}). We derive the electron density of the low-ionization gas in each stacked bin based on the observed \siib/\siir{} ratio using \texttt{pyneb} \citep{Luridiana+2015_pyneb} assuming an electron temperature for the ISM of $T_e = 15000\; K$, consistent with direct-$T_e$ measurements at high-redshift \citep[e.g.][]{Sanders+2024,Laseter+2024, Pollock+2026a}. To propagate the flux uncertainties, we draw 500 Gaussian realizations of the \siib{} and \siir{} fluxes for each bin, and convert each resulting ratio into $n_e$. Realizations with ratios beyond the range allowed by atomic physics ($0.44 < [\ion{S}{II}]\lambda6716/\lambda6731 < 1.43$, computed with \texttt{pyneb})
are assigned the critical density of \siib{} ($\sim1.9\times10^3\,\rm cm^{-3}$) when beyond the high-density limit and $10\,\rm cm^{-3}$ when beyond the low-density limit. Similarly to \cite{topping+2025_aurora}, we only consider bins in which the [\ion{S}{ii}] doublet is significantly detected, requiring both lines to be detected at $>5\sigma$, which excludes the two lowest-mass bins. 

In most bins, the $[\ion{S}{II}]$ ratio is close to the low-density limit, so that the individual densities are only weakly constrained. We therefore adopt as reference the median across bins, $n_e = 184^{+226}_{-169}\,\rm cm^{-3}$, with uncertainties derived from the Monte Carlo realizations, which is consistent with the low-density limit within $1\sigma$. This density is below the critical densities of the $[\ion{S}{II}]$ lines ($\sim2-5\times10^3\,\rm cm^{-3}$; computed with \texttt{pyneb} at $T_e = 15000$~K), the lowest among the lines used in this work. At this density, collisional de-excitation changes S2 and N2S2 by $\lesssim0.02$~dex relative to the low-density limit, as estimated by comparing the line emissivities computed with \texttt{pyneb} at $T_e = 15000$~K, so our metallicity and N/O diagnostics are not significantly affected.

Our median density is consistent with previous measurements of the low-ionization gas in star-forming galaxies at high redshift and, in particular, with those based on stacked spectra. \cite{Liu+2026} measured $[\ion{S}{II}]$-based densities from stacked JWST/NIRSpec spectra, finding $n_e = 124^{+37}_{-33}\,\rm cm^{-3}$ at $z\sim4.5$, in excellent agreement with our value. \cite{Stephenson+2026_ne} stacked NIRCam/grism samples of \ha{} emitters from FRESCO and CONGRESS, partially overlapping with our sample, and found median electron densities of $n_e([\ion{S}{ii}])= 463 ^{+281}_{-203} \,\rm cm^{-3}$ at $z\sim4.3$ and $301 ^{+388}_{-206} \,\rm cm^{-3}$ at $z\sim5.3$, consistent with our value within uncertainties, as well as $n_e([\ion{O}{ii}]) = 202^{+145}_{-102}\,\rm cm^{-3}$ from NIRSpec spectra of 12 galaxies at $z\sim5.7$. They also find that the median density deviates from a monotonic increase with redshift beyond $z\sim4$. \cite{topping+2025_aurora} found a median density of $n_e([\ion{S}{ii}]) = 480^{+390}_{-310}\,\rm cm^{-3}$ for individual galaxies at $z\sim5.3$, consistent with our measurement within the uncertainties, and derived a redshift evolution $n_e \propto (1+z)^{1.5\pm0.6}$, which would predict higher densities at our mean redshift. As also found by \cite{Liu+2026}, stacked spectra generally yield systematically lower densities than individual galaxies with high-S/N $[\ion{S}{II}]$ detections, possibly reflecting the selection of objects with bright $[\ion{S}{II}]$ emission. Finally, \cite{Isobe+2023} reported a median $[\ion{O}{II}]$-based density of $\sim300\,\rm cm^{-3}$ for 14 star-forming galaxies at $z\sim4-6$. However, \cite{Rong+2026_desi} showed that $[\ion{O}{II}]$-based densities are systematically higher than $[\ion{S}{II}]$-based ones, with a median offset of 0.23~dex in DESI DR1 galaxies at $z<0.5$ \footnote{$[\ion{O}{II}]$ also traces the intermediate-ionization zone, whereas $[\ion{S}{II}]$ traces only the low-ionization one \citep[see][]{Berg+2021}.}, so that the value of \cite{Isobe+2023} would correspond to $\sim175\,\rm cm^{-3}$ in terms of $[\ion{S}{II}]$, in agreement with our measurement.

\subsection{Strong line-based gas abundances}
\label{sec:derived_quantities}
The \niir{} emission line is commonly used as a metallicity diagnostic because of its proximity to the bright H$\alpha$ line, which implies that dust corrections are negligible \citep[e.g.,][]{Cardelli+1989}. The monotonic trend of the N2 diagnostics is useful to break degeneracies and avoid plateaus present in other strong-line calibrations \citep[see e.g.,][]{Sanders+2021_mosdef, curti+2020_calib, nakajima+2022_calib, Scholte+2025_excels, Cataldi+2025_calib, sanders+2026_aurora_calib}. A drawback of nitrogen based diagnostics is their sensitivity to variation in the nitrogen to oxygen abundance ratio (N/O). N/O increases with gas-phase metallicity and shows a large scatter at fixed O/H \citep[e.g.][]{Izotov+2006, Perez-Montero_Contini2009, Cataldi+2026_NOcalib}. Moreover, literature studies have reported nitrogen enhancement at fixed O/H in high-redshift galaxies \citep[e.g.,][]{Strom+2018,Cameron+2023_NO-enhancement, ArellanoCordova+2024, Zhu+2026_nitrogen}, although \cite{Schaerer+2026_no} recently found no evolution of the N/O -- O/H relation between $z\sim0$ to $z\sim3$ for typical star forming galaxies. If present, such an evolution would bias nitrogen-based calibrations derived at $z\sim 0$ when applied at high redshift \citep{Scholte+2025_excels}. To mitigate this issue, we adopt the N2 calibration of \cite{sanders+2026_aurora_calib}, derived from a sample of high-redshift galaxies, which accounts for the typical N/O of galaxies at these epochs.

Similarly to the N2 diagnostic, S2 exhibits a monotonic dependence on gas-phase metallicity over the range probed by our sample. Since sulfur, like oxygen, is an $\alpha$-element, its abundance is expected to broadly trace that of oxygen  \citep[e.g.][]{ArellanoCordova+2024, Stanton+2025, Esteban+2025}, so that biases arising from the evolution of abundance ratios should be less pronounced than for nitrogen-based diagnostics. However, recent JWST studies suggest that S/O may be sub-solar at high redshift and may vary with metallicity \citep{Rogers+2026, Cataldi+2026_chemical_abundances}. Although less widely adopted than oxygen-based diagnostics, S2 is increasingly used at high redshift, since sulfur lines are routinely detected in JWST spectra and, being close in wavelength to H$\alpha$, are largely insensitive to dust attenuation. For instance, \cite{Lewis+2026_mzr} combined S2 with R3 (defined as $\rm R3 = \log([\ion{O}{iii}]\lambda5007/H\beta)$) to break the degeneracy of the latter in a sample of galaxies at $z\sim3-6$ from the RUBIES survey . Using JWST/NIRSpec medium-resolution spectra from the EXCELS survey \citep{Carnall+2024_excels}, \cite{Scholte+2025_excels} found that S2 yields metallicities consistent with those derived from R2 (defined as $\rm R2 = \log([\ion{O}{ii}]\lambda\lambda3726,3729/H\beta)$), with a strong correlation between the two diagnostics. Following the procedure adopted for N2, we estimate the S2-based metallicity using \cite{sanders+2026_aurora_calib} calibrations. In Appendix \ref{app:calib}, we compare the metallicity estimates using calibrations from \cite{curti+2020_calib}, \cite{nakajima+2022_calib}, \cite{sanders+2026_aurora_calib} and \cite{Isobe+2026}.

To estimate the N/O ratio for our stacked sample, we adopted the high-redshift calibration proposed by \cite{Cataldi+2026_NOcalib}. This calibration is based on a sample of $\sim 660$ star forming galaxies observed with JWST/NIRSpec at a median redshift of $ \langle z \rangle = 3$. The detection of auroral emission lines in the NIRSpec spectra allowed \cite{Cataldi+2026_NOcalib} to derive direct estimates of nitrogen abundance. By comparing these measurements with the N2S2 strong-line ratio, they established an empirical calibration linking N2S2 to the N/O abundance ratio. Their best-fit relation is:

\begin{equation}
    \rm log(N/O) = (0.80 \pm 0.16) \cdot N2S2- (0.89 \pm 0.08)
\end{equation}
\noindent where $\rm N2S2 = log([\ion{N}{ii}]\lambda6584 / [\ion{S}{ii}]\,\lambda\lambda6716,6731)$ and with an intrinsic scatter of 0.1 dex.

The uncertainty on log(N/O) is estimated by summing in quadrature three independent sources of error: the measurement uncertainty on the N2S2, uncertainties on the calibration coefficients and the intrinsic scatter of the calibration. See Sect. \ref{sec:NO_ratio} for the results on the N/O abundance ratio.

\section{Results}
\subsection{Emission-line ratios of [NII], [SII] and H$\alpha$}
\label{sec:mass_bins}
\begin{figure*}
   \centering
   \includegraphics[scale = 0.48]{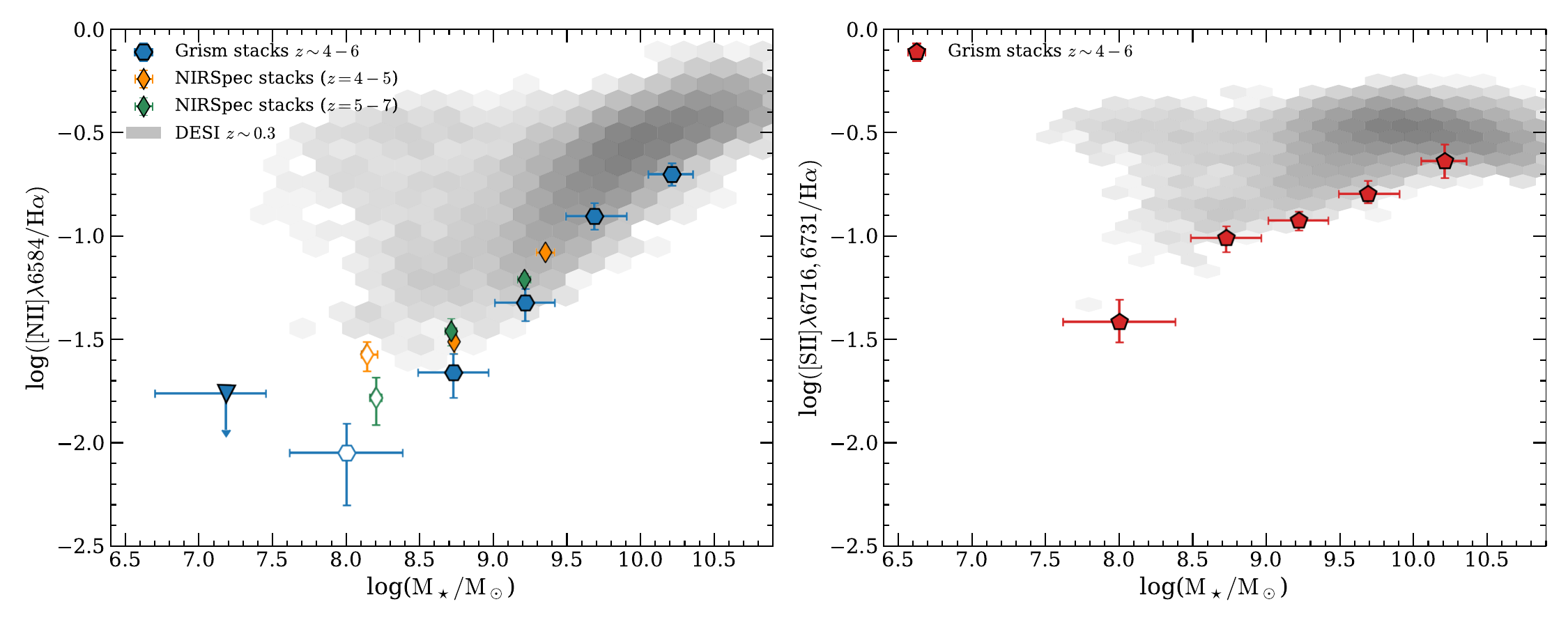}
   \caption{The \niiha{} (left) and \siiha{} (right) ratios as a function of stellar mass. We compare our sample to literature measurements at similar redshift and the $z\sim0.3$ baseline. \textbf{Left:}
   Blue hexagons show the stacked \niiha{} ratios in each mass bin. Filled markers indicate \niir{} detected at SNR$> 3$ in the stacks, while the open marker corresponds to a marginal detection with SNR$\sim 2$. For the lowest-mass bin we report a $2\sigma$ upper limit. Orange and green diamonds show the JWST/NIRSpec stacks at $z=4-5$ and $z=5-7$, respectively, from \cite{Lam+2026}, with open symbols highlighting values outside their representative mass range. \textbf{Right:} Red pentagons indicate the stacks with both \siir{} and \siib{} detected at SNR $>3$ in each mass bin. The only exception is the lowest-mass bin, where the \siir{} line has an SNR of $\sim2.6$. In gray we show star forming galaxies at $z\sim0.3$ from DESI DR1 \citep{DESI2026}.}
   \label{fig:n2ha_s2ha}
\end{figure*}

In this section, we present the results of our analysis in bins of stellar mass to investigate the trends between the N2 and S2 strong line calibrators and stellar masses. The stellar mass bins span the range $\rm log(M_\star/M_\odot) = 6 - 10.5$, with widths that vary according to the number of galaxies. Because higher-mass galaxies generally produce stronger emission lines, robust detections can be obtained with fewer objects, enabling our use of 0.5 dex bins. This fine mass binning enables us to trace the behavior of the relations with improved resolution. Table~\ref{tab:n2-s2} summarizes the sample properties in each stellar mass bin. The \niiha{} and \siiha{} ratios in each bin are derived using the emission-line fitting procedure described in Sect.~\ref{sec:stacking}. Figure~\ref{fig:n2ha_s2ha} shows the stacked \niiha{} and \siiha{} ratios for our high-redshift galaxy sample as a function of stellar mass. Furthermore, we show measurements from the DESI DR1 \citep{DESI2026}\footnote{\url{https://data.desi.lbl.gov/doc/releases/dr1/vac/stellar-mass-emline/}} for comparison. Star-forming galaxies at $z\sim0.3$ are identified using the $\rm [\ion{N}{ii}]/H\alpha$ vs. $\rm [\ion{O}{iii}]/H\beta$ and $\rm [\ion{S}{ii}]/H\alpha$ vs. $\rm [\ion{O}{iii}]/H\beta$ BPT-diagrams, following the starburst definition from \cite{Kewley+2001}. We selected DESI galaxies with SNR $\geq 3$ for $\rm [\ion{O}{iii}]\lambda5007$, H$\beta$, \ha, \niir, \siib{} and \siir, and with stellar masses and star formation rates comparable to those of our high-redshift sample.

Figure \ref{fig:n2ha_s2ha} shows an increasing trend for the N2 and S2 diagnostics with stellar mass. In the mass range $\rm log(M_\star/M_\odot) = 8.5-9.5$ where the samples overlap, the N2 ratio from our stacked sample is in agreement with the recent results of \cite{Lam+2026} based on stacked JWST/NIRSpec spectra of galaxies at $z=4-7$ from the JADES survey \citep{Eisenstein+2026_jades, deugenio+2025_jadesDR3}. At lower stellar masses, however, our N2 measurements tend to lie below their relation. This discrepancy may be partly related to the faintness of the \niir{} line in low-mass, highly star-forming galaxies \citep[see e.g.][]{Shapley+2023_neblines, Sanders+2023}, which makes the measurement particularly sensitive to the stacking procedure and sample selection. In addition, our \ha{}-selected galaxies lie above the SFMS at low masses and, if the FMR is in place, galaxies with enhanced SFRs at fixed mass are expected to have lower metallicities \citep[e.g.,][]{Lam+2026}. At higher masses, our stacks extend to median stellar masses of $\rm log(M_\star/M_\odot)\sim10.2$, around $\sim0.8-1$~dex higher than the highest-mass stacks of the JADES sample at these redshifts.

\begin{table*}[h]
\centering
\renewcommand{\arraystretch}{1.5}
\caption{Physical properties and line-diagnostics from the stacked spectra at $z\sim5$ in each mass bin.}
\begin{tabular}{c|c|c|c|c|c|c|c|c}  
\hline
\makecell{Mass bin} & \makecell{$ N$} & \makecell{$\rm log(M_\star)$} & \makecell{$\rm log(SFR)$}  & \makecell{N2} & \makecell{S2} & \makecell{$\rm 12+log(O/H)$} & \makecell{$\rm 12+log(O/H)$} & \makecell{$\rm log(N/O)$}  \\
\hline
 & & $\rm [M_\odot]$ & $\rm [M_\odot \; yr^{-1}]$ & & & N2-based & S2-based & \\
\hline
$6-7.5$& 295 & $7.2 ^{+0.3}_{-0.5}$ & $0.02 ^{+0.39}_{-0.37}$ & $-1.9$ & $-$ & 7.66 & $-$ & $-$ \\

$7.5-8.5$& 817 & $8.0 ^{+0.4}_{-0.4}$ &  $0.58 ^{+0.35}_{-0.35}$ &  $-2.05 ^{+0.14}_{-0.25}$  & $-1.41 ^{+0.11}_{-0.10}$ & $7.55 ^{+0.24}_{-0.22}$ & $7.62 ^{+0.36}_{-0.35}$ & $-1.40 \pm 0.24$ \\

$8.5-9$& 350 & $8.7 ^{+0.2}_{-0.2}$ & $0.86 ^{+0.37}_{-0.35}$ & $-1.66 ^{+0.09}_{-0.12}$ & $-1.01 ^{+0.06}_{-0.07}$ & $7.80 ^{+0.21}_{-0.22}$ & $8.18 ^{+0.34}_{-0.34}$ & $-1.41 \pm 0.19$ \\

$9-9.5$& 184 & $9.2 ^{+0.2}_{-0.2}$ &   $1.06 ^{+0.45}_{-0.40}$ & $-1.32 ^{+0.07}_{-0.09}$  & $-0.92 ^{+0.04}_{-0.05}$ & $8.02 ^{+0.21}_{-0.21}$ & $8.30 ^{+0.34}_{-0.34}$ & $-1.21 \pm 0.16$ \\

$9.5-10$& 76 & $9.7 ^{+0.2}_{-0.2}$ &   $1.34 ^{+0.33}_{-0.47}$ & $-0.90 ^{+0.06}_{-0.07}$  & $-0.79 ^{+0.06}_{-0.04}$ & $8.30 ^{+0.20}_{-0.21}$ & $8.48 ^{+0.34}_{-0.34}$ & $-0.98 \pm 0.15$ \\

$10-10.5$& 24 & $10.2 ^{+0.1}_{-0.2}$ &   $1.53 ^{+0.31}_{-0.31}$ & $-0.70 ^{+0.06}_{-0.06}$ & $-0.64 ^{+0.08}_{-0.08}$ & $8.43 ^{+0.20}_{-0.20}$ & $8.70 ^{+0.34}_{-0.35}$ & $-0.94 \pm 0.15$ \\
\hline
\end{tabular} 
\tablefoot{We list the mass bins, the number of galaxies in each bin and the median values for the stellar masses and SFRs with the errors corresponding to the 16th and 84th percentiles. We report the N2 (N2$=\rm log$(\niiha)) and S2 (S2$=\rm log$(\siiha)) diagnostics estimated from line-fitting as described in Sect.~\ref{sec:stacking}, the N2 value in the lowest mass bin has to be interpreted as a $2\sigma$ upper limit. For N2 and S2, we report the median and the 16th/84th percentiles as the best-fit value and its uncertainty, respectively. We list the oxygen abundance estimates based on the N2 and S2 diagnostics (Sect.~\ref{sec:derived_quantities}). The $\rm 12 + log(O/H)$ derived from the N2 diagnostic in the lower mass bin has to be interpreted as a $2\sigma$ upper limit. Finally, we list N/O abundances based on the N2S2 diagnostic from \cite{Cataldi+2026_NOcalib} high-redshift calibration. The $\rm log(N/O)$ uncertainties accounts for measurement uncertainties, uncertainties on the calibration coefficients and the intrinsic scatter of the calibration.}
\label{tab:n2-s2}
\end{table*}

\subsection{The mass-metallicity relation}
\label{sec:mzr}

\begin{figure*}
   \centering
   \includegraphics[scale = 0.47]{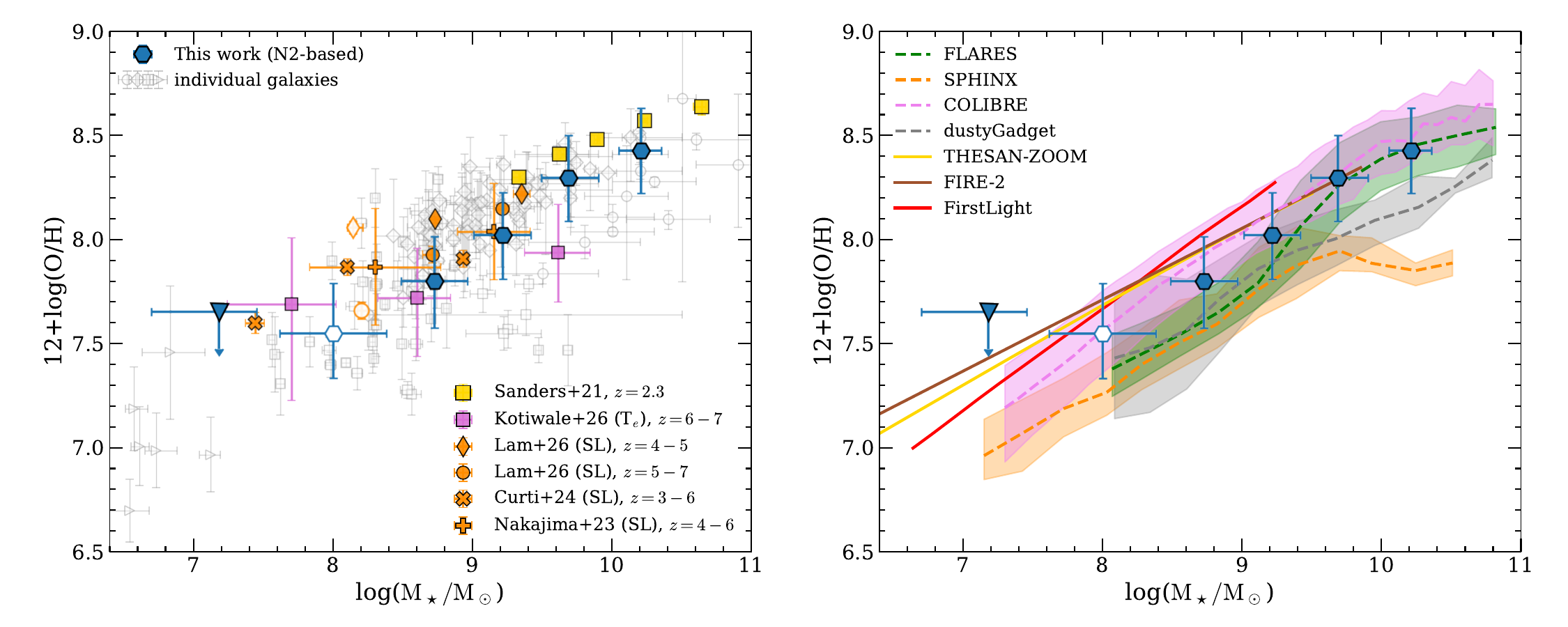}
   \caption{The mass-metallicity relation inferred from our measurements, comparison with observations and simulations. Blue hexagons show the metallicity estimates derived from the stacked galaxy sample in each mass bin using the N2-based diagnostic, the uncertainties account for measurement uncertainties and calibration systematics reported in \cite{sanders+2026_aurora_calib}. The open symbols corresponds to the gas-phase metallicity based on the \niir{} marginal detection with SNR$\sim 2$. \textbf{Left:} Together with our estimates, we report metallicity measurements for individual galaxies obtained using JWST/NIRSpec (gray open symbols). In particular, we show data from \cite{Chemerynska+2024_mzr} (triangles), \cite{Chakraborty+2025} (squares), \cite{Stanton+2026_excels} (diamonds) and \cite{faisst+2026_ifu} (circles). We also show results from stacked NIRCam/grism \citep[][pink]{Kotiwale+2026} and NIRSpec spectra \citep[][orange]{Lam+2026}, and average metallicity estimates \citep{Curti+2024_mzr, nakajima+2022_calib} obtained from NIRSpec spectroscopy (orange). Yellow squares show the stacked measurements at $z\sim2.3$ from \cite{Sanders+2021_mosdef}. \textbf{Right:} Comparison between our metallicity estimates and predictions from hydrodynamical simulations. Dashed lines show median values - with 16th and 84th percentiles - for SPHINX$^{20}$ \citep{Rosdahl+2018_sphinx, Katz+2023_sphinx} at $z = 4.64-6$, FLARES \citep{lovell+2021_FLARES, Vijayan+2021_FLARES} at $z = 5-7$, COLIBRE \citep{Schaye+2026_COLIBRE, Sharda+2026_mzrCOLIBRE} at $z = 5$ and \texttt{dustyGadget} \citep{Graziani+2020_dustyG, DiCesare+2023, DAngelo+2026} at $z=4.5$. Solid lines show the best-fit results from THESAN-ZOOM  \citep{Kannan+2025_thesanzoom, McClymont+2026_mzrTHESAN}, FIRE-2 \citep{Ma+2018_FIRE2sr, Marszewski+2024_mzrFIRE2} and FirstLight \citep{Ceverino+2017_firstlight,Langan+2020_firstlight} at $z = 5$.}
   \label{fig:mzr}
\end{figure*}

In this section, we present the mass-metallicity relation at $z = 3.8 - 6.6$. We estimate the metallicity of our stacked sample from the \niiha{} ratio, as described in Sect.~\ref{sec:mass_bins}, using the N2 strong-line calibration of \cite{sanders+2026_aurora_calib}. This calibration is based on a sample of 68 high-redshift star forming galaxies with auroral line detections and direct-method oxygen abundance measurements \footnote{\cite{sanders+2026_aurora_calib} report a validity range for the N2 calibration ($ 7.8 < \rm 12+log(O/H) < 8.6$). While we also show metallicity estimates outside this range for completeness, all the measurements used for the main results and conclusions of this work fall within the calibration's validity range.}. In App.~\ref{app:calib}, we compare the MZR relation obtained from different literature calibrations. Appendix~\ref{app:S2_met} shows the mass-metallicity relation using \siiha{} measurements, i.e. S2-based strong line calibrations. We report the gas-phase metallicity and stellar mass estimates of our sample in Tab.~\ref{tab:n2-s2}. 

Figure \ref{fig:mzr} presents our metallicity measurements, together with the $2\sigma$-upper limit in the lowest-mass bin and compares them with recent JWST results for both individual galaxies and stacked samples, as well as predictions from numerical simulations. Owing to the large statistical sample employed in this work, which spans a broad dynamic range in stellar mass, we extend the mass range probed by stacked analyses beyond that of previous studies at both the high- and low-mass ends. We caution, however, that below our mass completeness limit ($\rm log(M_\star/M_\odot) \sim 8.5$), our \ha{}-selected stacks lie above the H$\alpha$-based SFMS of \cite{Clarke+2025} and \cite{DiCesare+2026} by $\sim0.6$~dex (Sect.~\ref{sec:SFMS}) and are thus biased towards high sSFR. If the FMR is in place, this may bias our metallicities low at fixed stellar mass in this regime. At higher masses, our stacks are consistent with the SFMS. We provide metallicity measurements at the high-mass end ($\rm \log(M_\star/M_\odot) > 9.6$) and probe the low-mass regime ($\rm \log(M_\star/M_\odot) \lesssim 8$) through tentative metallicity measurements based on \niiha{} detection with SNR$\sim 2$, as well as upper limits. The metallicity implied by our estimates suggests that the oxygen abundance varies between $\rm 12+log(O/H)= 7.6 - 8.4$ (i.e. $0.1-0.50 \; Z_\odot$) at $z\sim5$.

Metallicity estimates for individual galaxies rely on JWST/NIRSpec observations using a variety of methods. \cite{Chakraborty+2025} report metallicity for galaxies at $z = 3 - 10$ based on detections of the $[\ion{O}{iii}]\,\lambda4363$ auroral line. \cite{Stanton+2026_excels} estimate metallicity values using strong line calibrations, from oxygen- and neon-based diagnostics, for galaxies in the JWST/EXCELS survey at $z =4 -8$. At $ z = 6-8$, \cite{Chemerynska+2024_mzr} derive metallicity for low-mass galaxies ($\rm log(M_\star/M_\odot) < 8$) using strong-line calibrators based on oxygen emission lines \footnote{The observations reported by \cite{Chemerynska+2024_mzr} are part of the UNCOVER survey \citep{Bezanson+2024_uncover} which targets the region behind the Abell2744 lensing cluster, enabling the detection of intrinsically faint galaxies through gravitational magnification.}. Finally, \cite{faisst+2026_ifu} report metallicity measurements for massive galaxies ($\rm log(M_\star/M_\odot) > 9.5$) at $z \sim 5$ from the ALPINE-CRISTAL-JWST survey, based on strong-line diagnostics. 

We find good overall agreement between our stacked estimates and these individual galaxy measurements across the explored stellar mass range. At higher masses, our estimates lie toward the upper end of the parameter space explored by ALPINE-CRISTAL-JWST, which currently provides the only sample probing this stellar mass regime, but which spans a diverse selection function.

Stacked and median measurements provide a more direct comparison with our results, as they also trace the average properties of galaxy populations. Fig.~\ref{fig:mzr} includes stacked measurements from \cite{Kotiwale+2026} and \cite{Lam+2026}, who use direct-$T_e$ and strong-line calibrations, respectively, to estimate the metallicity of galaxies observed with JWST/NIRCam grism and NIRSpec. We also report the average metallicity measurements obtained from NIRSpec spectroscopy at $z=3-6$ \citep[via strong-line diagnostics;][]{Curti+2024_mzr} and $z=4-6$ \citep[via direct-$T_e$ method and strong-line diagnostics;][]{nakajima+2023_mzr} as well as the stacked measurements at $z \sim 2.3$ from the MOSDEF survey \citep{Sanders+2021_mosdef}. We find good agreement with the measurements of \cite{nakajima+2023_mzr}, \cite{Curti+2024_mzr} and \cite{Lam+2026} at $4 < z < 6$ over $\rm M_\star\sim 4\cdot10^{8} - 3\cdot10^{9} \; \rm M_\odot$. In contrast, we find a lower metallicity value in the mass bin at $\sim 10^8 \; \rm M_\odot$, which may be a consequence of i) the higher sSFR at this fixed stellar mass (see Fig. \ref{fig:sfms}) which results in biased low-metallicity estimates (see discussion in \citealt{Lam+2026}); ii) the faintness of the \niir{} line for low-metallicity, low-mass galaxies \citep[see e.g. ][]{Shapley+2023_neblines, Sanders+2023}. At stellar masses $\rm M_\star > 3\cdot10^{9} \; M_\odot$, our metallicity measurements lie above those of \cite{Kotiwale+2026}, which probe $z >6$ galaxies using the $T_e$-based metallicity method. Interestingly, at these high stellar masses, our measurements are consistent with the $z \sim 2.3$ stacks of \cite{Sanders+2021_mosdef}, which reach $\rm 12 + log(O/H) \sim 8.6$ at $\rm M_\star \sim 4\cdot10^{10} \; M_\odot$. This suggests little evolution of the high-mass end of the MZR from $z\sim2$ to $z \sim 5$, in contrast to the lower-mass regime; such a weak evolution of the high-mass end is also predicted by the COLIBRE simulations \citep{Sharda+2026_mzrCOLIBRE}. We further discuss the comparison with literature works in Sect.~\ref{sec:comp_literature}.

We parameterize the MZR using a linear relation between $\rm 12 + log(O/H)$ and $\rm log(M_\star/M_\odot)$. Since the median mass value of our stacked sample is lower than the typical MZR turnover mass (log($\rm M_\star/M_\odot$) = 10), we use $\rm M_\star = 10^9 \; M_\odot$ as the pivot mass. The relation is

\begin{equation}
\label{eq:mzr_fit}
    \rm 12 + log(O/H) = \gamma \cdot log \left( \frac{M_\star}{10^{9} \; M_\odot} \right) + Z_{m9}
\end{equation}
\noindent where $\rm Z_{m9}$ is the normalization at $\rm M_\star = 10^{9} \; M_\odot$ and $\gamma$ is the slope. The fit is performed using orthogonal distance regression (ODR, \texttt{scipy.odr}) on the five stacked data points, which accounts for uncertainties in both the stellar mass and gas-phase metallicity of each stack. Following the mass completeness discussion of \cite{DiCesare+2026}, we restrict the fit to galaxies with stellar masses $\rm M_\star \geq 10^{8.5} \;M_\odot$, but we extrapolate the best-fitting relation to lower masses (see Sect.~\ref{sec:comp_literature}). Owing to the relatively large measurement uncertainties on the metallicity estimates at $\rm M_\star \sim 10^8 \; M_\odot$, including this value has a negligible impact on the inferred slope and normalization. To propagate the uncertainties, we employ a Monte Carlo approach: we draw 500 realizations of each data point by sampling from the asymmetric error distributions, run ODR on each realization, and take the median of the resulting parameter distributions as the best-fit values. The 16th and 84th percentiles of the distributions define the uncertainties on slope and normalization. From the fit, we find $\rm Z_{m9} = 7.94 ^{+0.06}_{-0.09}$ and $\gamma=0.42 ^{+0.10}_{-0.07}$, we report our best-fitting values for the MZR based on N2- and S2-diagnostics in Tab.~\ref{tab:MZRfit}.
 
\begin{table}
    \centering
    \renewcommand{\arraystretch}{1.5}
    \caption{Best-fit parameters for the slope and normalization at $\rm M_\star = 10^9 \; M_\odot$ of the mass-metallicity relation from Eq. \ref{eq:mzr_fit}. }
    \begin{tabular}{c|c|c}
    \hline
     & \makecell{$ \gamma$} & \makecell{$\rm Z_{m9}$} \\
     \hline
     \hline
         N2-based & $0.42 ^{+0.10}_{-0.07}$   & $7.94^{+0.06}_{-0.09}$ \\
         S2-based & $0.35^{+0.13}_{-0.10}$  & $8.25^{+0.06}_{-0.09}$\\
    \hline
    \end{tabular}
    \tablefoot{We report the values for both N2- and S2-based MZR. The fit is performed on the stacks with $\rm log(M_\star/M_\odot) \geq 8.5$, above our mass completeness limit (see Sect.~\ref{sec:SFMS}). The fitting procedure is described in Sect.~\ref{sec:mzr}.}
    \label{tab:MZRfit}
\end{table}

In the right panel of Fig. \ref{fig:mzr} we compare the mass-metallicity relation from our stacked sample to that predicted using FLARES \citep{lovell+2021_FLARES, Vijayan+2021_FLARES}, SPHINX$^{20}$ \citep{Rosdahl+2018_sphinx, Rosdahl+2022_sphix, Katz+2023_sphinx}, COLIBRE \citep{Schaye+2026_COLIBRE, Sharda+2026_mzrCOLIBRE}, \texttt{dustyGadget} \citep{Graziani+2020_dustyG, DiCesare+2023}, THESAN-ZOOM \citep{Kannan+2025_thesanzoom, McClymont+2026_mzrTHESAN}, FIRE-2 \citep{Ma+2018_FIRE2sr, Marszewski+2024_mzrFIRE2} and FirstLight \citep{Ceverino+2017_firstlight, Langan+2020_firstlight} hydrodynamical simulations. Among these simulations, FLARES, THESAN-ZOOM, FIRE-2 and FirstLight are cosmological zoom-in simulations that offer improved mass and spatial resolution, at the cost of smaller sample sizes. We used mass-weighted gas-phase metallicity estimates from all simulations and show their best-fitting mass metallicity relation or median values at $z \sim 5$. Despite the heterogeneity of the simulation schemes and the derivation of galaxy properties, this compilation provides a useful benchmark for comparing theoretical predictions with our observational results. A description of the aspects of the simulations relevant to this work is provided in App.~\ref{app:sims}.

The average metallicity increases with stellar mass in all simulations. Figure \ref{fig:mzr} shows that zoom-in simulations, such as THESAN-ZOOM, FIRE-2, and FirstLight, probe a lower-mass range extending to $\rm M_\star < 10^7\;M_\odot$, whereas cosmological simulations sample higher stellar masses, up to $\rm M_\star > 10^{10}\;M_\odot$. We find good agreement over the full mass range, both in normalization and slope, between our observational results and the predictions from the COLIBRE simulation. Unlike \texttt{dustyGadget} and SPHINX$^{20}$, COLIBRE does not impose a pressure floor and instead explicitly models the cold ISM down to $T=10$ K, allowing it to resolve the cold, dense gas phase in which metal enrichment and mixing occurs \citep{Sharda+2026_mzrCOLIBRE}. This suggests that explicitly modeling the cold ISM, rather than imposing an effective pressure floor, may be important for reproducing the physical conditions that govern chemical enrichment at high redshift. The zoom-in simulations agree well with our data for $\rm M_\star < 10^9\;M_\odot$, suggesting that their more explicit implementation of bursty star formation, metal transport and feedback better captures enrichment processes in low-mass galaxies. FLARES, by contrast, reproduces the high-mass end of the MZR, but we lie systematically above its predictions at the low-mass end. Its S-shaped trend suggests a transition in the feedback mechanisms regulating the slope of the MZR around $\rm M_\star \sim 10^9\,M_\odot$, consistent with the $z\sim0$ DESI results of \citet{Scholte+2024}. Both \texttt{dustyGadget} and SPHINX reproduce a slope similar to the observations but with systematically lower normalization.

The broad agreement among simulations with different feedback implementations at low and intermediate stellar masses reflects the degeneracy of the MZR, which can be reproduced by different combinations of inflows, outflows and star formation efficiencies. The MZR is therefore a necessary, but not sufficient, test of galaxy formation models and its constraining power is greatest in regimes where the predictions diverge, such as the high-mass end. It is precisely this mass range where our data add new observational constraints compared to previous results (left panel, Fig. $\ref{fig:mzr}$). At $\rm log(M_\star/M_\odot) \gtrsim 9.5$, only FLARES, SPHINX, \texttt{dustyGadget} and COLIBRE extend their predictions, and our measurements lie systematically closer to the upper envelope defined by COLIBRE and FLARES, within their $1\sigma$ uncertainties. In contrast, SPHINX and \texttt{dustyGadget} predict lower metallicities, by up to $\sim0.3-0.5$~dex at $\rm log(M_\star/M_\odot) \sim 10$.

Interestingly, this distinction correlates with the presence of BH growth and AGN feedback in the sub-grid physics of these simulations: FLARES and COLIBRE both incorporate BH accretion and AGN feedback, whereas \texttt{dustyGadget} and SPHINX do not. This suggests that AGN feedback, rather than stellar feedback alone, may help regulate chemical enrichment in massive galaxies at these redshifts, favoring the COLIBRE and FLARES implementations. 

These conclusions are robust against the choice of metallicity calibration. Adopting S2-based metallicities, which yield a normalization higher by $\sim0.3$~dex but a consistent slope (App.~\ref{app:S2_met}), our measurements remain in good agreement with COLIBRE, and agree slightly better with THESAN-ZOOM, FirstLight and FIRE-2 at intermediate masses.

\subsection{The N/O ratio}
\label{sec:NO_ratio}

\begin{figure}
   \centering
   \includegraphics[scale = 0.47]{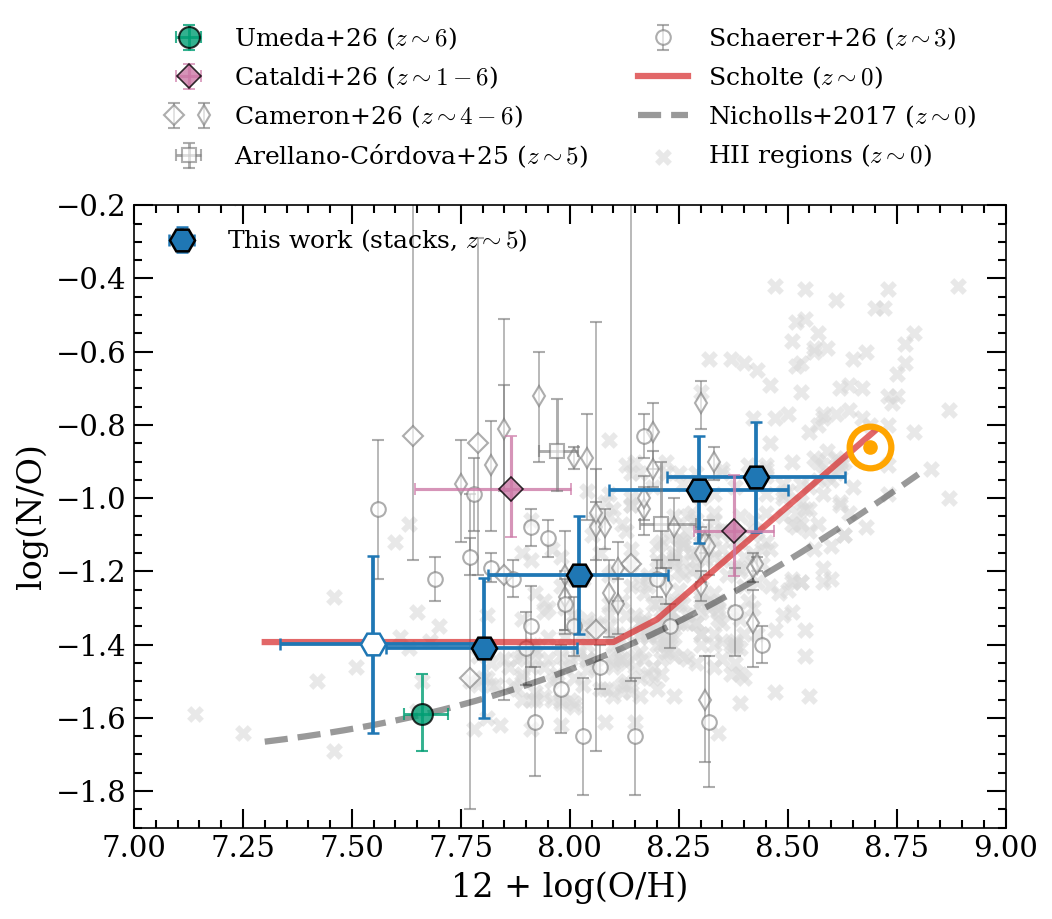}
   \caption{The nitrogen-to-oxygen (N/O) abundance ratio as a function of gas-phase metallicity estimated based on N2-diagnostic. Blue hexagons show the results from our stacked sample in each mass bin. The blue open symbol shows the gas-phase metallicity based on the \niir{} marginal detection with SNR$\sim 2$. For comparison, we show local H\textsc{ii} regions from \cite{Izotov+2006, Berg+2020} (gray), together with the local N/O-O/H relations of \cite{Nicholls+2017} (gray dashed line) and \cite{Scholte+2026_DESIabundances} (red solid line). 
   We also include measurements of individual high-$z$ galaxies \citep{Arellano-Cordova+2025_NO,Cameron+2026_NO, Schaerer+2026_no}, stacked sample at $z\sim6$ \citep{Umeda+2026_NO} and median values at $z\sim 1-6$ \citep{Cataldi+2026_NOcalib}. Large and small diamonds indicate measurements from \cite{Cameron+2026_NO}, based on direct-$T_e$ and strong-line metallicity calibrations, respectively. All high-$z$ N/O estimates are based on detections of the optical $[\ion{N}{II}]$ line. The orange Sun symbol marks the solar abundance values from \cite{Asplund+2021}.}
   \label{fig:no-oh}
\end{figure}

In this section, we present the N/O abundance ratio of our stacked sample of galaxies using the N2S2 diagnostic, which has been widely adopted as a tracer of N/O in multiple studies \citep[e.g.][]{Perez-Montero_Contini2009, Strom+2018, Hayden-Pawson+2022, Cataldi+2026_NOcalib}. Since sulfur, like oxygen, is an $\alpha$-element \citep{Kobayashi+2020}, the S/O ratio is expected to be roughly constant, so that N2S2 primarily traces the abundance of nitrogen relative to oxygen.

Owing to the small wavelength separation between \niir{} and $[\ion{S}{II}]\,\lambda\lambda6716,6731$, N2S2 is largely insensitive to dust attenuation.  Its interpretation, however,  is affected by variations in the S/O ratio and by differences in the ionization structure of N and S \citep{Perez-Montero_Contini2009, Hayden-Pawson+2022}. In particular, recent JWST studies suggest that S/O may be sub-solar at $z\sim2-3$ \citep{Rogers+2026} and that it decreases with metallicity below $\rm 12+log(O/H) \approx 8.1$ before flattening to values consistent with solar \citep{Cataldi+2026_chemical_abundances}. Our lower mass stacks, with $\rm 12+\log(O/H) < 8.1$, therefore fall in the regime where S/O may vary with metallicity. With these caveats in mind, we follow the prescriptions of \cite{Cataldi+2026_NOcalib} and adopt N2S2 as a proxy for N/O.

Figure~\ref{fig:no-oh} shows the relation between the N/O ratio and oxygen abundance for our stacked sample, with metallicity estimates based on N2-diagnostics, as discussed in Sect. \ref{sec:derived_quantities}. Additionally, we show N/O -- O/H measurements from high-$z$ observations of individual galaxies \citep{Arellano-Cordova+2025_NO, Cameron+2026_NO, Schaerer+2026_no}, stacked sample \citep{Umeda+2026_NO} and median values at $z\sim 1-6$ \citep{Cataldi+2026_NOcalib}, where the N/O estimates are based on optical emission lines. For reference, we show the N/O -- O/H trend for a local sample of H\textsc{ii} regions by \cite{Izotov+2006, Berg+2020} and star-forming galaxies from DESI DR2 \citep{Scholte+2026_DESIabundances}, together with the local relation from \cite{Nicholls+2017}. In App.~\ref{app:S2_met}, we report the same relation using S2-based metallicities.

Figure~\ref{fig:no-oh} shows that our high-$z$ stacks show a similar trend seen in the local sample of H\,\textsc{ii} regions and star-forming galaxies at $z 
\sim 0$: the N/O abundance remains approximately constant at low oxygen abundance, consistent with primary nitrogen production, and then increases steeply for $\rm 12 + log(O/H) \sim 8$, where the nitrogen production from low- and intermediate-mass stars becomes significant \citep[e.g.,][]{Vila-Costas+1993,Belfiore+2015,Vincenzo+2016,VincenzoKobayashi2018}. Our results are consistent with the individual high-redshift from the JADES \citep{Cameron+2026_NO} and EXCELS \citep{Arellano-Cordova+2025_NO} surveys that do not show nitrogen enhancement, as well as with the $z\sim3$ sample of \cite{Schaerer+2026_no}. They are also broadly consistent with the stacked optical measurements of \cite{Umeda+2026_NO} and with the median $\rm log(N/O)$ value derived by \cite{Cataldi+2026_NOcalib} at $\rm 12 + log(O/H) \sim 8.4$. At $\rm 12 + log(O/H) > 8$ , our two most massive stacks lie at lower O/H for a given N/O than star-forming galaxies at $z\sim0$ and local H\textsc{ii} regions when N2-based metallicities are adopted, although this offset is only marginally significant given the uncertainties. This offset disappears when the higher S2-based metallicities are adopted, in which case our stacks are fully consistent with the local relation (App.~\ref{app:S2_met}). Since N/O is the same in both cases, the offset likely reflects a difference between the N2 and S2 metallicity calibrations in this regime rather than a genuine difference in N/O. This highlights the need for further tests of the strong-line metallicity methods based these line ratios at high redshift.

We report the N/O ratios and metallicities of our stacks in Tab.~\ref{tab:n2-s2}. As shown in Fig.~\ref{fig:no-oh}, our stacks, which trace typical star-forming galaxies at $z\sim4-6$, show no evidence for moderate ($\log(\rm N/O) > -1.1$) or extreme ($\log(\rm N/O) > -0.6$) nitrogen enhancement at $\rm 12+\log(O/H) < 8$ \citep[see][]{Cameron+2026_NO}, the latter being characteristic of \ion{N}{III}]- and \ion{N}{IV}]-emitting galaxies at $z>3$ \citep[e.g.][]{Berg+2026}. Notably, the low-mass bins probing this metallicity regime lie above the SFMS (Sect.~\ref{sec:SFMS}), and are therefore, if anything, biased towards galaxies in a bursty phase. Since our measurements trace the average N/O of hundreds of galaxies per bin, they do not exclude a minority of nitrogen-enhanced objects but indicate that such enhancement is not common among star-forming galaxies at these redshifts, even during moderately bursty phases \citep[see also][]{Cameron+2026_NO, Schaerer+2026_no}. This suggests that extreme N/O ratios are reached only in rare conditions, such as the most intense starburst phases \citep[e.g.][]{Topping+2024_no}, and the classical primary/secondary enrichment mechanisms dominate the typical galaxy at $z\sim5$.

\section{Discussion}
\label{sec:discussion}

\subsection{Comparison to other work}
\label{sec:comp_literature}

\begin{figure}
   \centering
   \includegraphics[scale = 0.5]{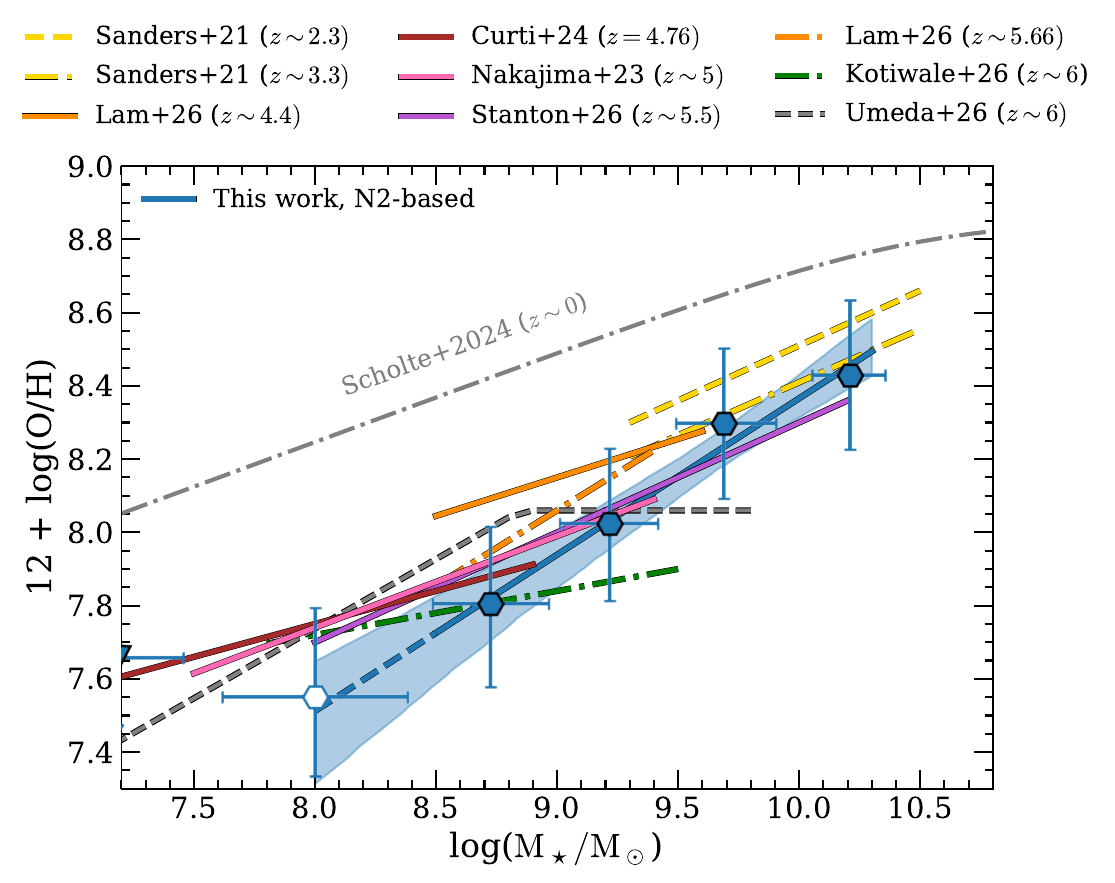}
   \caption{The redshift evolution of the mass-metallicity relation. Blue hexagons show our N2-based stacked measurements; the blue solid line and shaded region show our best-git MZR and its 1$\sigma$ (16th–84th percentile) uncertainty, derived from the Monte Carlo realizations described in Sect.~\ref{sec:mzr}. For comparison, we show literature MZRs at $z \sim 0$ \citep{Scholte+2024}, at $z < 2-3$ \citep{Sanders+2021_mosdef} and at $z > 4$ \citep{nakajima+2023_mzr,Curti+2024_mzr,Lam+2026, Stanton+2026_excels, Kotiwale+2026, Umeda+2026_NO}.}
   \label{fig:MZR_z}
\end{figure}

In this section, we discuss the redshift evolution of the slope and normalization of the MZR and its implications for the chemical evolution of galaxies. As discussed in Sect.~\ref{sec:mzr}, we measure an MZR slope of $\gamma=0.42 ^{+0.10}_{-0.07}$ (based on N2-diagnostic) and $\gamma=0.35 ^{+0.13}_{-0.10}$ (based on S2-diagnostic) at $z \sim 5$. Figure~\ref{fig:MZR_z} compares our mass-metallicity relation with those reported in the literature. In general, for $\rm log(M_\star/M_\odot) < 10.5$ the slope of the MZR remains relatively constant from $z\sim 0$ to $z\sim 3$ \citep[e.g.,][$\gamma \sim 0.29$]{Sanders+2021_mosdef}. Some studies report indications of flattening toward lower stellar masses at $z<3$ \citep[e.g.,][]{He+2024}, as well as higher redshifts $z \gtrsim 6$ \citep[e.g.][]{Curti+2024_mzr, Kotiwale+2026}. In contrast, \cite{Chemerynska+2024_mzr} report a steeper slope in the low-mass regime for $z\sim 7$ galaxies. From our stacked sample, we find that the N2-based metallicity estimates show an MZR slope that is consistent with that of the S2-based MZR slope and other literature works at $z \sim 5$. The N2-based MZR slope ($\gamma \approx 0.42$) agrees with that reported \cite{Lam+2026} ($\gamma = 0.41 \pm 0.09$) at $z \sim 5.66$ as well as with the measurements of \cite{Umeda+2026_NO} ($\gamma = 0.38 \pm 0.06$) for galaxies with stellar masses $\rm M_\star \lesssim 10^9 \; M_\odot$. However, we do not find the flattening they report at higher stellar masses. Our MZR slope is also consistent within $2\sigma$ with that derived by \cite{Stanton+2026_excels} at $z \sim 5.5$ ($\gamma = 0.30 \pm 0.03$) \footnote{We note that our S2-based slope ($\gamma \approx 0.35$) is fully consistent within $1\sigma$ with the result of \cite{Stanton+2026_excels}, as well as with the measurements of \cite{Lam+2026} at $z \sim 5.66$ and \cite{Umeda+2026_NO}.}. In contrast, both our measured slopes are in tension with the shallower values ($\gamma \sim 0.2$) reported by \cite{Curti+2024_mzr} at $3 < z < 6$, \cite{Lam+2026} at $4 < z < 5 $, \cite{Kotiwale+2026} at $z \sim 6$ and \cite{nakajima+2023_mzr} at $4 < z < 10$.

Given that MZRs from the literature are obtained using different strong-line diagnostics \citep[e.g.][]{Curti+2024_mzr,nakajima+2023_mzr,Lam+2026, Stanton+2026_excels}, calibrations \citep[e.g.][]{nakajima+2022_calib,Scholte+2025_excels, sanders+2026_aurora_calib} and sample selection functions, comparisons between MZR fits from the literature should be interpreted with caution, even though several recent measurements yield consistent slopes. 

At first order, the MZR slope is thought to be regulated by feedback-driven outflows \citep[e.g.,][]{PeeplesShankar2011,Sanders+2021_mosdef}, with steeper slopes favoring the energy-driven winds scenario \citep[$\gamma \sim 0.33$;][]{Dave+2012, Lilly+2013} and shallower slopes favoring momentum-driven winds \cite[$\gamma \sim 0.17$;][]{Guo+2016}. Both our MZR fits show a slope consistent with energy-driven winds scenario for low-mass galaxies. Moreover, the shape of our MZR is consistent with that predicted by the COLIBRE simulations over the full mass range (Sect.~\ref{sec:mzr}). This agreement suggests that the outflows that emerge in the COLIBRE simulation, together with its explicit modeling of the cold ISM, capture the processes that set the MZR slope at these redshifts.

The slope of the MZR may also be affected by observational biases. As discussed in \cite{Kotiwale+2026}, if an FMR is in place \citep{Sanders+2021_mosdef, nakajima+2023_mzr, Sarkar+2025}, selection cuts on emission-line luminosity preferentially include low-mass galaxies with enhanced SFRs and hence lower metallicities, making the observed MZR steeper than the intrinsic one. This effect is likely to apply to our sample, as our stacks below the mass completeness limit lie above the SFMS by $\sim0.6$~dex (Sect.~\ref{sec:SFMS}). These bins are excluded from our MZR fit, but the lowest-mass bin included in the fit still lies $\sim0.3$~dex above the SFMS and may therefore bias our measured slope. The magnitude of this effect also depends on whether, and in which form, the FMR holds at these redshifts and stellar masses, which remains unclear (Sect.~\ref{sec:fmr}). For instance, some simulations predict no correlation between MZR offsets and specific SFR at low stellar masses, as a result of bursty star formation in shallow potential wells \citep{McClymont+2026_mzrTHESAN}. In this case, the selection of galaxies with enhanced SFRs would not bias their metallicities, and the observed MZR slope would reflect the intrinsic one.

The N2 and S2 diagnostics have the advantage of having a monotonic dependence on metallicity over the range relevant for our sample, avoiding the degeneracies and plateaus that affect other diagnostics such as R3 and R23 \footnote{$\rm R23 = log(([\ion{O}{ii}]\lambda\lambda3726,3729 + [\ion{O}{iii}]\lambda\lambda4959,5007)/H\beta)$.} \citep[see e.g.,][]{Scholte+2025_excels, sanders+2026_aurora_calib}. However, N2-based metallicities may be biased if the N/O ratio of high-redshift galaxies at fixed O/H differs from that of the calibration sample \citep[see e.g.][]{Scholte+2025_excels}. At fixed O/H, an enhanced N/O ratio increases N2, leading to overestimated metallicities. If this effect were important in low-mass galaxies, it would bias their metallicities high and flatten the low-mass end of the N2-based MZR. Using N2S2 as a proxy for N/O \citep{Cataldi+2026_NOcalib}, we find no evidence for systematic nitrogen enhancement in our sample (Sect.~\ref{sec:NO_ratio}), although this tracer also depends on the assumed S/O ratio, which may differ from solar at high redshift \citep{Rogers+2026, Cataldi+2026_chemical_abundances}. 
Independently of S/O, the S2-based MZR, which is insensitive to N/O, has a slope consistent with the N2-based one (App.~\ref{app:S2_met}), suggesting that variations in N/O do not significantly shape the observed N2-based MZR. The S2 diagnostic, on the other hand, is expected to be more sensitive to variations in the ionization parameter \citep{KewleyDopita2002,nakajima+2022_calib}, since the S$^+$ ionization zone is narrower than that of N$^+$ \citep{Berg+2021}. For this reason, we adopt calibrations derived from a high-redshift galaxy sample, as calibrations based on lower-redshift galaxies may not accurately capture the ionization conditions of our galaxies \citep{sanders+2026_aurora_calib}. Both the $\rm [\ion{N}{II}]$ and $\rm [\ion{S}{II}]$ emission lines become intrinsically faint at low-metallicity ($\rm 12 + log(O/H) \lesssim 7.7$, i.e. $ Z \lesssim 0.1\; Z_\odot$), limiting the reliability of N2- and S2-based calibrations in this regime \citep[see e.g.,][]{sanders+2026_aurora_calib}. At such low metallicity, diagnostics based on stronger emission lines (e.g. $\rm [\ion{O}{III}]$/H$\beta$; \citealt{Chemerynska+2024_mzr}) or, ideally, direct-method abundance measurements \citep[e.g.,][]{Umeda+2026_NO} may provide more robust metallicity estimates.

At $\rm M_\star = 10^9 \; M_\odot$, our N2-based MZR lies $\sim 0.5$~dex below the local relation \citep{Scholte+2024}, corresponding to a factor $\simeq 3$ decrease of O/H, and $\sim 0.2$~dex below the $z\sim3.3$ relation of \cite{Sanders+2020}. Our normalization is consistent with other measurements at $z\sim 5-6$ \citep{nakajima+2023_mzr, Curti+2024_mzr, Lam+2026,Stanton+2026_excels}, suggesting a continued, decline of the gas-phase metallicity beyond $z\sim3$. We note, however, that the absolute normalization depends on the adopted calibration: our S2-based relation is $\sim0.3$~dex higher (App.~\ref{app:S2_met}), comparable to the inferred evolution between $z\sim3$ and $z\sim5$. The lower metallicities of high-redshift galaxies are often attributed to higher gas fractions sustained by the accretion of metal-poor gas \citep[e.g.,][]{Sanders+2021_mosdef, Heintz+2023, Pollock+2026a, Wang+2026}. This interpretation is supported by the offset of our stacks below the local FMR (Sect.~\ref{sec:fmr}), which persists after accounting for their SFRs. In contrast, the THESAN-ZOOM simulations, which reproduce the slow evolution of the MZR at $z>3$, attribute it mainly to the efficiency with which galaxies retain their metals rather than to their gas fractions \citep{McClymont+2026_mzrTHESAN}. Distinguishing between these scenarios will require independent constraints on the gas content and outflow properties of high-redshift galaxies.
 
\subsection{The Fundamental Metallicity Relation}
\label{sec:fmr}

\begin{figure}
   \centering
   \includegraphics[scale = 0.5]{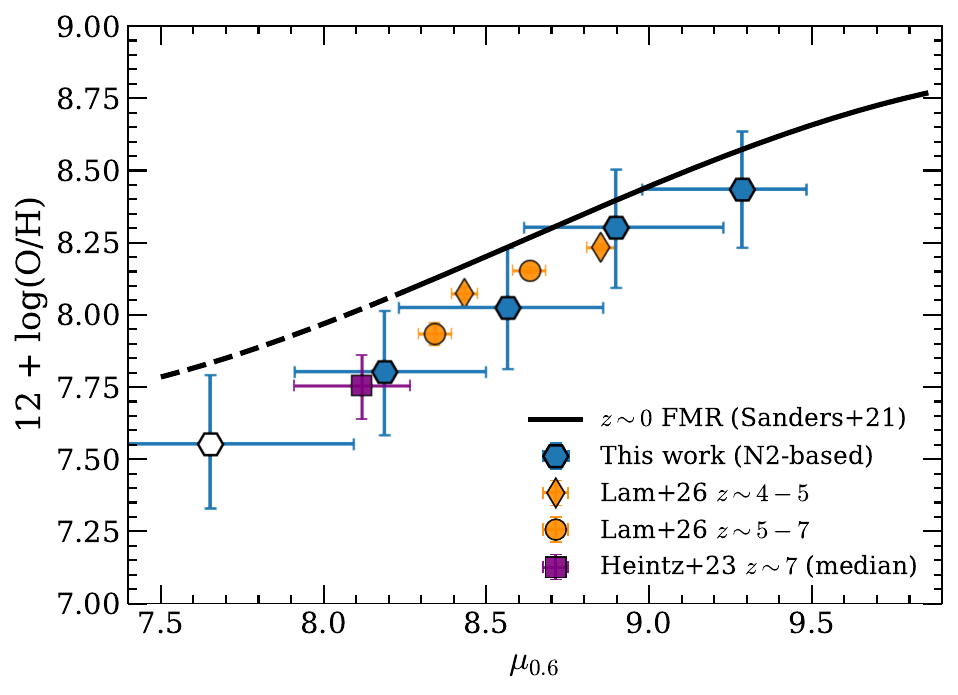}
   \caption{Metallicity as a function of $\mu _{0.6}$. Our stacked measurements derived using the N2-diagnostic (blue), compared with high-redshift measurements from \cite{Lam+2026} (orange) at $z=4-7$ and the median metallicity values of galaxies at $z<8$ from \cite{Heintz+2023} (purple). The solid black line shows the local FMR from \cite{Sanders+2021_mosdef}, while the dashed line its extrapolation. }
   \label{fig:fmr}
\end{figure}

In this section, we explore how our high-redshift measurements compare to the FMR, which describes the dependence of metallicity on both stellar mass and SFR established in the low-redshift Universe \citep[e.g.][]{Lara-Lopez+2010,Mannucci+2010}. At fixed stellar mass, galaxies with higher SFR have lower gas-phase metallicities, commonly interpreted as a consequence of self-regulation of galaxies through the interplay between stellar feedback, inflows and outflows \citep[e.g.,][]{Lilly+2013, Dave+2012, curti+2020_calib, Torrey+2018}, a picture supported by observations linking gas inflows to enhanced SFRs and diluted metallicities \citep[e.g.,][]{SanchezAlmeida+2014,Langan+2023}. \cite{Sanders+2021_mosdef} showed that star-forming galaxies at $z\sim2.2 $ and $z\sim 3.3$ lie on the same FMR plane as low-redshift galaxies. This suggests that both the existence of a three-dimensional correlation structure among these properties, as well as the location of the FMR in 3D space remain largely unchanged out to $z\sim3.3$. Earlier JWST studies, however, have indicated that galaxies at $z>4$ may be offset toward lower metallicities at fixed stellar mass and SFR \citep[e.g.,][]{Heintz+2023, Curti+2023_z8, nakajima+2023_mzr, Lam+2026}. 

We now test this in our measurements that are based on a large sample with a simple SFR-limited selection. Following \cite{Lam+2026}, we adopt the FMR projection proposed by \cite{Mannucci+2010} defining the $\mu_\alpha$ parameter as $\mu_\alpha = \rm log(M_\star/M_\odot) - \alpha \cdot log(SFR/M_\odot \, yr^{-1})$. We set $\alpha = 0.6$ which has been shown to minimize the scatter of SDSS stacks at $z\sim0$ \citep{Sanders+2021_mosdef}. Figure \ref{fig:fmr} presents the metallicity as a function of $\mu_{0.6}$, with the FMR at $z\sim0$ as parametrized by \cite{Sanders+2021_mosdef}. We also include high-redshift JWST/NIRSpec observations \citep{Heintz+2023,Lam+2026}. For \cite{Heintz+2023} we only consider galaxies at $z<8$, recompute $\mu_{0.6}$ for each galaxy following the same procedure and show the median metallicity and $\mu_{0.6}$ of this sample.

By computing the residuals with respect to the $z\sim0$ relation, we find that all our stellar mass bins lie below the local FMR. Since $\mu_{0.6}$ includes the measured SFR, the comparison already accounts for the elevated SFRs of our low-mass stacks relative to the SFMS (Sect.~\ref{sec:SFMS}), provided that the SFR dependence of the local FMR also holds at these redshifts. Over the $\mu_{0.6}$ range constrained by the SDSS stacks ($\mu_{0.6}\geq8.25$), our measurements are offset by $0.10-0.21$~dex (mean $0.15$~dex), comparable to the offsets found by \citet{Lam+2026}. As our \ha{}-selected sample does not require the detection of faint lines in individual galaxies, this agreement suggests that the offsets reported by NIRSpec studies are not driven by their sample selection. Moreover, since our measurements are based on stacked spectra, the systematic offset across all mass bins suggests that the equilibrium metallicity itself may be lower at $z\sim4-6$ than in the local Universe, as expected in gas-regulator models for the higher gas fractions and accretion rates of early galaxies \citep[e.g.,][]{Dave+2012, Lilly+2013}.

At lower $\mu_{0.6}$, where the local FMR is extrapolated, our two lowest bins show larger offsets of $0.25-0.28$~dex, consistent with the median of the \citet{Heintz+2023} sample, suggesting that such offsets are typical for the population. This may indicate a stronger evolution of the equilibrium in low-mass galaxies, which are more sensitive to inflows and feedback \citep[see also][]{Pollock+2026b}, implying that the FMR is intrinsically steeper at low $\mu_{0.6}$ than predicted by the local parametrization. However, our low-mass bins lie above the SFMS (Sect.~\ref{sec:SFMS}), suggesting that at these masses our \ha{} selection favors galaxies caught during bursts of star formation. Since the SFR may vary on shorter timescales than the metallicity \citep{Torrey+2018, Tacchella+2020_sfhII}, $\mu_{0.6}$ may not trace the equilibrium state of these galaxies. The redshift evolution of the FMR plane provides valuable new insights into the interplay between gas inflows, star formation and feedback, and how these processes jointly regulate galaxy evolution across cosmic time.

\section{Summary}
\label{sec:summary}

In this work, we measure the relation between stellar mass and gas-phase metallicity for \ha{} selected galaxies (with SNR$_{H\alpha}$ > 5) in the redshift range $z = 3.8 - 6.6$. The spectra of these galaxies are drawn from deep JWST/NIRCam WFSS surveys: ALT \citepalias{Naidu+2024_alt}, CONGRESS \citep{Sun+2026}, COLA1 \citep{Torralba+2024}, EIGER \citep{Kashino+2023} and FRESCO \citep{Oesch+2023_fresco}. Generally, grism spectroscopy enables observations of large statistical samples, benefits from a well-understood, flux-limited selection function and are not subject to slit losses; however, they cover a narrower wavelength range and are less sensitive than NIRSpec spectra. For this reason and in order not to be biased by sources with individual detections of metal lines, we stack spectra in bins of stellar mass to detect faint emission lines such as \niir{} and $[\ion{S}{ii}]\,\lambda\lambda6716,6731$. We stack the spectra of 1748 star-forming galaxies, spanning a stellar mass range $\rm log(M_\star/M_\odot) = 6 - 10.8$. Stellar mass for individual galaxies are derived from \textsc{Bagpipes} SED-fitting of all available photometry in each field. Our main results are summarized below.

\begin{itemize}
    \item We detect \ha{}, $[\ion{N}{ii}]$ and $[\ion{S}{ii}]$ emission in stacks of galaxies binned by stellar mass, enabling us to probe the MZR over the range $\rm log(M_\star/M_\odot)=8-10.2$, while placing an upper limit on the metallicity at $\rm log(M_\star/M_\odot)\sim7.2$. Compared to previous NIRSpec based studies, our sample extends the measured MZR to an order of magnitude higher stellar masses. We adopt the N2 and S2 diagnostics to estimate the gas-phase metallicity using strong-line calibrations by \cite{sanders+2026_aurora_calib}, which are derived from a sample of high-redshift galaxies. Our measurements suggest that the oxygen abundance varies between $\rm 12+log(O/H) \sim 7.6 - 8.4$ (i.e. $Z_{gas}\sim 0.10-0.50 \; Z_\odot$) at $z\sim5$. 
    
    \item We fit a linear relation to the MZR using the N2-based (S2-based) gas-phase metallicity for masses above $>3\cdot10^{8}$~M$_{\odot}$ where we are likely complete for star-forming galaxies. We measure a slope of $\gamma = 0.42 ^{+0.10}_{-0.07}$ ($\gamma = 0.35 ^{+0.13}_{-0.10}$) and normalization $Z_{m9} = 7.9 ^{+0.06}_{-0.09}$ ($Z_{m9} = 8.25 ^{+0.06}_{-0.09}$) at $\rm M_\star = 10^9 \; M_\odot$. Our N2-based MZR lies $\sim 0.5$~dex below that of galaxies in the local Universe \citep{Scholte+2024} and $\sim 0.2$~dex below the $z\sim3$ relation of \cite{Sanders+2021_mosdef}. Its slope is robust against the choice of strong-line diagnostic and consistent with an energy-driven feedback scenario shaping the relation.
    
    \item Both $\rm [\ion{N}{II}]$ and $[\ion{S}{II}]$ emission lines become intrinsically faint at low-metallicity \citep[$\rm 12 + log(O/H) \lesssim 7.7$, i.e. $ Z \lesssim 0.1\; Z_\odot$; see][]{sanders+2026_aurora_calib}. In addition, the N2 and S2 calibrations are less well constrained in this regime, where few calibrating galaxies have detected [\ion{N}{II}] and [\ion{S}{II}] lines \citep{sanders+2026_aurora_calib}. Extending these calibrations to lower metallicities, using direct-method abundances of metal-poor galaxies, will be necessary to probe the low-mass end of the MZR with these diagnostics.
    
    \item We estimate the N/O ratio from optical emission lines using the N2S2-N/O relation proposed by \cite{Cataldi+2026_NOcalib}, calibrated on a sample of high-$z$ galaxies. We find no evidence for N-enhancement in our sample, as reported for individual sources in previous studies, suggesting that N-enhanced sources may represent a relatively rare fraction of the overall population. Comparing to literature N/O measurements from optical emission lines in local and high-$z$ galaxies, our sample follows the trend expected from primary and secondary nitrogen production for $\rm 12+log(O/H) \lesssim 8$ and $\rm 12+log(O/H) \gtrsim 8$, respectively, in agreement with recent results from \cite{Schaerer+2026_no}.
    
    \item The grism spectral resolution (R $\sim 1600$) allows us to resolve the $[\ion{S}{ii}]\,\lambda\lambda6716,6731$ doublet, whose ratio constrains the physical conditions of the ISM. Using \texttt{pyneb} and assuming $T_e = 15000$ K, we find a median electron density of $n_e = 184^{+226}_{-169}\,\rm cm^{-3}$ in agreement with recent based on stacked spectra of high-$z$ galaxies \citep[e.g.,][]{Isobe+2023}.
    
    \item We present the FMR relation at $z=4-7$. Our stacked sample lies systematically below the local FMR with an offset $\sim 0.1-0.2$~dex over the $\mu_{0.6}$ range constrained by the SDSS stacks, consistent with previous JWST results \citep[][]{Lam+2026, Heintz+2023}. Since our measurements are based on stacked spectra of an \ha{}-selected sample, this suggests that the offset is not driven by the selection of NIRSpec samples and that the equilibrium metallicity itself may be lower at high redshift than in the local Universe. The offset increases to $0.25-0.28$~dex in our lowest-mass bins, which may reflect a stronger evolution of the equilibrium metallicity in low-mass galaxies, implying a steeper FMR at low $\mu_{0.6}$.
\end{itemize}

While this work provides an important benchmark for the MZR relation at $z\sim 4-6$ and substantially extends the probed mass range by spectroscopic studies, deeper observations will be essential to determine whether the MZR slope and the ISM conditions reported here hold at the lowest stellar masses, where current measurements are limited by sensitivity. Future deep spectroscopic surveys targeting statistically significant and representative samples of low-mass galaxies ($\rm log(M_\star/M_\odot)<8$) at these redshifts, unbiased with respect to the star-forming main sequence, will enable metallicity measurements using the direct-$T_e$ method, providing more robust constraints on the mass-metallicity relation and its evolution. These measurements will ultimately lead to a more detailed picture of metallicity scaling relations and improve our understanding of baryon cycling, galaxy growth, and formation in the early Universe.   

\begin{acknowledgements}
We thank A.~D'Angelo and A.~P.~Vijayan for sharing the mass-metallicity predictions of \texttt{dustyGadget} and FLARES, respectively. 

Funded by the European Union (ERC, AGENTS,  101076224). Views and opinions expressed are however those of the author(s) only and do not necessarily reflect those of the European Union or the European Research Council. Neither the European Union nor the granting authority can be held responsible for them. This work is based on observations made with the NASA/ESA/CSA James Webb Space Telescope. The data were obtained from the Mikulski Archive for Space Telescopes at the Space Telescope Science Institute, which is operated by the Association of Universities for Research in Astronomy, Inc., under NASA contract NAS 5-03127 for JWST. These observations are associated with programs \#1243, \#1933, \#1985, \#3516 and \#3577. We acknowledge funding from {\it JWST} program GO-3516. Generative AI tools (Claude, Anthropic) were used to improve the language and readability of the manuscript and to assist with coding. All scientific content was reviewed and verified by the authors, who take full responsibility for it.

Software used in developing this work includes: matplotlib \citep{Hunter2007}, numpy \citep{Oliphant2007}, scipy \citep{Virtanen2020}, TOPCAT \citep{topcat2005}, and Astropy \citep{astropy2013}.        

\end{acknowledgements}

\bibliographystyle{aa} 
\bibliography{biblio} 

\begin{thebibliography}{206}
\expandafter\ifx\csname natexlab\endcsname\relax\def\natexlab#1{#1}\fi

\bibitem[{{Abazajian} {et~al.}(2009){Abazajian}, {Adelman-McCarthy}, {Ag{\"u}eros}, {Allam}, {Allende Prieto}, {An}, {Anderson}, {Anderson}, {Annis}, {Bahcall}, {Bailer-Jones}, {Barentine}, {Bassett}, {Becker}, {Beers}, {Bell}, {Belokurov}, {Berlind}, {Berman}, {Bernardi}, {Bickerton}, {Bizyaev}, {Blakeslee}, {Blanton}, {Bochanski}, {Boroski}, {Brewington}, {Brinchmann}, {Brinkmann}, {Brunner}, {Budav{\'a}ri}, {Carey}, {Carliles}, {Carr}, {Castander}, {Cinabro}, {Connolly}, {Csabai}, {Cunha}, {Czarapata}, {Davenport}, {de Haas}, {Dilday}, {Doi}, {Eisenstein}, {Evans}, {Evans}, {Fan}, {Friedman}, {Frieman}, {Fukugita}, {G{\"a}nsicke}, {Gates}, {Gillespie}, {Gilmore}, {Gonzalez}, {Gonzalez}, {Grebel}, {Gunn}, {Gy{\"o}ry}, {Hall}, {Harding}, {Harris}, {Harvanek}, {Hawley}, {Hayes}, {Heckman}, {Hendry}, {Hennessy}, {Hindsley}, {Hoblitt}, {Hogan}, {Hogg}, {Holtzman}, {Hyde}, {Ichikawa}, {Ichikawa}, {Im}, {Ivezi{\'c}}, {Jester}, {Jiang}, {Johnson}, {Jorgensen}, {Juri{\'c}}, {Kent}, {Kessler}, {Kleinman}, {Knapp},
  {Konishi}, {Kron}, {Krzesinski}, {Kuropatkin}, {Lampeitl}, {Lebedeva}, {Lee}, {Lee}, {French Leger}, {L{\'e}pine}, {Li}, {Lima}, {Lin}, {Long}, {Loomis}, {Loveday}, {Lupton}, {Magnier}, {Malanushenko}, {Malanushenko}, {Mandelbaum}, {Margon}, {Marriner}, {Mart{\'\i}nez-Delgado}, {Matsubara}, {McGehee}, {McKay}, {Meiksin}, {Morrison}, {Mullally}, {Munn}, {Murphy}, {Nash}, {Nebot}, {Neilsen}, {Newberg}, {Newman}, {Nichol}, {Nicinski}, {Nieto-Santisteban}, {Nitta}, {Okamura}, {Oravetz}, {Ostriker}, {Owen}, {Padmanabhan}, {Pan}, {Park}, {Pauls}, {Peoples}, {Percival}, {Pier}, {Pope}, {Pourbaix}, {Price}, {Purger}, {Quinn}, {Raddick}, {Re Fiorentin}, {Richards}, {Richmond}, {Riess}, {Rix}, {Rockosi}, {Sako}, {Schlegel}, {Schneider}, {Scholz}, {Schreiber}, {Schwope}, {Seljak}, {Sesar}, {Sheldon}, {Shimasaku}, {Sibley}, {Simmons}, {Sivarani}, {Allyn Smith}, {Smith}, {Smol{\v{c}}i{\'c}}, {Snedden}, {Stebbins}, {Steinmetz}, {Stoughton}, {Strauss}, {SubbaRao}, {Suto}, {Szalay}, {Szapudi}, {Szkody}, {Tanaka},
  {Tegmark}, {Teodoro}, {Thakar}, {Tremonti}, {Tucker}, {Uomoto}, {Vanden Berk}, {Vandenberg}, {Vidrih}, {Vogeley}, {Voges}, {Vogt}, {Wadadekar}, {Watters}, {Weinberg}, {West}, {White}, {Wilhite}, {Wonders}, {Yanny}, \& {Yocum}}]{Abazajian+2009}
{Abazajian}, K.~N., {Adelman-McCarthy}, J.~K., {Ag{\"u}eros}, M.~A., {et~al.} 2009, \apjs, 182, 543

\bibitem[{{Abell} {et~al.}(1989){Abell}, {Corwin}, \& {Olowin}}]{Abell+1989}
{Abell}, G.~O., {Corwin}, Jr., H.~G., \& {Olowin}, R.~P. 1989, \apjs, 70, 1

\bibitem[{{Andrews} \& {Martini}(2013)}]{AndrewsMartini2013}
{Andrews}, B.~H. \& {Martini}, P. 2013, \apj, 765, 140

\bibitem[{{Arellano-C{\'o}rdova} {et~al.}(2024){Arellano-C{\'o}rdova}, {Berg}, {Mingozzi}, {James}, {Rogers}, {Skillman}, {Cullen}, {Alexander}, {Amor{\'\i}n}, {Chisholm}, {Hayes}, {Heckman}, {Hernandez}, {Kumari}, {Leitherer}, {Martin}, {Maseda}, {Nanayakkara}, {Parker}, {Ravindranath}, {Strom}, {Vincenzo}, \& {Wofford}}]{ArellanoCordova+2024}
{Arellano-C{\'o}rdova}, K.~Z., {Berg}, D.~A., {Mingozzi}, M., {et~al.} 2024, \apj, 968, 98

\bibitem[{{Arellano-C{\'o}rdova} {et~al.}(2025){Arellano-C{\'o}rdova}, {Cullen}, {Carnall}, {Scholte}, {Stanton}, {Kobayashi}, {Martinez}, {Berg}, {Barrufet}, {Begley}, {Donnan}, {Dunlop}, {Hamadouche}, {McLeod}, {McLure}, {Rowlands}, \& {Shapley}}]{Arellano-Cordova+2025_NO}
{Arellano-C{\'o}rdova}, K.~Z., {Cullen}, F., {Carnall}, A.~C., {et~al.} 2025, \mnras, 540, 2991

\bibitem[{{Arellano-C{\'o}rdova} {et~al.}(2026){Arellano-C{\'o}rdova}, {M{\'e}ndez-Delgado}, {Flury}, {Esteban}, {Kreckel}, {Garc{\'\i}a-Rojas}, {Cullen}, {Carigi}, {Morisset}, {Rosales-Ortega}, {Peimbert}, {Stanton}, \& {Scholte}}]{Arellano-Cordova+2026}
{Arellano-C{\'o}rdova}, K.~Z., {M{\'e}ndez-Delgado}, J.~E., {Flury}, S.~R., {et~al.} 2026, \mnras, 547, stag380

\bibitem[{{Asano} {et~al.}(2013){Asano}, {Takeuchi}, {Hirashita}, \& {Inoue}}]{Asano+2013_dust}
{Asano}, R.~S., {Takeuchi}, T.~T., {Hirashita}, H., \& {Inoue}, A.~K. 2013, Earth, Planets and Space, 65, 213

\bibitem[{{Asplund} {et~al.}(2021){Asplund}, {Amarsi}, \& {Grevesse}}]{Asplund+2021}
{Asplund}, M., {Amarsi}, A.~M., \& {Grevesse}, N. 2021, \aap, 653, A141

\bibitem[{{Asplund} {et~al.}(2009){Asplund}, {Grevesse}, {Sauval}, \& {Scott}}]{Asplund+2009_sun}
{Asplund}, M., {Grevesse}, N., {Sauval}, A.~J., \& {Scott}, P. 2009, \araa, 47, 481

\bibitem[{{Astropy Collaboration} {et~al.}(2013){Astropy Collaboration}, {Robitaille}, {Tollerud}, {Greenfield}, {Droettboom}, {Bray}, {Aldcroft}, {Davis}, {Ginsburg}, {Price-Whelan}, {Kerzendorf}, {Conley}, {Crighton}, {Barbary}, {Muna}, {Ferguson}, {Grollier}, {Parikh}, {Nair}, {Unther}, {Deil}, {Woillez}, {Conseil}, {Kramer}, {Turner}, {Singer}, {Fox}, {Weaver}, {Zabalza}, {Edwards}, {Azalee Bostroem}, {Burke}, {Casey}, {Crawford}, {Dencheva}, {Ely}, {Jenness}, {Labrie}, {Lim}, {Pierfederici}, {Pontzen}, {Ptak}, {Refsdal}, {Servillat}, \& {Streicher}}]{astropy2013}
{Astropy Collaboration}, {Robitaille}, T.~P., {Tollerud}, E.~J., {et~al.} 2013, \aap, 558, A33

\bibitem[{{Barnes} {et~al.}(2017){Barnes}, {Kay}, {Henson}, {McCarthy}, {Schaye}, \& {Jenkins}}]{Barnes+2017}
{Barnes}, D.~J., {Kay}, S.~T., {Henson}, M.~A., {et~al.} 2017, \mnras, 465, 213

\bibitem[{{Belfiore} {et~al.}(2015){Belfiore}, {Maiolino}, {Bundy}, {Thomas}, {Maraston}, {Wilkinson}, {S{\'a}nchez}, {Bershady}, {Blanc}, {Bothwell}, {Cales}, {Coccato}, {Drory}, {Emsellem}, {Fu}, {Gelfand}, {Law}, {Masters}, {Parejko}, {Tremonti}, {Wake}, {Weijmans}, {Yan}, {Xiao}, {Zhang}, {Zheng}, {Bizyaev}, {Kinemuchi}, {Oravetz}, \& {Simmons}}]{Belfiore+2015}
{Belfiore}, F., {Maiolino}, R., {Bundy}, K., {et~al.} 2015, \mnras, 449, 867

\bibitem[{{Berg} {et~al.}(2021){Berg}, {Chisholm}, {Erb}, {Skillman}, {Pogge}, \& {Olivier}}]{Berg+2021}
{Berg}, D.~A., {Chisholm}, J., {Erb}, D.~K., {et~al.} 2021, \apj, 922, 170

\bibitem[{{Berg} {et~al.}(2026){Berg}, {Naidu}, {Chisholm}, {Atek}, {Fujimoto}, {Kokorev}, {Furtak}, {Kobayashi}, {Schaerer}, {Adamo}, {Fei}, {Korber}, {Matthee}, {Marques-Chaves}, {Martinez}, {McQuinn}, {Mu{\~n}oz}, {Oesch}, {Saldana-Lopez}, {Stark}, {Stephenson}, \& {Hsiao}}]{Berg+2026}
{Berg}, D.~A., {Naidu}, R.~P., {Chisholm}, J., {et~al.} 2026, \apj, 1003, 112

\bibitem[{{Berg} {et~al.}(2020){Berg}, {Pogge}, {Skillman}, {Croxall}, {Moustakas}, {Rogers}, \& {Sun}}]{Berg+2020}
{Berg}, D.~A., {Pogge}, R.~W., {Skillman}, E.~D., {et~al.} 2020, \apj, 893, 96

\bibitem[{{Bezanson} {et~al.}(2024){Bezanson}, {Labbe}, {Whitaker}, {Leja}, {Price}, {Franx}, {Brammer}, {Marchesini}, {Zitrin}, {Wang}, {Weaver}, {Furtak}, {Atek}, {Coe}, {Cutler}, {Dayal}, {van Dokkum}, {Feldmann}, {F{\"o}rster Schreiber}, {Fujimoto}, {Geha}, {Glazebrook}, {de Graaff}, {Greene}, {Juneau}, {Kassin}, {Kriek}, {Khullar}, {Maseda}, {Mowla}, {Muzzin}, {Nanayakkara}, {Nelson}, {Oesch}, {Pacifici}, {Pan}, {Papovich}, {Setton}, {Shapley}, {Smit}, {Stefanon}, {Taylor}, \& {Williams}}]{Bezanson+2024_uncover}
{Bezanson}, R., {Labbe}, I., {Whitaker}, K.~E., {et~al.} 2024, \apj, 974, 92

\bibitem[{{Bian} {et~al.}(2016){Bian}, {Kewley}, {Dopita}, \& {Juneau}}]{Bian+2016_localanalogs}
{Bian}, F., {Kewley}, L.~J., {Dopita}, M.~A., \& {Juneau}, S. 2016, \apj, 822, 62

\bibitem[{{Brammer}(2019)}]{Brammer+2019_grizli}
{Brammer}, G. 2019, {Grizli: Grism redshift and line analysis software}, Astrophysics Source Code Library, record ascl:1905.001

\bibitem[{{Bruzual} \& {Charlot}(2003)}]{BruzualCharlot2003}
{Bruzual}, G. \& {Charlot}, S. 2003, \mnras, 344, 1000

\bibitem[{{Byler} {et~al.}(2017){Byler}, {Dalcanton}, {Conroy}, \& {Johnson}}]{Byler+2017}
{Byler}, N., {Dalcanton}, J.~J., {Conroy}, C., \& {Johnson}, B.~D. 2017, \apj, 840, 44

\bibitem[{{Calzetti} {et~al.}(2000){Calzetti}, {Armus}, {Bohlin}, {Kinney}, {Koornneef}, \& {Storchi-Bergmann}}]{Calzetti+2000}
{Calzetti}, D., {Armus}, L., {Bohlin}, R.~C., {et~al.} 2000, \apj, 533, 682

\bibitem[{{Cameron} {et~al.}(2026){Cameron}, {Carreira}, {Simmonds}, {Bunker}, {Saxena}, {Carniani}, {Charlot}, {Chevallard}, {Curtis-Lake}, {Hainline}, {Hausen}, {Ji}, {Ji}, {Johnson}, {Rinaldi}, {Robertson}, {Scholtz}, {Silcock}, {Tacchella}, {Trussler}, {{\"U}bler}, {Williams}, {Willmer}, {Willott}, \& {Witstok}}]{Cameron+2026_NO}
{Cameron}, A.~J., {Carreira}, C., {Simmonds}, C., {et~al.} 2026, \mnras, 552, stag1575

\bibitem[{{Cameron} {et~al.}(2023){Cameron}, {Katz}, {Rey}, \& {Saxena}}]{Cameron+2023_NO-enhancement}
{Cameron}, A.~J., {Katz}, H., {Rey}, M.~P., \& {Saxena}, A. 2023, \mnras, 523, 3516

\bibitem[{{Cardelli} {et~al.}(1989){Cardelli}, {Clayton}, \& {Mathis}}]{Cardelli+1989}
{Cardelli}, J.~A., {Clayton}, G.~C., \& {Mathis}, J.~S. 1989, \apj, 345, 245

\bibitem[{{Carnall} {et~al.}(2024){Carnall}, {Cullen}, {McLure}, {McLeod}, {Begley}, {Donnan}, {Dunlop}, {Shapley}, {Rowlands}, {Almaini}, {Arellano-C{\'o}rdova}, {Barrufet}, {Cimatti}, {Ellis}, {Grogin}, {Hamadouche}, {Illingworth}, {Koekemoer}, {Leung}, {Lovell}, {P{\'e}rez-Gonz{\'a}lez}, {Santini}, {Stanton}, \& {Wild}}]{Carnall+2024_excels}
{Carnall}, A.~C., {Cullen}, F., {McLure}, R.~J., {et~al.} 2024, \mnras, 534, 325

\bibitem[{{Carnall} {et~al.}(2018){Carnall}, {McLure}, {Dunlop}, \& {Dav{\'e}}}]{Carnall+2018_bagpipes}
{Carnall}, A.~C., {McLure}, R.~J., {Dunlop}, J.~S., \& {Dav{\'e}}, R. 2018, \mnras, 480, 4379

\bibitem[{{Cataldi} {et~al.}(2025){Cataldi}, {Belfiore}, {Curti}, {Moreschini}, {Mannucci}, {D'Amato}, {Cresci}, {Feltre}, {Ginolfi}, {Marconi}, {Amiri}, {Arnaboldi}, {Bertola}, {Bracci}, {Carniani}, {Ceci}, {Chakraborty}, {Cirasuolo}, {Cullen}, {Kobayashi}, {Kumari}, {Maiolino}, {Marconcini}, {Scialpi}, \& {Ulivi}}]{Cataldi+2025_calib}
{Cataldi}, E., {Belfiore}, F., {Curti}, M., {et~al.} 2025, \aap, 703, A208

\bibitem[{{Cataldi} {et~al.}(2026{\natexlab{a}}){Cataldi}, {Belfiore}, {Curti}, {Moreschini}, {Marconi}, {Maiolino}, {Feltre}, {Ginolfi}, {Mannucci}, {Cresci}, {Ji}, {Amiri}, {Arnaboldi}, {Bertola}, {Bracci}, {Ceci}, {Chakraborty}, {Cullen}, {D'Amato}, {Kobayashi}, {Lamperti}, {Marconcini}, {Scialpi}, {Ulivi}, \& {Zanchettin}}]{Cataldi+2026_NOcalib}
{Cataldi}, E., {Belfiore}, F., {Curti}, M., {et~al.} 2026{\natexlab{a}}, \aap, 709, A199

\bibitem[{{Cataldi} {et~al.}(2026{\natexlab{b}}){Cataldi}, {Belfiore}, {Curti}, {Moreschini}, {Marconi}, {Mannucci}, {Cresci}, {Feltre}, {Ginolfi}, {Amiri}, {Arnaboldi}, {Bhattacharya}, {Bertola}, {Bracci}, {Ceci}, {Chakraborty}, {D'Amato}, {Kobayashi}, {Marconcini}, {Maiolino}, {Querci}, {Scialpi}, {Ulivi}, {Venturi}, \& {Zanchettin}}]{Cataldi+2026_chemical_abundances}
{Cataldi}, E., {Belfiore}, F., {Curti}, M., {et~al.} 2026{\natexlab{b}}, arXiv e-prints, arXiv:2610.28717

\bibitem[{{Ceverino} {et~al.}(2017){Ceverino}, {Glover}, \& {Klessen}}]{Ceverino+2017_firstlight}
{Ceverino}, D., {Glover}, S. C.~O., \& {Klessen}, R.~S. 2017, \mnras, 470, 2791

\bibitem[{{Chakraborty} {et~al.}(2025){Chakraborty}, {Sarkar}, {Smith}, {Ferland}, {McDonald}, {Forman}, {Vogelsberger}, {Torrey}, {Garcia}, {Bautz}, {Foster}, {Miller}, \& {Grant}}]{Chakraborty+2025}
{Chakraborty}, P., {Sarkar}, A., {Smith}, R., {et~al.} 2025, \apj, 985, 24

\bibitem[{{Chemerynska} {et~al.}(2024){Chemerynska}, {Atek}, {Dayal}, {Furtak}, {Feldmann}, {Greene}, {Maseda}, {Nanayakkara}, {Oesch}, {Fujimoto}, {Labb{\'e}}, {Bezanson}, {Brammer}, {Cutler}, {Leja}, {Pan}, {Price}, {Wang}, {Weaver}, \& {Whitaker}}]{Chemerynska+2024_mzr}
{Chemerynska}, I., {Atek}, H., {Dayal}, P., {et~al.} 2024, \apjl, 976, L15

\bibitem[{{Chiappini} {et~al.}(2006){Chiappini}, {Hirschi}, {Meynet}, {Ekstr{\"o}m}, {Maeder}, \& {Matteucci}}]{Chiappini+2006}
{Chiappini}, C., {Hirschi}, R., {Meynet}, G., {et~al.} 2006, \aap, 449, L27

\bibitem[{{Clarke} {et~al.}(2025){Clarke}, {Shapley}, {Lam}, {Topping}, {Brammer}, {Sanders}, {Reddy}, \& {Karthikeyan}}]{Clarke+2025}
{Clarke}, L., {Shapley}, A.~E., {Lam}, N., {et~al.} 2025, subm. to \apj, arXiv:2510.06681

\bibitem[{{Clarke} {et~al.}(2024){Clarke}, {Shapley}, {Sanders}, {Topping}, {Brammer}, {Bento}, {Reddy}, \& {Kehoe}}]{Clarke+2024}
{Clarke}, L., {Shapley}, A.~E., {Sanders}, R.~L., {et~al.} 2024, \apj, 977, 133

\bibitem[{{Cochrane} {et~al.}(2025){Cochrane}, {Katz}, {Begley}, {Hayward}, \& {Best}}]{Cochrane+2025_SEDfitting}
{Cochrane}, R.~K., {Katz}, H., {Begley}, R., {Hayward}, C.~C., \& {Best}, P.~N. 2025, \apjl, 978, L42

\bibitem[{{Covelo-Paz} {et~al.}(2025){Covelo-Paz}, {Giovinazzo}, {Oesch}, {Meyer}, {Weibel}, {Brammer}, {Fudamoto}, {Kerutt}, {Lin}, {Matharu}, {Naidu}, {Velichko}, {Bollo}, {Bouwens}, {Chisholm}, {Illingworth}, {Kramarenko}, {Magee}, {Maseda}, {Matthee}, {Nelson}, {Reddy}, {Schaerer}, {Stefanon}, \& {Xiao}}]{Covelo-Paz+2025}
{Covelo-Paz}, A., {Giovinazzo}, E., {Oesch}, P.~A., {et~al.} 2025, \aap, 694, A178

\bibitem[{{Crain} {et~al.}(2015){Crain}, {Schaye}, {Bower}, {Furlong}, {Schaller}, {Theuns}, {Dalla Vecchia}, {Frenk}, {McCarthy}, {Helly}, {Jenkins}, {Rosas-Guevara}, {White}, \& {Trayford}}]{Crain+2015_eagle}
{Crain}, R.~A., {Schaye}, J., {Bower}, R.~G., {et~al.} 2015, \mnras, 450, 1937

\bibitem[{{Curti}(2025)}]{Curti2025}
{Curti}, M. 2025, arXiv e-prints, arXiv:2504.08933

\bibitem[{{Curti} {et~al.}(2017){Curti}, {Cresci}, {Mannucci}, {Marconi}, {Maiolino}, \& {Esposito}}]{Curti+2017}
{Curti}, M., {Cresci}, G., {Mannucci}, F., {et~al.} 2017, \mnras, 465, 1384

\bibitem[{{Curti} {et~al.}(2023){Curti}, {D'Eugenio}, {Carniani}, {Maiolino}, {Sandles}, {Witstok}, {Baker}, {Bennett}, {Piotrowska}, {Tacchella}, {Charlot}, {Nakajima}, {Maheson}, {Mannucci}, {Amiri}, {Arribas}, {Belfiore}, {Bonaventura}, {Bunker}, {Chevallard}, {Cresci}, {Curtis-Lake}, {Hayden-Pawson}, {Jones}, {Kumari}, {Laseter}, {Looser}, {Marconi}, {Maseda}, {Scholtz}, {Smit}, {{\"U}bler}, \& {Wallace}}]{Curti+2023_z8}
{Curti}, M., {D'Eugenio}, F., {Carniani}, S., {et~al.} 2023, \mnras, 518, 425

\bibitem[{{Curti} {et~al.}(2024){Curti}, {Maiolino}, {Curtis-Lake}, {Chevallard}, {Carniani}, {D'Eugenio}, {Looser}, {Scholtz}, {Charlot}, {Cameron}, {{\"U}bler}, {Witstok}, {Boyett}, {Laseter}, {Sandles}, {Arribas}, {Bunker}, {Giardino}, {Maseda}, {Rawle}, {Rodr{\'\i}guez Del Pino}, {Smit}, {Willott}, {Eisenstein}, {Hausen}, {Johnson}, {Rieke}, {Robertson}, {Tacchella}, {Williams}, {Willmer}, {Baker}, {Bhatawdekar}, {Egami}, {Helton}, {Ji}, {Kumari}, {Perna}, {Shivaei}, \& {Sun}}]{Curti+2024_mzr}
{Curti}, M., {Maiolino}, R., {Curtis-Lake}, E., {et~al.} 2024, \aap, 684, A75

\bibitem[{{Curti} {et~al.}(2020){Curti}, {Mannucci}, {Cresci}, \& {Maiolino}}]{curti+2020_calib}
{Curti}, M., {Mannucci}, F., {Cresci}, G., \& {Maiolino}, R. 2020, \mnras, 491, 944

\bibitem[{{Curtis-Lake} {et~al.}(2026){Curtis-Lake}, {Cameron}, {Bunker}, {Scholtz}, {Carniani}, {Parlanti}, {D'Eugenio}, {Jakobsen}, {Willmer}, {Arribas}, {Baker}, {Charlot}, {Chevallard}, {Circosta}, {Curti}, {Duan}, {Eisenstein}, {Hainline}, {Ji}, {Johnson}, {Jones}, {Maiolino}, {Maseda}, {Perna}, {P{\'e}rez-Gonz{\'a}lez}, {Rawle}, {Rieke}, {Rinaldi}, {Robertson}, {Rodr{\'\i}guez Del Pino}, {Saxena}, {Shivaei}, {Smit}, {Tacchella}, {{\"U}bler}, {Venturi}, {Williams}, \& {Willott}}]{Curtis-Lake+2026_jadesDR4}
{Curtis-Lake}, E., {Cameron}, A.~J., {Bunker}, A.~J., {et~al.} 2026, \mnras, 549, stag836

\bibitem[{{D'Angelo} {et~al.}(2026){D'Angelo}, {Schneider}, {Graziani}, {Di Cesare}, {Valiante}, {Popping}, {Rodighiero}, {Bisigello}, {Burgarella}, {Gandolfi}, {Otaki}, \& {Zana}}]{DAngelo+2026}
{D'Angelo}, A., {Schneider}, R., {Graziani}, L., {et~al.} 2026, arXiv e-prints, arXiv:2610.40790

\bibitem[{{Dav{\'e}} {et~al.}(2012){Dav{\'e}}, {Finlator}, \& {Oppenheimer}}]{Dave+2012}
{Dav{\'e}}, R., {Finlator}, K., \& {Oppenheimer}, B.~D. 2012, \mnras, 421, 98

\bibitem[{{DESI Collaboration} {et~al.}(2026){DESI Collaboration}, {Abdul Karim}, {Adame}, {Aguado}, {Aguilar}, {Ahlen}, {Alam}, {Aldering}, {Alexander}, {Alfarsy}, {Allen}, {Allende Prieto}, {Alves}, {Anand}, {Andrade}, {Armengaud}, {Avila}, {Aviles}, {Awan}, {Bailey}, {Baleato Lizancos}, {Ballester}, {Bault}, {Bautista}, {Bean}, {Behera}, {BenZvi}, {Beraldo e Silva}, {Bermejo-Climent}, {Beutler}, {Bianchi}, {Blake}, {Blum}, {Bolton}, {Bonici}, {Brieden}, {Brodzeller}, {Brooks}, {Buckley-Geer}, {Burtin}, {Bystr{\"o}m}, {Canning}, {Carnero Rosell}, {Carr}, {Carrilho}, {Casas}, {Castander}, {Cereskaite}, {Cervantes-Cota}, {Chaussidon}, {Chaves-Montero}, {Chen}, {Chen}, {Circosta}, {Claybaugh}, {Cole}, {Cooper}, {Cousinou}, {Cuceu}, {Davis}, {Dawson}, {de Belsunce}, {de la Cruz}, {de la Macorra}, {de Mattia}, {Deiosso}, {Della Costa}, {Demina}, {Demirbozan}, {DeRose}, {Dey}, {Dey}, {Ding}, {Ding}, {Doel}, {Douglass}, {Dowicz}, {Ebina}, {Edelstein}, {Eisenstein}, {Elbers}, {Emas}, {Escoffier}, {Fagrelius},
  {Fan}, {Fanning}, {Favole}, {Fawcett}, {Fern{\'a}ndez-Garc{\'\i}a}, {Ferraro}, {Findlay}, {Font-Ribera}, {Forero-Romero}, {Forero-S{\'a}nchez}, {Frenk}, {G{\"a}nsicke}, {Galbany}, {Garc{\'\i}a-Bellido}, {Garcia-Quintero}, {Garrison}, {Gazta{\~n}aga}, {Gil-Mar{\'\i}n}, {Gloudemans}, {Gnedin}, {Gontcho A Gontcho}, {Gonzalez}, {Gonzalez-Morales}, {Gonzalez-Perez}, {Gordon}, {Graur}, {Green}, {Gruen}, {Gsponer}, {Guandalin}, {Gutierrez}, {Guy}, {Hahn}, {Han}, {Han}, {He}, {Herrera-Alcantar}, {Heydenreich}, {Honscheid}, {Hou}, {Howlett}, {Huterer}, {Ir{\v{s}}i{\v{c}}}, {Ishak}, {Jacques}, {Jiang}, {Jimenez}, {Jing}, {Joachimi}, {Joudaki}, {Joyce}, {Jullo}, {Juneau}, {Kara{\c{c}}ayl{\i}}, {Karim}, {Kehoe}, {Kent}, {Khederlarian}, {Kirkby}, {Kisner}, {Kitaura}, {Kizhuprakkat}, {Kong}, {Koposov}, {Kremin}, {Krolewski}, {Lahav}, {Lai}, {Lamman}, {Lan}, {Landriau}, {Lang}, {Lange}, {Lasker}, {Le Goff}, {Le Guillou}, {Leauthaud}, {Levi}, {Li}, {Li}, {Liu}, {Lodha}, {Lokken}, {Luo}, {Magneville}, {Manera}, {Manser},
  {Margala}, {Martini}, {Maus}, {McCullough}, {McDonald}, {Medina}, {Medina-Varela}, {Meisner}, {Mena-Fern{\'a}ndez}, {Menegas}, {Meneses-Rizo}, {Mezcua}, {Miquel}, {Montero-Camacho}, {Moon}, {Moustakas}, {Mu{\~n}oz-Guti{\'e}rrez}, {Mu noz-Santos}, {Myers}, {Myles}, {Nadathur}, {Najita}, {Napolitano}, {Newman}, {Nikakhtar}, {Nikutta}, {Niz}, {Noriega}, \& {Nugent}}]{DESI2026}
{DESI Collaboration}, {Abdul Karim}, M., {Adame}, A.~G., {et~al.} 2026, \aj, 171, 285

\bibitem[{{D'Eugenio} {et~al.}(2025){D'Eugenio}, {Cameron}, {Scholtz}, {Carniani}, {Willott}, {Curtis-Lake}, {Bunker}, {Parlanti}, {Maiolino}, {Willmer}, {Jakobsen}, {Robertson}, {Johnson}, {Tacchella}, {Cargile}, {Rawle}, {Arribas}, {Chevallard}, {Curti}, {Egami}, {Eisenstein}, {Kumari}, {Looser}, {Rieke}, {Rodr{\'\i}guez Del Pino}, {Saxena}, {{\"U}bler}, {Venturi}, {Witstok}, {Baker}, {Bhatawdekar}, {Bonaventura}, {Boyett}, {Charlot}, {Danhaive}, {Hainline}, {Hausen}, {Helton}, {Ji}, {Ji}, {Jones}, {Juod{\v{z}}balis}, {Maseda}, {P{\'e}rez-Gonz{\'a}lez}, {Perna}, {Pusk{\'a}s}, {Shivaei}, {Silcock}, {Simmonds}, {Smit}, {Sun}, {Villanueva}, {Williams}, \& {Zhu}}]{deugenio+2025_jadesDR3}
{D'Eugenio}, F., {Cameron}, A.~J., {Scholtz}, J., {et~al.} 2025, \apjs, 277, 4

\bibitem[{{D'Eugenio} {et~al.}(2026){D'Eugenio}, {Nelson}, {Eisenstein}, {Maiolino}, {Carniani}, {Scholtz}, {Curti}, {Willmer}, {Bunker}, {Helton}, {Juod{\v{z}}balis}, {Sun}, {Tacchella}, {Arribas}, {Cameron}, {Charlot}, {Curtis-Lake}, {Hainline}, {Johnson}, {Robertson}, {Williams}, {Willott}, {Baker}, {Chevallard}, {Danhaive}, {Isobe}, {Ji}, {Ji}, {Jones}, {Kumari}, {Looser}, {Lyu}, {Parlanti}, {Perna}, {Pusk{\'a}s}, {Rinaldi}, {Simmonds}, {Sun}, {{\"U}bler}, {Venturi}, {Witstok}, {Wu}, \& {Zhu}}]{DEugenio+2026_jadesDH}
{D'Eugenio}, F., {Nelson}, E.~J., {Eisenstein}, D.~J., {et~al.} 2026, \mnras, 549, stag824

\bibitem[{{Di Cesare} {et~al.}(2023){Di Cesare}, {Graziani}, {Schneider}, {Ginolfi}, {Venditti}, {Santini}, \& {Hunt}}]{DiCesare+2023}
{Di Cesare}, C., {Graziani}, L., {Schneider}, R., {et~al.} 2023, \mnras, 519, 4632

\bibitem[{{Di Cesare} {et~al.}(2026){Di Cesare}, {Matthee}, {Naidu}, {Torralba}, {Kotiwale}, {Kramarenko}, {Blaizot}, {Rosdahl}, {Leja}, {Iani}, {Adamo}, {Covelo-Paz}, {Furtak}, {Heintz}, {Mascia}, {Navarrete}, {Oesch}, {Romano}, {Shivaei}, \& {Tacchella}}]{DiCesare+2026}
{Di Cesare}, C., {Matthee}, J., {Naidu}, R.~P., {et~al.} 2026, \aap, 707, A129

\bibitem[{{Dopita} {et~al.}(2016){Dopita}, {Kewley}, {Sutherland}, \& {Nicholls}}]{Dopita+2016}
{Dopita}, M.~A., {Kewley}, L.~J., {Sutherland}, R.~S., \& {Nicholls}, D.~C. 2016, \apss, 361, 61

\bibitem[{{Egami} {et~al.}(2023){Egami}, {Sun}, {Alberts}, {Baum}, {Boyett}, {Bunker}, {Cameron}, {Carniani}, {Charlot}, {Chen}, {Chevallard}, {Curti}, {D'Eugenio}, {Danhaive}, {DeCoursey}, {Dudzeviciute}, {Eisenstein}, {Hainline}, {Helton}, {Ji}, {Johnson}, {Kumari}, {Looser}, {Lyu}, {Ma}, {Maiolino}, {Maseda}, {Nelson}, {Rawle}, {Rieke}, {Robertson}, {Sandles}, {Shivaei}, {Smit}, {Suess}, {Tacchella}, {Uebler}, {Whitler}, {Williams}, {Willmer}, {Willott}, {Witstok}, \& {de Graaff}}]{Egami+2023jwst.prop}
{Egami}, E., {Sun}, F., {Alberts}, S., {et~al.} 2023, {Complete NIRCam Grism Redshift Survey (CONGRESS)}, JWST Proposal. Cycle 2, ID. \#3577

\bibitem[{{Eisenstein} {et~al.}(2025){Eisenstein}, {Johnson}, {Robertson}, {Tacchella}, {Hainline}, {Jakobsen}, {Maiolino}, {Bonaventura}, {Bunker}, {Cameron}, {Cargile}, {Curtis-Lake}, {Hausen}, {Pusk{\'a}s}, {Rieke}, {Sun}, {Willmer}, {Willott}, {Alberts}, {Arribas}, {Baker}, {Baum}, {Bhatawdekar}, {Carniani}, {Charlot}, {Chen}, {Chevallard}, {Curti}, {DeCoursey}, {D'Eugenio}, {de Graaff}, {Egami}, {Helton}, {Ji}, {Jones}, {Kumari}, {L{\"u}tzgendorf}, {Laseter}, {Looser}, {Lyu}, {Maseda}, {Nelson}, {Parlanti}, {Rauscher}, {Rawle}, {Rieke}, {Rix}, {Rujopakarn}, {Sandles}, {Saxena}, {Scholtz}, {Sharpe}, {Shivaei}, {Simmonds}, {Smit}, {Topping}, {{\"U}bler}, {Venturi}, {Williams}, {Witstok}, \& {Woodrum}}]{Eisenstein+2025_jof}
{Eisenstein}, D.~J., {Johnson}, B.~D., {Robertson}, B., {et~al.} 2025, \apjs, 281, 50

\bibitem[{{Eisenstein} {et~al.}(2026){Eisenstein}, {Willott}, {Alberts}, {Arribas}, {Bonaventura}, {Bunker}, {Cameron}, {Carniani}, {Charlot}, {Curtis-Lake}, {D'Eugenio}, {Ferruit}, {Giardino}, {Hainline}, {Hausen}, {Jakobsen}, {Johnson}, {Maiolino}, {Rauscher}, {Rieke}, {Rieke}, {Rix}, {Robertson}, {Stark}, {Tacchella}, {Williams}, {Willmer}, {Baker}, {Baum}, {Bhatawdekar}, {Boyett}, {Chen}, {Chevallard}, {Circosta}, {Curti}, {Danhaive}, {DeCoursey}, {Endsley}, {de Graaff}, {Dressler}, {Egami}, {Helton}, {Hviding}, {Ji}, {Jones}, {Kumari}, {L{\"u}tzgendorf}, {Laseter}, {Looser}, {Lyu}, {Maseda}, {Nelson}, {Parlanti}, {Perna}, {Pusk{\'a}s}, {Rawle}, {Rodr{\'\i}guez Del Pino}, {Rujopakarn}, {Sandles}, {Saxena}, {Scholtz}, {Sharpe}, {Shivaei}, {Silcock}, {Simmonds}, {Skarbinski}, {Smit}, {Stone}, {Suess}, {Sun}, {Tang}, {Topping}, {{\"U}bler}, {Villanueva}, {Wallace}, {Whitler}, {Witstok}, \& {Woodrum}}]{Eisenstein+2026_jades}
{Eisenstein}, D.~J., {Willott}, C., {Alberts}, S., {et~al.} 2026, \apjs, 283, 6

\bibitem[{{Endsley} {et~al.}(2025){Endsley}, {Chisholm}, {Stark}, {Topping}, \& {Whitler}}]{Endsley+2024}
{Endsley}, R., {Chisholm}, J., {Stark}, D.~P., {Topping}, M.~W., \& {Whitler}, L. 2025, \apj, 987, 189

\bibitem[{{Esteban} {et~al.}(2025){Esteban}, {M{\'e}ndez-Delgado}, {Garc{\'\i}a-Rojas}, {Arellano-C{\'o}rdova}, {Rosales-Ortega}, {Orte-Garc{\'\i}a}, {Reyes-Rodr{\'\i}guez}, {Carigi}, \& {Amayo}}]{Esteban+2025}
{Esteban}, C., {M{\'e}ndez-Delgado}, J.~E., {Garc{\'\i}a-Rojas}, J., {et~al.} 2025, \aap, 697, A61

\bibitem[{{Faisst} {et~al.}(2026){Faisst}, {Liu}, {Dubois}, {Nakazato}, {Osman}, {Pallottini}, {Vallini}, {Fujimoto}, {Mobasher}, {Wang}, {Lin}, {Amor{\'\i}n}, {Aravena}, {Assef}, {Battisti}, {B{\'e}thermin}, {Boquien}, {Cassata}, {da Cunha}, {Dam}, {de Lucia}, {De Looze}, {Dessauges-Zavadsky}, {Ferrara}, {Finner}, {Fontanot}, {Ginolfi}, {G{\'o}mez-Espinoza}, {Gruppioni}, {Guti{\'e}rrez-Vera}, {Hadi}, {Herrera-Camus}, {Hirschmann}, {Ibar}, {Inami}, {Kartaltepe}, {Koekemoer}, {Kohandel}, {Lee}, {Li}, {Molina}, {Nanni}, {Narayanan}, {Pozzi}, {Relano}, {Romano}, {Sanders}, {Silverman}, {Sommovigo}, {Spilker}, {Tsujita}, {{\"U}bler}, {G.~C.}, {Veraldi}, {Villanueva}, {Xie}, \& {Zamorani}}]{faisst+2026_ifu}
{Faisst}, A.~L., {Liu}, L.-J., {Dubois}, Y., {et~al.} 2026, \apj, 1004, 22

\bibitem[{{Feldmann} {et~al.}(2023){Feldmann}, {Quataert}, {Faucher-Gigu{\`e}re}, {Hopkins}, {{\c{C}}atmabacak}, {Kere{\v{s}}}, {Bassini}, {Bernardini}, {Bullock}, {Cenci}, {Gensior}, {Liang}, {Moreno}, \& {Wetzel}}]{Feldmann+2023}
{Feldmann}, R., {Quataert}, E., {Faucher-Gigu{\`e}re}, C.-A., {et~al.} 2023, \mnras, 522, 3831

\bibitem[{{Ferland} {et~al.}(2017){Ferland}, {Chatzikos}, {Guzm{\'a}n}, {Lykins}, {van Hoof}, {Williams}, {Abel}, {Badnell}, {Keenan}, {Porter}, \& {Stancil}}]{Ferland+2017}
{Ferland}, G.~J., {Chatzikos}, M., {Guzm{\'a}n}, F., {et~al.} 2017, \rmxaa, 53, 385

\bibitem[{{Finkelstein} {et~al.}(2022){Finkelstein}, {Bagley}, {Song}, {Larson}, {Papovich}, {Dickinson}, {Finkelstein}, {Koekemoer}, {Pirzkal}, {Somerville}, {Yung}, {Behroozi}, {Ferguson}, {Giavalisco}, {Grogin}, {Hathi}, {Hutchison}, {Jung}, {Kocevski}, {Kawinwanichakij}, {Rojas-Ruiz}, {Ryan}, {Snyder}, \& {Tacchella}}]{Finkelstein+2022}
{Finkelstein}, S.~L., {Bagley}, M., {Song}, M., {et~al.} 2022, \apj, 928, 52

\bibitem[{{Furtak} {et~al.}(2023){Furtak}, {Zitrin}, {Weaver}, {Atek}, {Bezanson}, {Labb{\'e}}, {Whitaker}, {Leja}, {Price}, {Brammer}, {Wang}, {Marchesini}, {Pan}, {Dayal}, {van Dokkum}, {Feldmann}, {Fujimoto}, {Franx}, {Khullar}, {Nelson}, \& {Mowla}}]{Furtak+2023}
{Furtak}, L.~J., {Zitrin}, A., {Weaver}, J.~R., {et~al.} 2023, \mnras, 523, 4568

\bibitem[{{Giavalisco} {et~al.}(2004){Giavalisco}, {Ferguson}, {Koekemoer}, {Dickinson}, {Alexander}, {Bauer}, {Bergeron}, {Biagetti}, {Brandt}, {Casertano}, {Cesarsky}, {Chatzichristou}, {Conselice}, {Cristiani}, {Da Costa}, {Dahlen}, {de Mello}, {Eisenhardt}, {Erben}, {Fall}, {Fassnacht}, {Fosbury}, {Fruchter}, {Gardner}, {Grogin}, {Hook}, {Hornschemeier}, {Idzi}, {Jogee}, {Kretchmer}, {Laidler}, {Lee}, {Livio}, {Lucas}, {Madau}, {Mobasher}, {Moustakas}, {Nonino}, {Padovani}, {Papovich}, {Park}, {Ravindranath}, {Renzini}, {Richardson}, {Riess}, {Rosati}, {Schirmer}, {Schreier}, {Somerville}, {Spinrad}, {Stern}, {Stiavelli}, {Strolger}, {Urry}, {Vandame}, {Williams}, \& {Wolf}}]{Giavalisco+2004}
{Giavalisco}, M., {Ferguson}, H.~C., {Koekemoer}, A.~M., {et~al.} 2004, \apjl, 600, L93

\bibitem[{{Graziani} {et~al.}(2020){Graziani}, {Schneider}, {Ginolfi}, {Hunt}, {Maio}, {Glatzle}, \& {Ciardi}}]{Graziani+2020_dustyG}
{Graziani}, L., {Schneider}, R., {Ginolfi}, M., {et~al.} 2020, \mnras, 494, 1071

\bibitem[{{Grogin} {et~al.}(2011){Grogin}, {Kocevski}, {Faber}, {Ferguson}, {Koekemoer}, {Riess}, {Acquaviva}, {Alexander}, {Almaini}, {Ashby}, {Barden}, {Bell}, {Bournaud}, {Brown}, {Caputi}, {Casertano}, {Cassata}, {Castellano}, {Challis}, {Chary}, {Cheung}, {Cirasuolo}, {Conselice}, {Roshan Cooray}, {Croton}, {Daddi}, {Dahlen}, {Dav{\'e}}, {de Mello}, {Dekel}, {Dickinson}, {Dolch}, {Donley}, {Dunlop}, {Dutton}, {Elbaz}, {Fazio}, {Filippenko}, {Finkelstein}, {Fontana}, {Gardner}, {Garnavich}, {Gawiser}, {Giavalisco}, {Grazian}, {Guo}, {Hathi}, {H{\"a}ussler}, {Hopkins}, {Huang}, {Huang}, {Jha}, {Kartaltepe}, {Kirshner}, {Koo}, {Lai}, {Lee}, {Li}, {Lotz}, {Lucas}, {Madau}, {McCarthy}, {McGrath}, {McIntosh}, {McLure}, {Mobasher}, {Moustakas}, {Mozena}, {Nandra}, {Newman}, {Niemi}, {Noeske}, {Papovich}, {Pentericci}, {Pope}, {Primack}, {Rajan}, {Ravindranath}, {Reddy}, {Renzini}, {Rix}, {Robaina}, {Rodney}, {Rosario}, {Rosati}, {Salimbeni}, {Scarlata}, {Siana}, {Simard}, {Smidt}, {Somerville}, {Spinrad},
  {Straughn}, {Strolger}, {Telford}, {Teplitz}, {Trump}, {van der Wel}, {Villforth}, {Wechsler}, {Weiner}, {Wiklind}, {Wild}, {Wilson}, {Wuyts}, {Yan}, \& {Yun}}]{Grogin+2011}
{Grogin}, N.~A., {Kocevski}, D.~D., {Faber}, S.~M., {et~al.} 2011, \apjs, 197, 35

\bibitem[{{Guo} {et~al.}(2016){Guo}, {Koo}, {Lu}, {Forbes}, {Rafelski}, {Trump}, {Amor{\'\i}n}, {Barro}, {Dav{\'e}}, {Faber}, {Hathi}, {Yesuf}, {Cooper}, {Dekel}, {Guhathakurta}, {Kirby}, {Koekemoer}, {P{\'e}rez-Gonz{\'a}lez}, {Lin}, {Newman}, {Primack}, {Rosario}, {Willmer}, \& {Yan}}]{Guo+2016}
{Guo}, Y., {Koo}, D.~C., {Lu}, Y., {et~al.} 2016, \apj, 822, 103

\bibitem[{{Hayden-Pawson} {et~al.}(2022){Hayden-Pawson}, {Curti}, {Maiolino}, {Cirasuolo}, {Belfiore}, {Cappellari}, {Concas}, {Cresci}, {Cullen}, {Kobayashi}, {Mannucci}, {Marconi}, {Meneghetti}, {Mercurio}, {Peng}, {Swinbank}, \& {Vincenzo}}]{Hayden-Pawson+2022}
{Hayden-Pawson}, C., {Curti}, M., {Maiolino}, R., {et~al.} 2022, \mnras, 512, 2867

\bibitem[{{He} {et~al.}(2024){He}, {Wang}, {Jones}, {Treu}, {Glazebrook}, {Malkan}, {Vulcani}, {Metha}, {Brada{\v{c}}}, {Brammer}, {Roberts-Borsani}, {Strait}, {Bonchi}, {Castellano}, {Fontana}, {Mason}, {Merlin}, {Morishita}, {Paris}, {Santini}, {Trenti}, {Boyett}, \& {Grasha}}]{He+2024}
{He}, X., {Wang}, X., {Jones}, T., {et~al.} 2024, \apjl, 960, L13

\bibitem[{{Heintz} {et~al.}(2023){Heintz}, {Brammer}, {Gim{\'e}nez-Arteaga}, {Strait}, {Lagos}, {Vijayan}, {Matthee}, {Watson}, {Mason}, {Hutter}, {Toft}, {Fynbo}, \& {Oesch}}]{Heintz+2023}
{Heintz}, K.~E., {Brammer}, G.~B., {Gim{\'e}nez-Arteaga}, C., {et~al.} 2023, Nature Astronomy, 7, 1517

\bibitem[{{Henry} {et~al.}(2000){Henry}, {Edmunds}, \& {K{\"o}ppen}}]{Henry+2000}
{Henry}, R.~B.~C., {Edmunds}, M.~G., \& {K{\"o}ppen}, J. 2000, \apj, 541, 660

\bibitem[{{Hopkins} {et~al.}(2018){Hopkins}, {Wetzel}, {Kere{\v{s}}}, {Faucher-Gigu{\`e}re}, {Quataert}, {Boylan-Kolchin}, {Murray}, {Hayward}, {Garrison-Kimmel}, {Hummels}, {Feldmann}, {Torrey}, {Ma}, {Angl{\'e}s-Alc{\'a}zar}, {Su}, {Orr}, {Schmitz}, {Escala}, {Sanderson}, {Grudi{\'c}}, {Hafen}, {Kim}, {Fitts}, {Bullock}, {Wheeler}, {Chan}, {Elbert}, \& {Narayanan}}]{HOpkins2018_FIRE2}
{Hopkins}, P.~F., {Wetzel}, A., {Kere{\v{s}}}, D., {et~al.} 2018, \mnras, 480, 800

\bibitem[{{Hsiao} {et~al.}(2026){Hsiao}, {Berg}, {Finkelstein}, {Gupta}, {Martinez}, {Taylor}, {Akins}, {Chavez Ortiz}, {Chisholm}, {Furtak}, \& {Kokorev}}]{Hsiao+2026_ne}
{Hsiao}, T. Y.-Y., {Berg}, D.~A., {Finkelstein}, S.~L., {et~al.} 2026, arXiv e-prints, arXiv:2608.20339

\bibitem[{{Hunter}(2007)}]{Hunter2007}
{Hunter}, J.~D. 2007, Computing in Science and Engineering, 9, 90

\bibitem[{{Isobe} {et~al.}(2026){Isobe}, {Curti}, {Maiolino}, {Duan}, {McClymont}, {Pusk{\'a}s}, {D'Eugenio}, {Rinaldi}, {Trussler}, {Scholtz}, {Looser}, {Nelson}, {Ji}, {Langeroodi}, {Tacchella}, {Jones}, {Juod{\v{z}}balis}, {Pascalau}, {Hsiao}, {{\"U}bler}, {Baker}, {Bunker}, {Carniani}, {Charlot}, {Curtis-Lake}, {Geris}, {Koller}, {Lyu}, {Robertson}, {Williams}, \& {Wu}}]{Isobe+2026}
{Isobe}, Y., {Curti}, M., {Maiolino}, R., {et~al.} 2026, arXiv e-prints, arXiv:2606.11345

\bibitem[{{Isobe} {et~al.}(2021){Isobe}, {Ouchi}, {Kojima}, {Shibuya}, {Hayashi}, {Rauch}, {Kikuchihara}, {Zhang}, {Ono}, {Fujimoto}, {Harikane}, {Kim}, {Komiyama}, {Kusakabe}, {Lee}, {Mawatari}, {Onodera}, {Sugahara}, \& {Yabe}}]{Isobe+2021}
{Isobe}, Y., {Ouchi}, M., {Kojima}, T., {et~al.} 2021, \apj, 918, 54

\bibitem[{{Isobe} {et~al.}(2023{\natexlab{a}}){Isobe}, {Ouchi}, {Nakajima}, {Harikane}, {Ono}, {Xu}, {Zhang}, \& {Umeda}}]{Isobe+2023}
{Isobe}, Y., {Ouchi}, M., {Nakajima}, K., {et~al.} 2023{\natexlab{a}}, \apj, 956, 139

\bibitem[{{Isobe} {et~al.}(2023{\natexlab{b}}){Isobe}, {Ouchi}, {Tominaga}, {Watanabe}, {Nakajima}, {Umeda}, {Yajima}, {Harikane}, {Fukushima}, {Xu}, {Ono}, \& {Zhang}}]{Isobe+2023_Nemitter}
{Isobe}, Y., {Ouchi}, M., {Tominaga}, N., {et~al.} 2023{\natexlab{b}}, \apj, 959, 100

\bibitem[{{Izotov} {et~al.}(2006){Izotov}, {Stasi{\'n}ska}, {Meynet}, {Guseva}, \& {Thuan}}]{Izotov+2006}
{Izotov}, Y.~I., {Stasi{\'n}ska}, G., {Meynet}, G., {Guseva}, N.~G., \& {Thuan}, T.~X. 2006, \aap, 448, 955

\bibitem[{{Johnson} {et~al.}(2021){Johnson}, {Leja}, {Conroy}, \& {Speagle}}]{Johnson+2021}
{Johnson}, B.~D., {Leja}, J., {Conroy}, C., \& {Speagle}, J.~S. 2021, \apjs, 254, 22

\bibitem[{{Kaasinen} {et~al.}(2017){Kaasinen}, {Bian}, {Groves}, {Kewley}, \& {Gupta}}]{Kaasinen+2017}
{Kaasinen}, M., {Bian}, F., {Groves}, B., {Kewley}, L.~J., \& {Gupta}, A. 2017, \mnras, 465, 3220

\bibitem[{{Kannan} {et~al.}(2022){Kannan}, {Garaldi}, {Smith}, {Pakmor}, {Springel}, {Vogelsberger}, \& {Hernquist}}]{Kannan+2022-thesan}
{Kannan}, R., {Garaldi}, E., {Smith}, A., {et~al.} 2022, \mnras, 511, 4005

\bibitem[{{Kannan} {et~al.}(2025){Kannan}, {Puchwein}, {Smith}, {Borrow}, {Garaldi}, {Keating}, {Vogelsberger}, {Zier}, {McClymont}, {Shen}, {Popovic}, {Tacchella}, {Hernquist}, \& {Springel}}]{Kannan+2025_thesanzoom}
{Kannan}, R., {Puchwein}, E., {Smith}, A., {et~al.} 2025, The Open Journal of Astrophysics, 8, 153

\bibitem[{{Kapoor} {et~al.}(2026){Kapoor}, {Matthee}, {Torralba}, {Kramarenko}, {Bordoloi}, {Greene}, {Iani}, {Kashino}, {Liu}, {Mackenzie}, {Mascia}, {Naidu}, \& {Simcoe}}]{Kapoor+2026_lrds}
{Kapoor}, S., {Matthee}, J., {Torralba}, A., {et~al.} 2026, arXiv e-prints, arXiv:2607.00084

\bibitem[{{Karthikeyan} {et~al.}(2026){Karthikeyan}, {Clarke}, {Shapley}, {Lam}, {Sanders}, {Reddy}, {Topping}, \& {Brammer}}]{Karthikeyan+2026}
{Karthikeyan}, S., {Clarke}, L., {Shapley}, A.~E., {et~al.} 2026, \apj, 1009, 83

\bibitem[{{Kashino} {et~al.}(2023){Kashino}, {Lilly}, {Matthee}, {Eilers}, {Mackenzie}, {Bordoloi}, \& {Simcoe}}]{Kashino+2023}
{Kashino}, D., {Lilly}, S.~J., {Matthee}, J., {et~al.} 2023, \apj, 950, 66

\bibitem[{{Kashino} {et~al.}(2026){Kashino}, {Lilly}, {Matthee}, {Mackenzie}, {Eilers}, {Bordoloi}, {Simcoe}, {Naidu}, {Yue}, \& {Liu}}]{Kashino2026}
{Kashino}, D., {Lilly}, S.~J., {Matthee}, J., {et~al.} 2026, \apj, 997, 280

\bibitem[{{Kashino} {et~al.}(2017){Kashino}, {Silverman}, {Sanders}, {Kartaltepe}, {Daddi}, {Renzini}, {Valentino}, {Rodighiero}, {Juneau}, {Kewley}, {Zahid}, {Arimoto}, {Nagao}, {Chu}, {Sugiyama}, {Civano}, {Ilbert}, {Kajisawa}, {Le F{\`e}vre}, {Maier}, {Masters}, {Miyaji}, {Onodera}, {Puglisi}, \& {Taniguchi}}]{Kashino+2017}
{Kashino}, D., {Silverman}, J.~D., {Sanders}, D., {et~al.} 2017, \apj, 835, 88

\bibitem[{{Katz} {et~al.}(2023){Katz}, {Rosdahl}, {Kimm}, {Blaizot}, {Choustikov}, {Farcy}, {Garel}, {Haehnelt}, {Michel-Dansac}, \& {Ocvirk}}]{Katz+2023_sphinx}
{Katz}, H., {Rosdahl}, J., {Kimm}, T., {et~al.} 2023, The Open Journal of Astrophysics, 6, 44

\bibitem[{{Kewley} \& {Dopita}(2002)}]{KewleyDopita2002}
{Kewley}, L.~J. \& {Dopita}, M.~A. 2002, \apjs, 142, 35

\bibitem[{{Kewley} {et~al.}(2001){Kewley}, {Dopita}, {Sutherland}, {Heisler}, \& {Trevena}}]{Kewley+2001}
{Kewley}, L.~J., {Dopita}, M.~A., {Sutherland}, R.~S., {Heisler}, C.~A., \& {Trevena}, J. 2001, \apj, 556, 121

\bibitem[{{Kewley} {et~al.}(2019){Kewley}, {Nicholls}, \& {Sutherland}}]{Kewley+2019}
{Kewley}, L.~J., {Nicholls}, D.~C., \& {Sutherland}, R.~S. 2019, \araa, 57, 511

\bibitem[{{Kobayashi} {et~al.}(2020){Kobayashi}, {Karakas}, \& {Lugaro}}]{Kobayashi+2020}
{Kobayashi}, C., {Karakas}, A.~I., \& {Lugaro}, M. 2020, \apj, 900, 179

\bibitem[{{Koekemoer} {et~al.}(2011){Koekemoer}, {Faber}, {Ferguson}, {Grogin}, {Kocevski}, {Koo}, {Lai}, {Lotz}, {Lucas}, {McGrath}, {Ogaz}, {Rajan}, {Riess}, {Rodney}, {Strolger}, {Casertano}, {Castellano}, {Dahlen}, {Dickinson}, {Dolch}, {Fontana}, {Giavalisco}, {Grazian}, {Guo}, {Hathi}, {Huang}, {van der Wel}, {Yan}, {Acquaviva}, {Alexander}, {Almaini}, {Ashby}, {Barden}, {Bell}, {Bournaud}, {Brown}, {Caputi}, {Cassata}, {Challis}, {Chary}, {Cheung}, {Cirasuolo}, {Conselice}, {Roshan Cooray}, {Croton}, {Daddi}, {Dav{\'e}}, {de Mello}, {de Ravel}, {Dekel}, {Donley}, {Dunlop}, {Dutton}, {Elbaz}, {Fazio}, {Filippenko}, {Finkelstein}, {Frazer}, {Gardner}, {Garnavich}, {Gawiser}, {Gruetzbauch}, {Hartley}, {H{\"a}ussler}, {Herrington}, {Hopkins}, {Huang}, {Jha}, {Johnson}, {Kartaltepe}, {Khostovan}, {Kirshner}, {Lani}, {Lee}, {Li}, {Madau}, {McCarthy}, {McIntosh}, {McLure}, {McPartland}, {Mobasher}, {Moreira}, {Mortlock}, {Moustakas}, {Mozena}, {Nandra}, {Newman}, {Nielsen}, {Niemi}, {Noeske}, {Papovich},
  {Pentericci}, {Pope}, {Primack}, {Ravindranath}, {Reddy}, {Renzini}, {Rix}, {Robaina}, {Rosario}, {Rosati}, {Salimbeni}, {Scarlata}, {Siana}, {Simard}, {Smidt}, {Snyder}, {Somerville}, {Spinrad}, {Straughn}, {Telford}, {Teplitz}, {Trump}, {Vargas}, {Villforth}, {Wagner}, {Wandro}, {Wechsler}, {Weiner}, {Wiklind}, {Wild}, {Wilson}, {Wuyts}, \& {Yun}}]{Koekemoer+2011}
{Koekemoer}, A.~M., {Faber}, S.~M., {Ferguson}, H.~C., {et~al.} 2011, \apjs, 197, 36

\bibitem[{{Kojima} {et~al.}(2020){Kojima}, {Ouchi}, {Rauch}, {Ono}, {Nakajima}, {Isobe}, {Fujimoto}, {Harikane}, {Hashimoto}, {Hayashi}, {Komiyama}, {Kusakabe}, {Kim}, {Lee}, {Mukae}, {Nagao}, {Onodera}, {Shibuya}, {Sugahara}, {Umemura}, \& {Yabe}}]{Kojima+2020}
{Kojima}, T., {Ouchi}, M., {Rauch}, M., {et~al.} 2020, \apj, 898, 142

\bibitem[{{Kotiwale} {et~al.}(2026){Kotiwale}, {Matthee}, {Kashino}, {Vijayan}, {Torralba}, {Di Cesare}, {Iani}, {Bordoloi}, {Leja}, {Maseda}, {Tacchella}, {Shivaei}, {Heintz}, {Danhaive}, {Mascia}, {Kramarenko}, {Navarrete}, {Mackenzie}, {Naidu}, \& {Sobral}}]{Kotiwale+2026}
{Kotiwale}, G., {Matthee}, J., {Kashino}, D., {et~al.} 2026, \aap, 706, A165

\bibitem[{{Kramarenko} {et~al.}(2025){Kramarenko}, {Rosdahl}, {Blaizot}, {Matthee}, {Katz}, \& {Di Cesare}}]{Kramarenko+2025_sphinx}
{Kramarenko}, I.~G., {Rosdahl}, J., {Blaizot}, J., {et~al.} 2025, subm. to \aap, arXiv:2509.05403

\bibitem[{{Kroupa}(2001)}]{Kroupa2001}
{Kroupa}, P. 2001, \mnras, 322, 231

\bibitem[{{Lam} {et~al.}(2026){Lam}, {Clarke}, {Shapley}, {Sanders}, {Topping}, {Brammer}, {Reddy}, \& {Karthikeyan}}]{Lam+2026}
{Lam}, N., {Clarke}, L., {Shapley}, A.~E., {et~al.} 2026, arXiv e-prints, arXiv:2605.30513

\bibitem[{{Langan} {et~al.}(2020){Langan}, {Ceverino}, \& {Finlator}}]{Langan+2020_firstlight}
{Langan}, I., {Ceverino}, D., \& {Finlator}, K. 2020, \mnras, 494, 1988

\bibitem[{{Langan} {et~al.}(2023){Langan}, {Zabl}, {Bouch{\'e}}, {Ginolfi}, {Popping}, {Schroetter}, {Wendt}, {Schaye}, {Boogaard}, {Freundlich}, {Richard}, {Matthee}, {Mercier}, {Contini}, {Guo}, \& {Cherrey}}]{Langan+2023}
{Langan}, I., {Zabl}, J., {Bouch{\'e}}, N.~F., {et~al.} 2023, \mnras, 521, 546

\bibitem[{{Lara-L{\'o}pez} {et~al.}(2010){Lara-L{\'o}pez}, {Bongiovanni}, {Cepa}, {P{\'e}rez Garc{\'\i}a}, {S{\'a}nchez-Portal}, {Casta{\~n}eda}, {Fern{\'a}ndez Lorenzo}, \& {Povi{\'c}}}]{Lara-Lopez+2010}
{Lara-L{\'o}pez}, M.~A., {Bongiovanni}, A., {Cepa}, J., {et~al.} 2010, \aap, 519, A31

\bibitem[{{Laseter} {et~al.}(2024{\natexlab{a}}){Laseter}, {Maseda}, {Curti}, {Maiolino}, {D'Eugenio}, {Cameron}, {Looser}, {Arribas}, {Baker}, {Bhatawdekar}, {Boyett}, {Bunker}, {Carniani}, {Charlot}, {Chevallard}, {Curtis-lake}, {Egami}, {Eisenstein}, {Hainline}, {Hausen}, {Ji}, {Kumari}, {Perna}, {Rawle}, {Rix}, {Robertson}, {Rodr{\'\i}guez Del Pino}, {Sandles}, {Scholtz}, {Smit}, {Tacchella}, {{\"U}bler}, {Williams}, {Willott}, \& {Witstok}}]{Laseter+2024_calib}
{Laseter}, I.~H., {Maseda}, M.~V., {Curti}, M., {et~al.} 2024{\natexlab{a}}, \aap, 681, A70

\bibitem[{{Laseter} {et~al.}(2024{\natexlab{b}}){Laseter}, {Maseda}, {Curti}, {Maiolino}, {D'Eugenio}, {Cameron}, {Looser}, {Arribas}, {Baker}, {Bhatawdekar}, {Boyett}, {Bunker}, {Carniani}, {Charlot}, {Chevallard}, {Curtis-lake}, {Egami}, {Eisenstein}, {Hainline}, {Hausen}, {Ji}, {Kumari}, {Perna}, {Rawle}, {Rix}, {Robertson}, {Rodr{\'\i}guez Del Pino}, {Sandles}, {Scholtz}, {Smit}, {Tacchella}, {{\"U}bler}, {Williams}, {Willott}, \& {Witstok}}]{Laseter+2024}
{Laseter}, I.~H., {Maseda}, M.~V., {Curti}, M., {et~al.} 2024{\natexlab{b}}, \aap, 681, A70

\bibitem[{{Leja} {et~al.}(2019){Leja}, {Carnall}, {Johnson}, {Conroy}, \& {Speagle}}]{Leja+2019}
{Leja}, J., {Carnall}, A.~C., {Johnson}, B.~D., {Conroy}, C., \& {Speagle}, J.~S. 2019, \apj, 876, 3

\bibitem[{{Leja} {et~al.}(2015){Leja}, {van Dokkum}, {Franx}, \& {Whitaker}}]{Leja+2015}
{Leja}, J., {van Dokkum}, P.~G., {Franx}, M., \& {Whitaker}, K.~E. 2015, \apj, 798, 115

\bibitem[{{Lewis} {et~al.}(2026){Lewis}, {Maseda}, {de Graaff}, {Leja}, {Wang}, {Rix}, {McConachie}, {Cleri}, {Bezanson}, {Boogaard}, {Brammer}, {Greene}, {Hirschmann}, {Katz}, {Labb{\'e}}, {Matthee}, {Miller}, {Naidu}, {Oesch}, {Setton}, {Suess}, {Weibel}, {Whitaker}, \& {Williams}}]{Lewis+2026_mzr}
{Lewis}, Z., {Maseda}, M.~V., {de Graaff}, A., {et~al.} 2026, \apj, 1005, 159

\bibitem[{{Lilly} {et~al.}(2013){Lilly}, {Carollo}, {Pipino}, {Renzini}, \& {Peng}}]{Lilly+2013}
{Lilly}, S.~J., {Carollo}, C.~M., {Pipino}, A., {Renzini}, A., \& {Peng}, Y. 2013, \apj, 772, 119

\bibitem[{{Liu} \& {Rong}(2026)}]{Liu+2026}
{Liu}, S. \& {Rong}, Y. 2026, arXiv e-prints, arXiv:2606.31549

\bibitem[{{Lovell} {et~al.}(2021){Lovell}, {Vijayan}, {Thomas}, {Wilkins}, {Barnes}, {Irodotou}, \& {Roper}}]{lovell+2021_FLARES}
{Lovell}, C.~C., {Vijayan}, A.~P., {Thomas}, P.~A., {et~al.} 2021, \mnras, 500, 2127

\bibitem[{{Luridiana} {et~al.}(2015){Luridiana}, {Morisset}, \& {Shaw}}]{Luridiana+2015_pyneb}
{Luridiana}, V., {Morisset}, C., \& {Shaw}, R.~A. 2015, \aap, 573, A42

\bibitem[{{Ma} {et~al.}(2016){Ma}, {Hopkins}, {Faucher-Gigu{\`e}re}, {Zolman}, {Muratov}, {Kere{\v{s}}}, \& {Quataert}}]{Ma+2016_mzr}
{Ma}, X., {Hopkins}, P.~F., {Faucher-Gigu{\`e}re}, C.-A., {et~al.} 2016, \mnras, 456, 2140

\bibitem[{{Ma} {et~al.}(2018){Ma}, {Hopkins}, {Garrison-Kimmel}, {Faucher-Gigu{\`e}re}, {Quataert}, {Boylan-Kolchin}, {Hayward}, {Feldmann}, \& {Kere{\v{s}}}}]{Ma+2018_FIRE2sr}
{Ma}, X., {Hopkins}, P.~F., {Garrison-Kimmel}, S., {et~al.} 2018, \mnras, 478, 1694

\bibitem[{{Mainali} {et~al.}(2023){Mainali}, {Stark}, {Jones}, {Ellis}, {Hezaveh}, \& {Rigby}}]{Mainali+2023}
{Mainali}, R., {Stark}, D.~P., {Jones}, T., {et~al.} 2023, \mnras, 520, 4037

\bibitem[{{Maiolino} \& {Mannucci}(2019)}]{MaiolinoMannucci2019}
{Maiolino}, R. \& {Mannucci}, F. 2019, \aapr, 27, 3

\bibitem[{{Mannucci} {et~al.}(2010){Mannucci}, {Cresci}, {Maiolino}, {Marconi}, \& {Gnerucci}}]{Mannucci+2010}
{Mannucci}, F., {Cresci}, G., {Maiolino}, R., {Marconi}, A., \& {Gnerucci}, A. 2010, \mnras, 408, 2115

\bibitem[{{Marino} {et~al.}(2013){Marino}, {Rosales-Ortega}, {S{\'a}nchez}, {Gil de Paz}, {V{\'\i}lchez}, {Miralles-Caballero}, {Kehrig}, {P{\'e}rez-Montero}, {Stanishev}, {Iglesias-P{\'a}ramo}, {D{\'\i}az}, {Castillo-Morales}, {Kennicutt}, {L{\'o}pez-S{\'a}nchez}, {Galbany}, {Garc{\'\i}a-Benito}, {Mast}, {Mendez-Abreu}, {Monreal-Ibero}, {Husemann}, {Walcher}, {Garc{\'\i}a-Lorenzo}, {Masegosa}, {Del Olmo Orozco}, {Mour{\~a}o}, {Ziegler}, {Moll{\'a}}, {Papaderos}, {S{\'a}nchez-Bl{\'a}zquez}, {Gonz{\'a}lez Delgado}, {Falc{\'o}n-Barroso}, {Roth}, {van de Ven}, \& {CALIFA Team}}]{Marino+2013}
{Marino}, R.~A., {Rosales-Ortega}, F.~F., {S{\'a}nchez}, S.~F., {et~al.} 2013, \aap, 559, A114

\bibitem[{{Marszewski} {et~al.}(2024){Marszewski}, {Sun}, {Faucher-Gigu{\`e}re}, {Hayward}, \& {Feldmann}}]{Marszewski+2024_mzrFIRE2}
{Marszewski}, A., {Sun}, G., {Faucher-Gigu{\`e}re}, C.-A., {Hayward}, C.~C., \& {Feldmann}, R. 2024, \apjl, 967, L41

\bibitem[{{Martinez} {et~al.}(2025){Martinez}, {Berg}, {James}, {Arellano-C{\'o}rdova}, {Stark}, {Senchyna}, {Skillman}, {Rogers}, \& {Chisholm}}]{Martinez+2025}
{Martinez}, Z., {Berg}, D.~A., {James}, B.~L., {et~al.} 2025, \apj, 995, 204

\bibitem[{{Matthee} {et~al.}(2024){Matthee}, {Naidu}, {Brammer}, {Chisholm}, {Eilers}, {Goulding}, {Greene}, {Kashino}, {Labbe}, {Lilly}, {Mackenzie}, {Oesch}, {Weibel}, {Wuyts}, {Xiao}, {Bordoloi}, {Bouwens}, {van Dokkum}, {Illingworth}, {Kramarenko}, {Maseda}, {Mason}, {Meyer}, {Nelson}, {Reddy}, {Shivaei}, {Simcoe}, \& {Yue}}]{Matthee+2024_lrd}
{Matthee}, J., {Naidu}, R.~P., {Brammer}, G., {et~al.} 2024, \apj, 963, 129

\bibitem[{{Matthee} {et~al.}(2025){Matthee}, {Naidu}, {Kotiwale}, {Furtak}, {Kramarenko}, {Mackenzie}, {Greene}, {Adamo}, {Bouwens}, {Di Cesare}, {Eilers}, {de Graaff}, {Heintz}, {Kashino}, {Maseda}, {Tacchella}, \& {Torralba}}]{Matthee+2024_env}
{Matthee}, J., {Naidu}, R.~P., {Kotiwale}, G., {et~al.} 2025, \apj, 988, 246

\bibitem[{{Matthee} \& {Schaye}(2018)}]{Matthee+2018}
{Matthee}, J. \& {Schaye}, J. 2018, \mnras, 479, L34

\bibitem[{{Matthee} {et~al.}(2018){Matthee}, {Sobral}, {Gronke}, {Paulino-Afonso}, {Stefanon}, \& {R{\"o}ttgering}}]{Matthee+2018_cola1}
{Matthee}, J., {Sobral}, D., {Gronke}, M., {et~al.} 2018, \aap, 619, A136

\bibitem[{{McClymont} {et~al.}(2026){McClymont}, {Tacchella}, {Smith}, {Kannan}, {Garaldi}, {Puchwein}, {Isobe}, {Ji}, {Shen}, {Wang}, {Belokurov}, {Borrow}, {D'Eugenio}, {Keating}, {Maiolino}, {Monty}, {Vogelsberger}, \& {Zier}}]{McClymont+2026_mzrTHESAN}
{McClymont}, W., {Tacchella}, S., {Smith}, A., {et~al.} 2026, \mnras, 548, stag016

\bibitem[{{McClymont} {et~al.}(2025){McClymont}, {Tacchella}, {Smith}, {Kannan}, {Puchwein}, {Borrow}, {Garaldi}, {Keating}, {Vogelsberger}, {Zier}, {Shen}, {Popovic}, \& {Simmonds}}]{McClymont+2025_sfms}
{McClymont}, W., {Tacchella}, S., {Smith}, A., {et~al.} 2025, \mnras, 544, 513

\bibitem[{{Merten} {et~al.}(2011){Merten}, {Coe}, {Dupke}, {Massey}, {Zitrin}, {Cypriano}, {Okabe}, {Frye}, {Braglia}, {Jim{\'e}nez-Teja}, {Ben{\'\i}tez}, {Broadhurst}, {Rhodes}, {Meneghetti}, {Moustakas}, {Sodr{\'e}}, {Krick}, \& {Bregman}}]{Merten+2011_a2744}
{Merten}, J., {Coe}, D., {Dupke}, R., {et~al.} 2011, \mnras, 417, 333

\bibitem[{{Meynet} \& {Maeder}(2002)}]{MaynetMaeder2002}
{Meynet}, G. \& {Maeder}, A. 2002, \aap, 390, 561

\bibitem[{{Miller} {et~al.}(2025){Miller}, {Suess}, {Setton}, {Price}, {Labbe}, {Bezanson}, {Brammer}, {Cutler}, {Furtak}, {Leja}, {Pan}, {Wang}, {Weaver}, {Whitaker}, {Dayal}, {de Graaff}, {Feldmann}, {Greene}, {Fujimoto}, {Maseda}, {Nanayakkara}, {Nelson}, {van Dokkum}, \& {Zitrin}}]{Miller+2025}
{Miller}, T.~B., {Suess}, K.~A., {Setton}, D.~J., {et~al.} 2025, \apj, 988, 196

\bibitem[{{Mingozzi} {et~al.}(2022){Mingozzi}, {James}, {Arellano-C{\'o}rdova}, {Berg}, {Senchyna}, {Chisholm}, {Brinchmann}, {Aloisi}, {Amor{\'\i}n}, {Charlot}, {Feltre}, {Hayes}, {Heckman}, {Henry}, {Hernandez}, {Kumari}, {Leitherer}, {Llerena}, {Martin}, {Nanayakkara}, {Ravindranath}, {Skillman}, {Sugahara}, {Wofford}, \& {Xu}}]{Mingozzi+2022}
{Mingozzi}, M., {James}, B.~L., {Arellano-C{\'o}rdova}, K.~Z., {et~al.} 2022, \apj, 939, 110

\bibitem[{{Naidu} {et~al.}(2024){Naidu}, {Matthee}, {Kramarenko}, {Weibel}, {Brammer}, {Oesch}, {Lechner}, {Furtak}, {Di Cesare}, {Torralba}, {Kotiwale}, {Bezanson}, {Bouwens}, {Chandra}, {Claeyssens}, {Danhaive}, {Frebel}, {de Graaff}, {Greene}, {Heintz}, {Ji}, {Kashino}, {Katz}, {Labbe}, {Leja}, {Li}, {Maseda}, {Richard}, {Shivaei}, {Simcoe}, {Sobral}, {Suess}, {Tacchella}, \& {Williams}}]{Naidu+2024_alt}
{Naidu}, R.~P., {Matthee}, J., {Kramarenko}, I., {et~al.} 2024, subm. to the Open Journal of Astrophysics, arXiv:2410.01874

\bibitem[{{Nakajima} {et~al.}(2023){Nakajima}, {Ouchi}, {Isobe}, {Harikane}, {Zhang}, {Ono}, {Umeda}, \& {Oguri}}]{nakajima+2023_mzr}
{Nakajima}, K., {Ouchi}, M., {Isobe}, Y., {et~al.} 2023, \apjs, 269, 33

\bibitem[{{Nakajima} {et~al.}(2022){Nakajima}, {Ouchi}, {Xu}, {Rauch}, {Harikane}, {Nishigaki}, {Isobe}, {Kusakabe}, {Nagao}, {Ono}, {Onodera}, {Sugahara}, {Kim}, {Komiyama}, {Lee}, \& {Zahedy}}]{nakajima+2022_calib}
{Nakajima}, K., {Ouchi}, M., {Xu}, Y., {et~al.} 2022, \apjs, 262, 3

\bibitem[{{Newville} {et~al.}(2014){Newville}, {Stensitzki}, {Allen}, \& {Ingargiola}}]{Newville+2014_lmfit}
{Newville}, M., {Stensitzki}, T., {Allen}, D.~B., \& {Ingargiola}, A. 2014, {LMFIT: Non-Linear Least-Square Minimization and Curve-Fitting for Python}

\bibitem[{{Nicholls} {et~al.}(2017){Nicholls}, {Sutherland}, {Dopita}, {Kewley}, \& {Groves}}]{Nicholls+2017}
{Nicholls}, D.~C., {Sutherland}, R.~S., {Dopita}, M.~A., {Kewley}, L.~J., \& {Groves}, B.~A. 2017, \mnras, 466, 4403

\bibitem[{{Oesch} {et~al.}(2023){Oesch}, {Brammer}, {Naidu}, {Bouwens}, {Chisholm}, {Illingworth}, {Matthee}, {Nelson}, {Qin}, {Reddy}, {Shapley}, {Shivaei}, {van Dokkum}, {Weibel}, {Whitaker}, {Wuyts}, {Covelo-Paz}, {Endsley}, {Fudamoto}, {Giovinazzo}, {Herard-Demanche}, {Kerutt}, {Kramarenko}, {Labbe}, {Leonova}, {Lin}, {Magee}, {Marchesini}, {Maseda}, {Mason}, {Matharu}, {Meyer}, {Neufeld}, {Prieto Lyon}, {Schaerer}, {Sharma}, {Shuntov}, {Smit}, {Stefanon}, {Wyithe}, \& {Xiao}}]{Oesch+2023_fresco}
{Oesch}, P.~A., {Brammer}, G., {Naidu}, R.~P., {et~al.} 2023, \mnras, 525, 2864

\bibitem[{{Oke} \& {Gunn}(1983)}]{Oke+Gunn1983}
{Oke}, J.~B. \& {Gunn}, J.~E. 1983, \apj, 266, 713

\bibitem[{{Oliphant}(2007)}]{Oliphant2007}
{Oliphant}, T.~E. 2007, Computing in Science and Engineering, 9, 10

\bibitem[{{Osterbrock} \& {Ferland}(2006)}]{OsterbrockFerland2006}
{Osterbrock}, D.~E. \& {Ferland}, G.~J. 2006, {Astrophysics of gaseous nebulae and active galactic nuclei}

\bibitem[{{Peeples} \& {Shankar}(2011)}]{PeeplesShankar2011}
{Peeples}, M.~S. \& {Shankar}, F. 2011, \mnras, 417, 2962

\bibitem[{{P{\'e}rez-Montero} \& {Contini}(2009)}]{Perez-Montero_Contini2009}
{P{\'e}rez-Montero}, E. \& {Contini}, T. 2009, \mnras, 398, 949

\bibitem[{{P{\'e}roux} \& {Howk}(2020)}]{PerouxHowk2020}
{P{\'e}roux}, C. \& {Howk}, J.~C. 2020, \araa, 58, 363

\bibitem[{{Perry} {et~al.}(2025){Perry}, {Taylor}, {Ch{\'a}vez Ortiz}, {Finkelstein}, {C.~K. Leung}, {Bagley}, {Fern{\'a}ndez}, {Arrabal Haro}, {Chworowsky}, {Cleri}, {Dickinson}, {Ellis}, {Kartaltepe}, {Koekemoer}, {Pacucci}, {Papovich}, {Pirzkal}, \& {Tacchella}}]{Perry+2025_burstySF}
{Perry}, M.~N., {Taylor}, A.~J., {Ch{\'a}vez Ortiz}, {\'O}.~A., {et~al.} 2025, \apj, 994, 14

\bibitem[{{Planck Collaboration} {et~al.}(2020){Planck Collaboration}, {Aghanim}, {Akrami}, {Ashdown}, {Aumont}, {Baccigalupi}, {Ballardini}, {Banday}, {Barreiro}, {Bartolo}, {Basak}, {Battye}, {Benabed}, {Bernard}, {Bersanelli}, {Bielewicz}, {Bock}, {Bond}, {Borrill}, {Bouchet}, {Boulanger}, {Bucher}, {Burigana}, {Butler}, {Calabrese}, {Cardoso}, {Carron}, {Challinor}, {Chiang}, {Chluba}, {Colombo}, {Combet}, {Contreras}, {Crill}, {Cuttaia}, {de Bernardis}, {de Zotti}, {Delabrouille}, {Delouis}, {Di Valentino}, {Diego}, {Dor{\'e}}, {Douspis}, {Ducout}, {Dupac}, {Dusini}, {Efstathiou}, {Elsner}, {En{\ss}lin}, {Eriksen}, {Fantaye}, {Farhang}, {Fergusson}, {Fernandez-Cobos}, {Finelli}, {Forastieri}, {Frailis}, {Fraisse}, {Franceschi}, {Frolov}, {Galeotta}, {Galli}, {Ganga}, {G{\'e}nova-Santos}, {Gerbino}, {Ghosh}, {Gonz{\'a}lez-Nuevo}, {G{\'o}rski}, {Gratton}, {Gruppuso}, {Gudmundsson}, {Hamann}, {Handley}, {Hansen}, {Herranz}, {Hildebrandt}, {Hivon}, {Huang}, {Jaffe}, {Jones}, {Karakci}, {Keih{\"a}nen},
  {Keskitalo}, {Kiiveri}, {Kim}, {Kisner}, {Knox}, {Krachmalnicoff}, {Kunz}, {Kurki-Suonio}, {Lagache}, {Lamarre}, {Lasenby}, {Lattanzi}, {Lawrence}, {Le Jeune}, {Lemos}, {Lesgourgues}, {Levrier}, {Lewis}, {Liguori}, {Lilje}, {Lilley}, {Lindholm}, {L{\'o}pez-Caniego}, {Lubin}, {Ma}, {Mac{\'\i}as-P{\'e}rez}, {Maggio}, {Maino}, {Mandolesi}, {Mangilli}, {Marcos-Caballero}, {Maris}, {Martin}, {Martinelli}, {Mart{\'\i}nez-Gonz{\'a}lez}, {Matarrese}, {Mauri}, {McEwen}, {Meinhold}, {Melchiorri}, {Mennella}, {Migliaccio}, {Millea}, {Mitra}, {Miville-Desch{\^e}nes}, {Molinari}, {Montier}, {Morgante}, {Moss}, {Natoli}, {N{\o}rgaard-Nielsen}, {Pagano}, {Paoletti}, {Partridge}, {Patanchon}, {Peiris}, {Perrotta}, {Pettorino}, {Piacentini}, {Polastri}, {Polenta}, {Puget}, {Rachen}, {Reinecke}, {Remazeilles}, {Renzi}, {Rocha}, {Rosset}, {Roudier}, {Rubi{\~n}o-Mart{\'\i}n}, {Ruiz-Granados}, {Salvati}, {Sandri}, {Savelainen}, {Scott}, {Shellard}, {Sirignano}, {Sirri}, {Spencer}, {Sunyaev}, {Suur-Uski}, {Tauber}, {Tavagnacco},
  {Tenti}, {Toffolatti}, {Tomasi}, {Trombetti}, {Valenziano}, {Valiviita}, {Van Tent}, {Vibert}, {Vielva}, {Villa}, {Vittorio}, {Wandelt}, {Wehus}, {White}, {White}, {Zacchei}, \& {Zonca}}]{Planck+2020}
{Planck Collaboration}, {Aghanim}, N., {Akrami}, Y., {et~al.} 2020, \aap, 641, A6

\bibitem[{{Pollock} {et~al.}(2026{\natexlab{a}}){Pollock}, {Gottumukkala}, {Heintz}, {Brammer}, {Roberts-Borsani}, {Oesch}, {Witstok}, {Arellano-C{\'o}rdova}, {Cullen}, {Scholte}, {Terp}, {Rowland}, {Sneppen}, {Ito}, {Valentino}, {Matthee}, {Watson}, \& {Toft}}]{Pollock+2026a}
{Pollock}, C.~L., {Gottumukkala}, R., {Heintz}, K.~E., {et~al.} 2026{\natexlab{a}}, \aap, 708, A203

\bibitem[{{Pollock} {et~al.}(2026{\natexlab{b}}){Pollock}, {Heintz}, {Witstok}, {Gottumukkala}, {Brammer}, {Bose}, {Cameron}, {Dayal}, {van Dokkum}, {Fynbo}, {Gelli}, {Hayes}, {Inoue}, {Lagos}, {Laursen}, {Meyer}, {Naidu}, {Oesch}, {Rowland}, {Tanvir}, {Tacchella}, {Terp}, {Valentino}, {Walter}, {Weaver}, \& {Witten}}]{Pollock+2026b}
{Pollock}, C.~L., {Heintz}, K.~E., {Witstok}, J., {et~al.} 2026{\natexlab{b}}, \aap, 711, A298

\bibitem[{{Price} {et~al.}(2025){Price}, {Bezanson}, {Labbe}, {Furtak}, {de Graaff}, {Greene}, {Kokorev}, {Setton}, {Suess}, {Brammer}, {Cutler}, {Leja}, {Pan}, {Wang}, {Weaver}, {Whitaker}, {Atek}, {Burgasser}, {Chemerynska}, {Dayal}, {Feldmann}, {F{\"o}rster Schreiber}, {Fudamoto}, {Fujimoto}, {Glazebrook}, {Goulding}, {Khullar}, {Kriek}, {Marchesini}, {Maseda}, {Miller}, {Muzzin}, {Nanayakkara}, {Nelson}, {Oesch}, {Shipley}, {Smit}, {Taylor}, {Dokkum}, {Williams}, \& {Zitrin}}]{Price+2025}
{Price}, S.~H., {Bezanson}, R., {Labbe}, I., {et~al.} 2025, \apj, 982, 51

\bibitem[{{Reddy} {et~al.}(2015){Reddy}, {Kriek}, {Shapley}, {Freeman}, {Siana}, {Coil}, {Mobasher}, {Price}, {Sanders}, \& {Shivaei}}]{Reddy+2015}
{Reddy}, N.~A., {Kriek}, M., {Shapley}, A.~E., {et~al.} 2015, \apj, 806, 259

\bibitem[{{Reddy} {et~al.}(2012){Reddy}, {Pettini}, {Steidel}, {Shapley}, {Erb}, \& {Law}}]{Reddy+2012}
{Reddy}, N.~A., {Pettini}, M., {Steidel}, C.~C., {et~al.} 2012, \apj, 754, 25

\bibitem[{{Robertson} {et~al.}(2026){Robertson}, {Johnson}, {Tacchella}, {Eisenstein}, {Hainline}, {Alberts}, {Arribas}, {Baker}, {Bunker}, {Cameron}, \& et~al.}]{Robertson+2026}
{Robertson}, B.~E., {Johnson}, B.~D., {Tacchella}, S., {et~al.} 2026, arXiv e-prints, arXiv:2601.15956

\bibitem[{{Rogers} {et~al.}(2026){Rogers}, {Strom}, {Rudie}, {Trainor}, {von Raesfeld}, {Raptis}, {Korhonen Cuestas}, {Miller}, {Steidel}, {Maseda}, {Chen}, \& {Law}}]{Rogers+2026}
{Rogers}, N. S.~J., {Strom}, A.~L., {Rudie}, G.~C., {et~al.} 2026, \apjl, 997, L44

\bibitem[{{Rong} {et~al.}(2026){Rong}, {Liu}, \& {Zou}}]{Rong+2026_desi}
{Rong}, Y., {Liu}, S., \& {Zou}, H. 2026, arXiv e-prints, arXiv:2606.28129

\bibitem[{{Rosdahl} {et~al.}(2022){Rosdahl}, {Blaizot}, {Katz}, {Kimm}, {Garel}, {Haehnelt}, {Keating}, {Martin-Alvarez}, {Michel-Dansac}, \& {Ocvirk}}]{Rosdahl+2022_sphix}
{Rosdahl}, J., {Blaizot}, J., {Katz}, H., {et~al.} 2022, \mnras, 515, 2386

\bibitem[{{Rosdahl} {et~al.}(2018){Rosdahl}, {Katz}, {Blaizot}, {Kimm}, {Michel-Dansac}, {Garel}, {Haehnelt}, {Ocvirk}, \& {Teyssier}}]{Rosdahl+2018_sphinx}
{Rosdahl}, J., {Katz}, H., {Blaizot}, J., {et~al.} 2018, \mnras, 479, 994

\bibitem[{{S{\'a}nchez Almeida} {et~al.}(2014){S{\'a}nchez Almeida}, {Elmegreen}, {Mu{\~n}oz-Tu{\~n}{\'o}n}, \& {Elmegreen}}]{SanchezAlmeida+2014}
{S{\'a}nchez Almeida}, J., {Elmegreen}, B.~G., {Mu{\~n}oz-Tu{\~n}{\'o}n}, C., \& {Elmegreen}, D.~M. 2014, \aapr, 22, 71

\bibitem[{{Sanders} {et~al.}(2021){Sanders}, {Shapley}, {Jones}, {Reddy}, {Kriek}, {Siana}, {Coil}, {Mobasher}, {Shivaei}, {Dav{\'e}}, {Azadi}, {Price}, {Leung}, {Freeman}, {Fetherolf}, {de Groot}, {Zick}, \& {Barro}}]{Sanders+2021_mosdef}
{Sanders}, R.~L., {Shapley}, A.~E., {Jones}, T., {et~al.} 2021, \apj, 914, 19

\bibitem[{{Sanders} {et~al.}(2020){Sanders}, {Shapley}, {Reddy}, {Kriek}, {Siana}, {Coil}, {Mobasher}, {Shivaei}, {Freeman}, {Azadi}, {Price}, {Leung}, {Fetherolf}, {de Groot}, {Zick}, {Fornasini}, \& {Barro}}]{Sanders+2020}
{Sanders}, R.~L., {Shapley}, A.~E., {Reddy}, N.~A., {et~al.} 2020, \mnras, 491, 1427

\bibitem[{{Sanders} {et~al.}(2026){Sanders}, {Shapley}, {Topping}, {Reddy}, {Berg}, {Khostovan}, {Bouwens}, {Brammer}, {Carnall}, {Cullen}, {Dav{\'e}}, {Dunlop}, {Ellis}, {F{\"o}rster Schreiber}, {Furlanetto}, {Glazebrook}, {Illingworth}, {Jones}, {Kriek}, {McLeod}, {McLure}, {Narayanan}, {Oesch}, {Pahl}, {Pettini}, {Schaerer}, {Stark}, {Steidel}, {Tang}, {Clarke}, {Donnan}, \& {Kehoe}}]{sanders+2026_aurora_calib}
{Sanders}, R.~L., {Shapley}, A.~E., {Topping}, M.~W., {et~al.} 2026, \apj, 1003, 228

\bibitem[{{Sanders} {et~al.}(2023){Sanders}, {Shapley}, {Topping}, {Reddy}, \& {Brammer}}]{Sanders+2023}
{Sanders}, R.~L., {Shapley}, A.~E., {Topping}, M.~W., {Reddy}, N.~A., \& {Brammer}, G.~B. 2023, \apj, 955, 54

\bibitem[{{Sanders} {et~al.}(2024){Sanders}, {Shapley}, {Topping}, {Reddy}, \& {Brammer}}]{Sanders+2024}
{Sanders}, R.~L., {Shapley}, A.~E., {Topping}, M.~W., {Reddy}, N.~A., \& {Brammer}, G.~B. 2024, \apj, 962, 24

\bibitem[{{Sarkar} {et~al.}(2025){Sarkar}, {Chakraborty}, {Vogelsberger}, {McDonald}, {Torrey}, {Garcia}, {Khullar}, {Ferland}, {Forman}, {Wolk}, {Schneider}, {Bautz}, {Miller}, {Grant}, \& {ZuHone}}]{Sarkar+2025}
{Sarkar}, A., {Chakraborty}, P., {Vogelsberger}, M., {et~al.} 2025, \apj, 978, 136

\bibitem[{{Schaerer} {et~al.}(2026){Schaerer}, {Izotov}, {Marques-Chaves}, {Steidel}, {Reddy}, {Shapley}, {Mascia}, {Chisholm}, {Flury}, {Guseva}, {Heckman}, {Henry}, {Inoue}, {Jung}, {Kusakabe}, {Mawatari}, {Oesch}, {{\"O}stlin}, {Pentericci}, {Roy}, {Saldana-Lopez}, {Sato}, {Vanzella}, {Verhamme}, \& {Wang}}]{Schaerer+2026_no}
{Schaerer}, D., {Izotov}, Y.~I., {Marques-Chaves}, R., {et~al.} 2026, \aap, 708, A242

\bibitem[{{Schaye} {et~al.}(2026){Schaye}, {Chaikin}, {Schaller}, {Ploeckinger}, {Hu{\v{s}}ko}, {McGibbon}, {Trayford}, {Ben{\'\i}tez-Llambay}, {Correa}, {Frenk}, {Richings}, {Forouhar Moreno}, {Bah{\'e}}, {Borrow}, {Durrant}, {Gebek}, {Helly}, {Jenkins}, {Lacey}, {Ludlow}, \& {Nobels}}]{Schaye+2026_COLIBRE}
{Schaye}, J., {Chaikin}, E., {Schaller}, M., {et~al.} 2026, \mnras, 548, stag375

\bibitem[{{Schaye} {et~al.}(2015){Schaye}, {Crain}, {Bower}, {Furlong}, {Schaller}, {Theuns}, {Dalla Vecchia}, {Frenk}, {McCarthy}, {Helly}, {Jenkins}, {Rosas-Guevara}, {White}, {Baes}, {Booth}, {Camps}, {Navarro}, {Qu}, {Rahmati}, {Sawala}, {Thomas}, \& {Trayford}}]{Schaye+2015_eagle}
{Schaye}, J., {Crain}, R.~A., {Bower}, R.~G., {et~al.} 2015, \mnras, 446, 521

\bibitem[{{Schneider} \& {Maiolino}(2024)}]{SchneiderMaiolino2024}
{Schneider}, R. \& {Maiolino}, R. 2024, \aapr, 32, 2

\bibitem[{{Scholte} {et~al.}(2025){Scholte}, {Cullen}, {Carnall}, {Arellano-C{\'o}rdova}, {Stanton}, {Barrufet}, {Begley}, {Bondestam}, {Donnan}, {Dunlop}, {Leung}, {McLeod}, {McLure}, {Moustakas}, {Pollock}, {Shapley}, {Stevenson}, \& {Zou}}]{Scholte+2025_excels}
{Scholte}, D., {Cullen}, F., {Carnall}, A.~C., {et~al.} 2025, \mnras, 540, 1800

\bibitem[{{Scholte} {et~al.}(2026){Scholte}, {Cullen}, {Moustakas}, {Zou}, {Saintonge}, {Arellano-Cordova}, {Stanton}, {Andrews}, {Sui}, {Aguilar}, {Ahlen}, {Bianchi}, {Brooks}, {Castander}, {Cheng}, {Claybaugh}, {de la Macorra}, {Dey}, {Doel}, {Douglass}, {Ferraro}, {Forero-Romero}, {Gazta{\~n}aga}, {Gontcho}, {Gutierrez}, {Joyce}, {Kremin}, {Lahav}, {Landriau}, {Le Guillou}, {Martini}, {Meisner}, {Miquel}, {Percival}, {Poppett}, {Prada}, {P{\'e}rez-R{\`a}fols}, {Rossi}, {Sanchez}, {Schlegel}, {Shao}, {Silber}, {Sprayberry}, {Tarl{\'e}}, \& {Weaver}}]{Scholte+2026_DESIabundances}
{Scholte}, D., {Cullen}, F., {Moustakas}, J., {et~al.} 2026, \mnras, 551, stag1381

\bibitem[{{Scholte} {et~al.}(2024){Scholte}, {Saintonge}, {Moustakas}, {Catinella}, {Zou}, {Dey}, {Aguilar}, {Ahlen}, {Anand}, {Blum}, {Brooks}, {Circosta}, {Claybaugh}, {de la Macorra}, {Doel}, {Font-Ribera}, {F{\"o}rster}, {Forero-Romero}, {Gazta{\~n}aga}, {Gontcho A Gontcho}, {Juneau}, {Kehoe}, {Kisner}, {Koposov}, {Kremin}, {Lambert}, {Landriau}, {Maraston}, {Martini}, {Meisner}, {Mighty}, {Miquel}, {Myers}, {Nie}, {Poppett}, {Prada}, {Rezaie}, {Rossi}, {Sanchez}, {Schubnell}, {Silber}, {Sprayberry}, {Siudek}, {Speranza}, {Tarl{\'e}}, {Tojeiro}, \& {Weaver}}]{Scholte+2024}
{Scholte}, D., {Saintonge}, A., {Moustakas}, J., {et~al.} 2024, \mnras, 535, 2341

\bibitem[{{Scholtz} {et~al.}(2026){Scholtz}, {Carniani}, {Parlanti}, {D'Eugenio}, {Curtis-Lake}, {Jakobsen}, {Bunker}, {Cameron}, {Arribas}, {Baker}, {Charlot}, {Chevellard}, {Circosta}, {Curti}, {Duan}, {Eisenstein}, {Hainline}, {Ji}, {Johnson}, {Jones}, {Kumari}, {Maiolino}, {Maseda}, {Perna}, {P{\'e}rez-Gonz{\'a}lez}, {Rawle}, {Rieke}, {Rinaldi}, {Robertson}, {Saxena}, {Shivaei}, {Silcock}, {Sun}, {Rodr{\'\i}guez Del Pino}, {Tacchella}, {{\"U}bler}, {Venturi}, {Williams}, {Willmer}, {Willott}, \& {Witstok}}]{Scholtz+2026_jadesDR4}
{Scholtz}, J., {Carniani}, S., {Parlanti}, E., {et~al.} 2026, \mnras, 549, stag939

\bibitem[{{Shapley} {et~al.}(2015){Shapley}, {Reddy}, {Kriek}, {Freeman}, {Sanders}, {Siana}, {Coil}, {Mobasher}, {Shivaei}, {Price}, \& {de Groot}}]{Shapley+2015}
{Shapley}, A.~E., {Reddy}, N.~A., {Kriek}, M., {et~al.} 2015, \apj, 801, 88

\bibitem[{{Shapley} {et~al.}(2023){Shapley}, {Reddy}, {Sanders}, {Topping}, \& {Brammer}}]{Shapley+2023_neblines}
{Shapley}, A.~E., {Reddy}, N.~A., {Sanders}, R.~L., {Topping}, M.~W., \& {Brammer}, G.~B. 2023, \apjl, 950, L1

\bibitem[{{Shapley} {et~al.}(2025){Shapley}, {Sanders}, {Topping}, {Reddy}, {Berg}, {Bouwens}, {Brammer}, {Carnall}, {Cullen}, {Dav{\'e}}, {Dunlop}, {Ellis}, {F{\"o}rster Schreiber}, {Furlanetto}, {Glazebrook}, {Illingworth}, {Jones}, {Kriek}, {McLeod}, {McLure}, {Narayanan}, {Oesch}, {Pahl}, {Pettini}, {Schaerer}, {Stark}, {Steidel}, {Tang}, {Clarke}, {Donnan}, \& {Kehoe}}]{Shapley+2025_aurora}
{Shapley}, A.~E., {Sanders}, R.~L., {Topping}, M.~W., {et~al.} 2025, \apj, 980, 242

\bibitem[{{Sharda} {et~al.}(2026){Sharda}, {Schaye}, {McGibbon}, {Ben{\'\i}tez-Llambay}, {Chaikin}, {Frenk}, {Hodge}, {Hu{\v{s}}ko}, {Ploeckinger}, {Richings}, \& {Schaller}}]{Sharda+2026_mzrCOLIBRE}
{Sharda}, P., {Schaye}, J., {McGibbon}, R.~J., {et~al.} 2026, arXiv e-prints, arXiv:2606.25995

\bibitem[{{Shivaei} {et~al.}(2026){Shivaei}, {Naidu}, {Rodr{\'\i}guez Montero}, {Matsumoto}, {Leja}, {Matthee}, {Johnson}, {Oesch}, {Chevallard}, {Adamo}, {Bodansky}, {Bunker}, {Covelo Paz}, {Di Cesare}, {Egami}, {Furtak}, {Heintz}, {Kramarenko}, {Meyer}, {Reddy}, {Rinaldi}, {Tacchella}, {Torralba}, {Witstok}, {Wozniak}, \& {Xiao}}]{Shivaei+2026}
{Shivaei}, I., {Naidu}, R.~P., {Rodr{\'\i}guez Montero}, F., {et~al.} 2026, \aap, 712, A132

\bibitem[{{Speagle} {et~al.}(2014){Speagle}, {Steinhardt}, {Capak}, \& {Silverman}}]{Speagle+2014}
{Speagle}, J.~S., {Steinhardt}, C.~L., {Capak}, P.~L., \& {Silverman}, J.~D. 2014, \apjs, 214, 15

\bibitem[{{Stanton} {et~al.}(2025){Stanton}, {Cullen}, {Carnall}, {Scholte}, {Arellano-C{\'o}rdova}, {McLeod}, {Begley}, {Donnan}, {Dunlop}, {Hamadouche}, {McLure}, {Shapley}, {Bondestam}, \& {Stevenson}}]{Stanton+2025}
{Stanton}, T.~M., {Cullen}, F., {Carnall}, A.~C., {et~al.} 2025, \mnras, 537, 1735

\bibitem[{{Stanton} {et~al.}(2026){Stanton}, {Cullen}, {Carnall}, {Scholte}, {Arellano-C{\'o}rdova}, {Shapley}, {McLeod}, {Donnan}, {Begley}, {Dav{\'e}}, {Dunlop}, {McLure}, {Rowlands}, {Bondestam}, {Hamadouche}, {Leung}, {Stevenson}, \& {Taylor}}]{Stanton+2026_excels}
{Stanton}, T.~M., {Cullen}, F., {Carnall}, A.~C., {et~al.} 2026, \mnras, 547, stag449

\bibitem[{{Stephenson} {et~al.}(2026){Stephenson}, {Chisholm}, {Saldana-Lopez}, {Matthee}, {Henry}, {Naidu}, {Flury}, {Hutchison}, {Finkelstein}, {Guseva}, {Izotov}, {Marques-Chaves}, {Oesch}, {Reddy}, \& {Schaerer}}]{Stephenson+2026_ne}
{Stephenson}, M.~G., {Chisholm}, J., {Saldana-Lopez}, A., {et~al.} 2026, \apj, 1010, 184

\bibitem[{{Strom} {et~al.}(2023){Strom}, {Rudie}, {Trainor}, {Brammer}, {Maseda}, {Raptis}, {Rogers}, {Steidel}, {Chen}, \& {Law}}]{Strom+2023_cecilia}
{Strom}, A.~L., {Rudie}, G.~C., {Trainor}, R.~F., {et~al.} 2023, \apjl, 958, L11

\bibitem[{{Strom} {et~al.}(2018){Strom}, {Steidel}, {Rudie}, {Trainor}, \& {Pettini}}]{Strom+2018}
{Strom}, A.~L., {Steidel}, C.~C., {Rudie}, G.~C., {Trainor}, R.~F., \& {Pettini}, M. 2018, \apj, 868, 117

\bibitem[{{Suess} {et~al.}(2024){Suess}, {Weaver}, {Price}, {Pan}, {Wang}, {Bezanson}, {Brammer}, {Cutler}, {Labb{\'e}}, {Leja}, {Williams}, {Whitaker}, {Atek}, {Dayal}, {de Graaff}, {Feldmann}, {Franx}, {Fudamoto}, {Fujimoto}, {Furtak}, {Goulding}, {Greene}, {Khullar}, {Kokorev}, {Kriek}, {Lorenz}, {Marchesini}, {Maseda}, {Matthee}, {Miller}, {Mitsuhashi}, {Mowla}, {Muzzin}, {Naidu}, {Nanayakkara}, {Nelson}, {Oesch}, {Setton}, {Shipley}, {Smit}, {Spilker}, {van Dokkum}, \& {Zitrin}}]{Suess+2024_megascience}
{Suess}, K.~A., {Weaver}, J.~R., {Price}, S.~H., {et~al.} 2024, \apj, 976, 101

\bibitem[{{Sun} {et~al.}(2026){Sun}, {Eisenstein}, {D'Eugenio}, {Hainline}, {Helton}, {Johnson}, {Lin}, {Rieke}, {Robertson}, {Tacchella}, {Bunker}, {Chevallard}, {Curtis-Lake}, {Egami}, {Hausen}, {Ji}, {Lyu}, {Maiolino}, {Rinaldi}, {Sun}, {Trussler}, {Williams}, {Willmer}, {Witstok}, {Wu}, \& {Zhu}}]{Sun+2026}
{Sun}, F., {Eisenstein}, D.~J., {D'Eugenio}, F., {et~al.} 2026, arXiv e-prints, arXiv:2601.15961

\bibitem[{{Sun} {et~al.}(2023){Sun}, {Faucher-Gigu{\`e}re}, {Hayward}, {Shen}, {Wetzel}, \& {Cochrane}}]{Sun+2023_bursty}
{Sun}, G., {Faucher-Gigu{\`e}re}, C.-A., {Hayward}, C.~C., {et~al.} 2023, \apjl, 955, L35

\bibitem[{{Tacchella} {et~al.}(2022){Tacchella}, {Finkelstein}, {Bagley}, {Dickinson}, {Ferguson}, {Giavalisco}, {Graziani}, {Grogin}, {Hathi}, {Hutchison}, {Jung}, {Koekemoer}, {Larson}, {Papovich}, {Pirzkal}, {Rojas-Ruiz}, {Song}, {Schneider}, {Somerville}, {Wilkins}, \& {Yung}}]{Tacchella+2022}
{Tacchella}, S., {Finkelstein}, S.~L., {Bagley}, M., {et~al.} 2022, \apj, 927, 170

\bibitem[{{Tacchella} {et~al.}(2020){Tacchella}, {Forbes}, \& {Caplar}}]{Tacchella+2020_sfhII}
{Tacchella}, S., {Forbes}, J.~C., \& {Caplar}, N. 2020, \mnras, 497, 698

\bibitem[{{Taylor}(2005)}]{topcat2005}
{Taylor}, M.~B. 2005, in Astronomical Society of the Pacific Conference Series, Vol. 347, Astronomical Data Analysis Software and Systems XIV, ed. P.~{Shopbell}, M.~{Britton}, \& R.~{Ebert}, 29

\bibitem[{{Theios} {et~al.}(2019){Theios}, {Steidel}, {Strom}, {Rudie}, {Trainor}, \& {Reddy}}]{Theios+2019}
{Theios}, R.~L., {Steidel}, C.~C., {Strom}, A.~L., {et~al.} 2019, \apj, 871, 128

\bibitem[{{Topping} {et~al.}(2025){Topping}, {Sanders}, {Shapley}, {Pahl}, {Reddy}, {Stark}, {Berg}, {Clarke}, {Cullen}, {Dunlop}, {Ellis}, {Schreiber}, {Illingworth}, {Jones}, {Narayanan}, {Pettini}, \& {Schaerer}}]{topping+2025_aurora}
{Topping}, M.~W., {Sanders}, R.~L., {Shapley}, A.~E., {et~al.} 2025, \mnras, 541, 1707

\bibitem[{{Topping} {et~al.}(2020){Topping}, {Shapley}, {Reddy}, {Sanders}, {Coil}, {Kriek}, {Mobasher}, \& {Siana}}]{Topping+2020_ion}
{Topping}, M.~W., {Shapley}, A.~E., {Reddy}, N.~A., {et~al.} 2020, \mnras, 495, 4430

\bibitem[{{Topping} {et~al.}(2024{\natexlab{a}}){Topping}, {Stark}, {Senchyna}, {Plat}, {Zitrin}, {Endsley}, {Charlot}, {Furtak}, {Maseda}, {Smit}, {Mainali}, {Chevallard}, {Molyneux}, \& {Rigby}}]{Topping+2024_Nemitter}
{Topping}, M.~W., {Stark}, D.~P., {Senchyna}, P., {et~al.} 2024{\natexlab{a}}, \mnras, 529, 3301

\bibitem[{{Topping} {et~al.}(2024{\natexlab{b}}){Topping}, {Stark}, {Senchyna}, {Plat}, {Zitrin}, {Endsley}, {Charlot}, {Furtak}, {Maseda}, {Smit}, {Mainali}, {Chevallard}, {Molyneux}, \& {Rigby}}]{Topping+2024_no}
{Topping}, M.~W., {Stark}, D.~P., {Senchyna}, P., {et~al.} 2024{\natexlab{b}}, \mnras, 529, 3301

\bibitem[{{Torralba} {et~al.}(2026){Torralba}, {Matthee}, {Marques-Chaves}, {Martins}, {Navarrete}, {Pezzulli}, {Iani}, {Naidu}, {Brammer}, {Di Cesare}, {Hutter}, {Kotiwale}, {Mackenzie}, {Mascia}, {Meyer}, \& {Schaerer}}]{Torralba+2026_vms}
{Torralba}, A., {Matthee}, J., {Marques-Chaves}, R., {et~al.} 2026, arXiv e-prints, arXiv:2610.29887

\bibitem[{{Torralba} {et~al.}(2024){Torralba}, {Matthee}, {Naidu}, {Mackenzie}, {Pezzulli}, {Hutter}, {Arnalte-Mur}, {Gurung-L{\'o}pez}, {Tacchella}, {Oesch}, {Kashino}, {Conroy}, \& {Sobral}}]{Torralba+2024}
{Torralba}, A., {Matthee}, J., {Naidu}, R.~P., {et~al.} 2024, \aap, 689, A44

\bibitem[{{Torrey} {et~al.}(2018){Torrey}, {Vogelsberger}, {Hernquist}, {McKinnon}, {Marinacci}, {Simcoe}, {Springel}, {Pillepich}, {Naiman}, {Pakmor}, {Weinberger}, {Nelson}, \& {Genel}}]{Torrey+2018}
{Torrey}, P., {Vogelsberger}, M., {Hernquist}, L., {et~al.} 2018, \mnras, 477, L16

\bibitem[{{Tremonti} {et~al.}(2004){Tremonti}, {Heckman}, {Kauffmann}, {Brinchmann}, {Charlot}, {White}, {Seibert}, {Peng}, {Schlegel}, {Uomoto}, {Fukugita}, \& {Brinkmann}}]{Tremonti+2004}
{Tremonti}, C.~A., {Heckman}, T.~M., {Kauffmann}, G., {et~al.} 2004, \apj, 613, 898

\bibitem[{{Umeda} {et~al.}(2026){Umeda}, {Ouchi}, {Nakajima}, {Watanabe}, {Harikane}, {Isobe}, {Nishigaki}, {Yoshiaki}, {Yajima}, \& {Xu}}]{Umeda+2026_NO}
{Umeda}, H., {Ouchi}, M., {Nakajima}, K., {et~al.} 2026, arXiv e-prints, arXiv:2607.15515

\bibitem[{{Ventura} {et~al.}(2013){Ventura}, {Di Criscienzo}, {Carini}, \& {D'Antona}}]{Ventura+2013}
{Ventura}, P., {Di Criscienzo}, M., {Carini}, R., \& {D'Antona}, F. 2013, \mnras, 431, 3642

\bibitem[{{Vijayan} {et~al.}(2021){Vijayan}, {Lovell}, {Wilkins}, {Thomas}, {Barnes}, {Irodotou}, {Kuusisto}, \& {Roper}}]{Vijayan+2021_FLARES}
{Vijayan}, A.~P., {Lovell}, C.~C., {Wilkins}, S.~M., {et~al.} 2021, \mnras, 501, 3289

\bibitem[{{Vila-Costas} \& {Edmunds}(1993)}]{Vila-Costas+1993}
{Vila-Costas}, M.~B. \& {Edmunds}, M.~G. 1993, \mnras, 265, 199

\bibitem[{{Vincenzo} {et~al.}(2016){Vincenzo}, {Belfiore}, {Maiolino}, {Matteucci}, \& {Ventura}}]{Vincenzo+2016}
{Vincenzo}, F., {Belfiore}, F., {Maiolino}, R., {Matteucci}, F., \& {Ventura}, P. 2016, \mnras, 458, 3466

\bibitem[{{Vincenzo} \& {Kobayashi}(2018)}]{VincenzoKobayashi2018}
{Vincenzo}, F. \& {Kobayashi}, C. 2018, \mnras, 478, 155

\bibitem[{{Virtanen} {et~al.}(2020){Virtanen}, {Gommers}, {Oliphant}, {Haberland}, {Reddy}, {Cournapeau}, {Burovski}, {Peterson}, {Weckesser}, {Bright}, {van der Walt}, {Brett}, {Wilson}, {Millman}, {Mayorov}, {Nelson}, {Jones}, {Kern}, {Larson}, {Carey}, {Polat}, {Feng}, {Moore}, {VanderPlas}, {Laxalde}, {Perktold}, {Cimrman}, {Henriksen}, {Quintero}, {Harris}, {Archibald}, {Ribeiro}, {Pedregosa}, {van Mulbregt}, \& {SciPy 1. 0 Contributors}}]{Virtanen2020}
{Virtanen}, P., {Gommers}, R., {Oliphant}, T.~E., {et~al.} 2020, Nature Medicine, 17, 261

\bibitem[{{Wang} {et~al.}(2026){Wang}, {Baugh}, {Boardman}, {Bose}, {Cai}, {Cole}, {Frenk}, {Lacey}, {Norberg}, {Peng}, {Santos-Santos}, {Saintonge}, {Theuns}, \& {Wang}}]{Wang+2026}
{Wang}, K., {Baugh}, C., {Boardman}, N.~F., {et~al.} 2026, arXiv e-prints, arXiv:2608.04784

\bibitem[{{Weibel} {et~al.}(2024){Weibel}, {Oesch}, {Barrufet}, {Gottumukkala}, {Ellis}, {Santini}, {Weaver}, {Allen}, {Bouwens}, {Bowler}, {Brammer}, {Carnall}, {Cullen}, {Dayal}, {Dickinson}, {Donnan}, {Dunlop}, {Giavalisco}, {Grogin}, {Illingworth}, {Koekemoer}, {Labbe}, {Marchesini}, {McLeod}, {McLure}, {Naidu}, {P{\'e}rez-Gonz{\'a}lez}, {Shuntov}, {Stefanon}, {Toft}, \& {Xiao}}]{Weibel+2024}
{Weibel}, A., {Oesch}, P.~A., {Barrufet}, L., {et~al.} 2024, \mnras, 533, 1808

\bibitem[{{Whitler} {et~al.}(2023){Whitler}, {Stark}, {Endsley}, {Leja}, {Charlot}, \& {Chevallard}}]{Whitler+2023}
{Whitler}, L., {Stark}, D.~P., {Endsley}, R., {et~al.} 2023, \mnras, 519, 5859

\bibitem[{{Williams} {et~al.}(2023){Williams}, {Tacchella}, {Maseda}, {Robertson}, {Johnson}, {Willott}, {Eisenstein}, {Willmer}, {Ji}, {Hainline}, {Helton}, {Alberts}, {Baum}, {Bhatawdekar}, {Boyett}, {Bunker}, {Carniani}, {Charlot}, {Chevallard}, {Curtis-Lake}, {de Graaff}, {Egami}, {Franx}, {Kumari}, {Maiolino}, {Nelson}, {Rieke}, {Sandles}, {Shivaei}, {Simmonds}, {Smit}, {Suess}, {Sun}, {{\"U}bler}, \& {Witstok}}]{Williams+2023_jems}
{Williams}, C.~C., {Tacchella}, S., {Maseda}, M.~V., {et~al.} 2023, \apjs, 268, 64

\bibitem[{{Woodrum} {et~al.}(2025){Woodrum}, {Shivaei}, {Witstok}, {Saxena}, {Simmonds}, {Scholtz}, {Bhatawdekar}, {Bunker}, {Carniani}, {Charlot}, {Curti}, {Curtis-Lake}, {Chevallard}, {D'Eugenio}, {Hainline}, {Helton}, {Maiolino}, {Perna}, {Rinaldi}, {Robertson}, {Straughn}, {Sun}, {Tacchella}, {Williams}, {Willott}, \& {Zhu}}]{Woodrum+2025}
{Woodrum}, C., {Shivaei}, I., {Witstok}, J., {et~al.} 2025, arXiv e-prints, arXiv:2510.00235

\bibitem[{{Zhu} {et~al.}(2026){Zhu}, {Trussler}, \& {Kewley}}]{Zhu+2026_nitrogen}
{Zhu}, P., {Trussler}, J., \& {Kewley}, L.~J. 2026, \apj, 998, 5

\end{thebibliography}

\begin{appendix}

\section{Stacked [SII] line-fit}
\label{app:SII_linefit}

\begin{figure*}[ht]
   \centering
   \includegraphics[scale = 0.4]{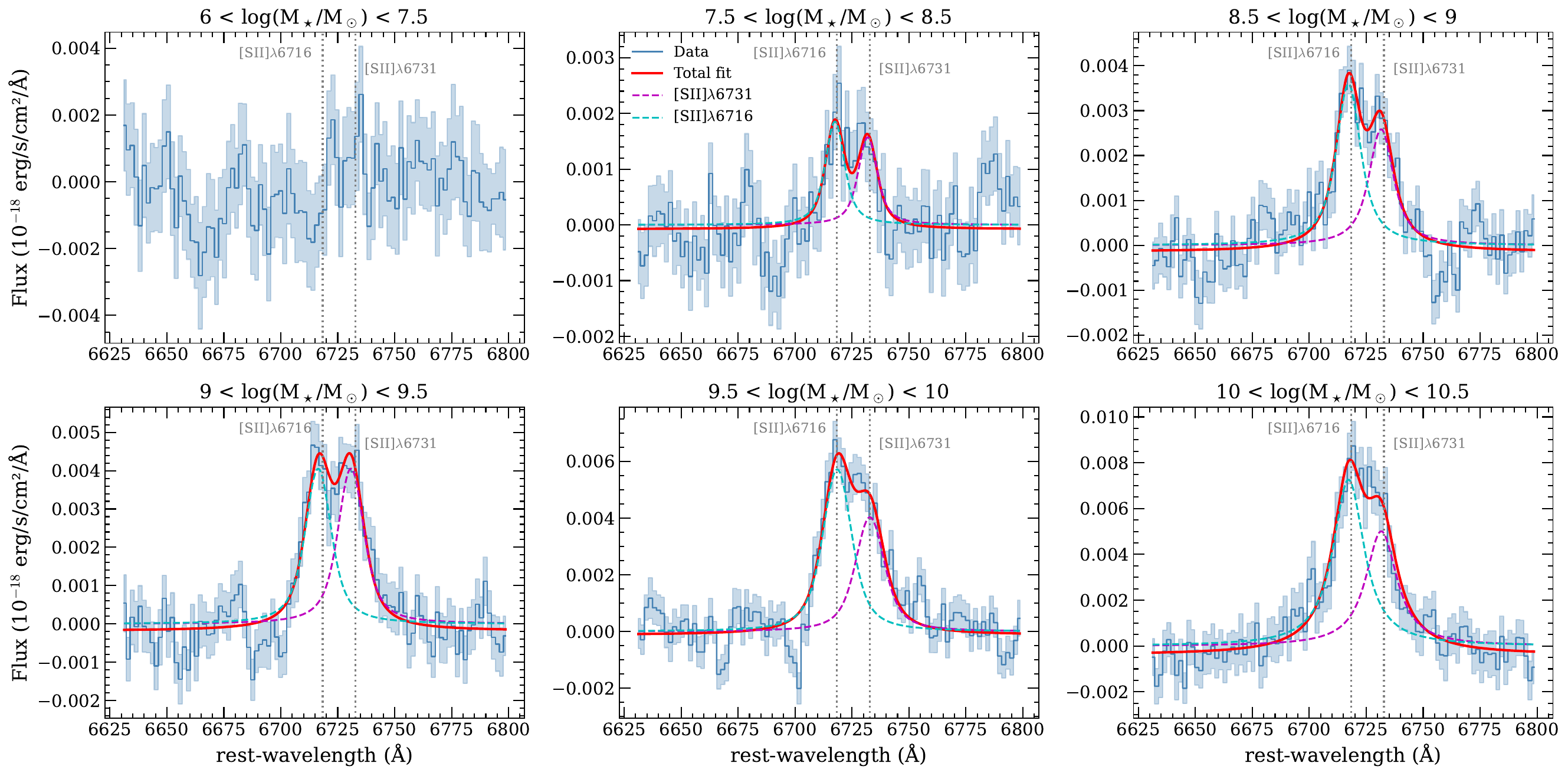}
   \caption{Line fit of  $\rm [\ion{S}{ii}]$-doublet emission lines. The fit is performed using a Voigt profile which Lorentzian parameter $\gamma$ is fixed by the \ha{} fit in the spatial direction (see Sect. \ref{sec:stacking}). We show the observational data together with the associated errors in blue, the fitting components with colored dashed lines and the resulting fit in red solid lines. Dotted vertical lines highlight the position of \siib{} and \siir.}
   \label{fig:SII_stacks}
\end{figure*}

We follow the procedure described in Sect. \ref{sec:stacking} to fit the $\rm[\ion{S}{ii}]$-doublet emission. Figure \ref{fig:SII_stacks} shows the results of our fit obtained using a Voigt profile. The spectrum in the lowest stellar mass bin ($\rm log(M_\star/M_\odot) = 6-7.5$) is too noisy to allow for a reliable identification of the $\rm[\ion{S}{ii}]$-doublet; therefore, we do not perform a line fit in this mass bin.

\section{N2 and S2 calibrations}
\label{app:calib}

In this section, we explore different calibrations proposed to estimate the gas-phase metallicity using the strong line method based on the N2 and S2 diagnostics. In particular, Fig. \ref{app:calib} shows the MZR obtained using \cite{curti+2020_calib}, \cite{nakajima+2022_calib}, \cite{Isobe+2026} and \cite{sanders+2026_aurora_calib} calibrations. The main difference among these calibrations lies in the underlying galaxy samples: the first two are based on observations of local galaxies, the third relies on stacked spectra of high-redshift galaxies and the latter is derived from a sample of individual high-redshift galaxies.

In particular, \cite{nakajima+2022_calib} derived their calibration on a sample of local extremely metal-poor and low-mass galaxies \citep{Kojima+2020, Isobe+2021}, with available auroral lines measurements. Their prescriptions also account for a dependence on the H$\beta$ equivalent width (EW(H$\beta$)), useful for studies of high-redshift galaxies. The authors tested their diagnostics against a sample of eight $z=2-3$ galaxies with metallicities independently determined through the electron temperature \citep{Sanders+2020}, finding good agreement between local calibrations and the observations up to $z\sim3$. Their results highlight the importance of accounting for ionization conditions when deriving metallicities from strong emission-line ratios and suggest that EW(H$\beta$)-based calibrations provide a practical approach that may be applicable to high-redshift galaxies. However, the limited sample size at $z\sim3$ prevents a robust assessment of the applicability of these indicators at higher redshifts. Both N2- and S2-based calibrations are derived over the metallicity range 12+log(O/H) = 7.1 - 8.9.

\cite{curti+2020_calib} based their calibrations on a local sample from SDSS DR7 \citep{Abazajian+2009} and therefore representative of typical $z\sim0$ star-forming galaxies. The authors derive metallicities using a combination of strong-line diagnostics calibrated on the $T_e$-based abundance scale, following \cite{Curti+2017}. Their calibration sample combines individual low-metallicity galaxies with auroral-line detections and stacked spectra of high-metallicity galaxies, providing a consistent $T_e$-based metallicity scale over a broad abundance range. However, since these calibrations are based on local galaxies, their application to higher redshift galaxies may be affected by the different ionization conditions of the ISM \citep[e.g.][]{Bian+2016_localanalogs, Kashino+2017, Kaasinen+2017}. Both N2- and S2-based diagnostics were calibrated against metallicity in the 12+log(O/H) = 7.6 - 8.9 range.

\cite{Isobe+2026} stacked $\sim 1500$ JWST/NIRSpec medium-resolution spectra of SF galaxies at $z\sim1$-$10$ from the JADES DR4 \citep{Curtis-Lake+2026_jadesDR4, Scholtz+2026_jadesDR4} and the JADES Dark Horse program \citep{DEugenio+2026_jadesDH}, binned in either strong-line metallicity or stellar mass. Stacking allowed them to detect the [\ion{O}{iii}]$\lambda4363$ auroral line down to $12+\log(\rm O/H)=7.0$ ($\sim2\%\, \rm Z_\odot$) and to derive stack-based strong-line calibrations from direct-$T_{\rm e}$ metallicities. Because the stacks are built from H$\beta$-detected galaxies, they do not require faint auroral emission to be detected in individual sources. This avoids biasing the sample towards high ionization parameters and large emission-line equivalent widths (see \cite{Sanders+2021_mosdef}). To constrain the metal-rich end, \cite{Isobe+2026} added emission-line ratios from stacks of $z\sim0$ galaxies \citep{AndrewsMartini2013}, keeping only those with sSFRs above the $z>1$ SFMS predicted by the \textsc{thesan-zoom} simulations \citep{McClymont+2025_sfms}. Combining the two datasets gives calibrations valid over $12+\rm log(O/H)=7.0-8.7$: the high-redshift stacks cover $12+\rm log(O/H)<8$, while the $z\sim0$ high-sSFR stacks cover $12+\rm log(O/H)>8$. 

\cite{sanders+2026_aurora_calib} based their metallicity calibrations on a sample of high-redshift galaxies from the AURORA survey \citep{Shapley+2025_aurora} supplemented by a sample of auroral-line detected star-forming galaxies compiled from literature. The combined sample spans a redshift range of $z=1.3-10.6$, with a median redshift of $z=3.80$. The N2- and S2-based diagnostics were calibrated against metallicity in the $\rm 12+log(O/H) = 7.8 - 8.6$ and $\rm 12+log(O/H) = 7.9 - 8.6$ ranges, respectively. Because these calibrations are derived from high-redshift star-forming galaxies, they account for typical ionization conditions of such systems. We therefore adopt these calibrations as the fiducial metallicity calibrations for our analysis.

\begin{figure*}
   \centering
   \includegraphics[scale = 0.46]{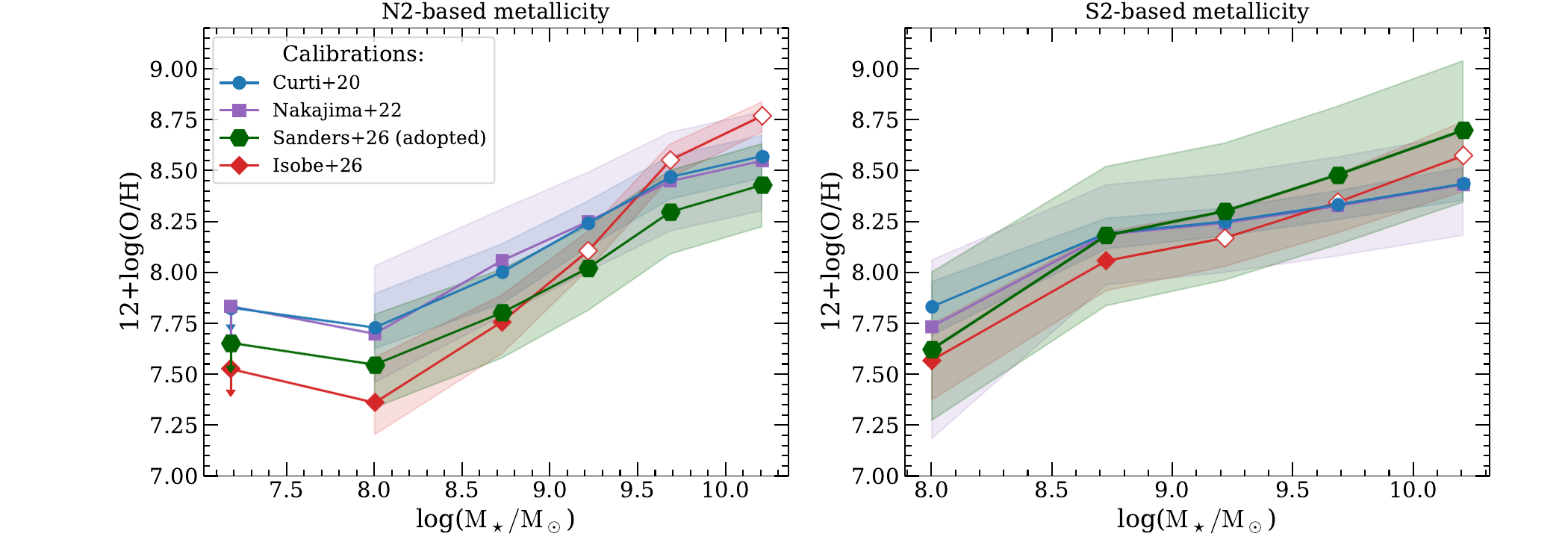}
   \caption{Mass-metallicity relation derived from the N2 (left) and S2 (right) ratios measured on stacked spectra in bins of stellar mass, obtained with the calibrations of \cite{curti+2020_calib} (blue), \cite{nakajima+2022_calib} (purple), \cite{sanders+2026_aurora_calib} (green) and \cite{Isobe+2026} (red). For the \cite{Isobe+2026} calibration, filled diamonds mark metallicities in the regime constrained by their high-$z$ JWST/NIRSpec stacks, while open diamonds those in the regime where the calibration is anchored to $z\sim0$ stacks with high sSFRs \citep{AndrewsMartini2013}. Downward arrows in the lowest-mass bin of the N2 panel indicate upper limits from the marginal $2\sigma$ detection of \niir{} in the stacked spectrum. Shaded regions show the uncertainty on the metallicity, obtained by propagating the measurement error on the stacked line ratio through each calibration with a Monte Carlo approach and adding in quadrature the intrinsic scatter of the calibration.}
   \label{fig:calib}
\end{figure*}

\section{S2 based metallicity}
\label{app:S2_met}

In this appendix, we present the MZR derived using the S2-based strong-line calibration of \cite{sanders+2026_aurora_calib}, complementing the N2-based measurements presented in Sect.~\ref{sec:mzr}. Unlike nitrogen, sulfur is an $\alpha$-element and, therefore, its abundance evolves in tandem with that of oxygen \citep[e.g.][]{ArellanoCordova+2024, Stanton+2025, Esteban+2025}. Sulfur-based diagnostics are consequently less sensitive to variations in the N/O ratio, which can affect N-based metallicity estimates \citep[see e.g.][]{Scholte+2025_excels, sanders+2026_aurora_calib}. In addition, \cite{Scholte+2025_excels} showed that \siiha{} diagnostic yields metallicity estimates consistent with those derived from the $[\ion{O}{ii}]\,\lambda\lambda3726,3729/\rm H\beta$ one.

\begin{figure*}[ht!]
   \centering
   \includegraphics[scale = 0.35]{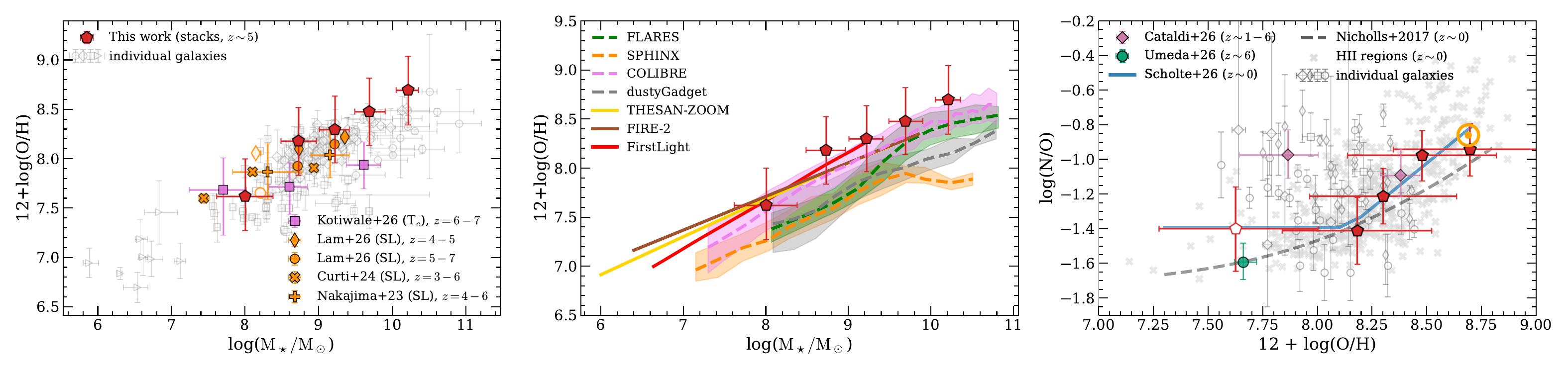}
   \caption{Left and Middle: the mass-metallicity relation inferred from our measurements, compared with observations and simulations. Same as Fig.~\ref{fig:mzr}, but using the S2-based diagnostic from \cite{sanders+2026_aurora_calib} for our stacked galaxy sample. Right: The N/O ratio as a function of gas-phase metallicity. Same as Fig.~\ref{fig:no-oh}, but with metallicity estimated based on the S2-diagnostic. Open markers show individual galaxy measurements from \cite{Arellano-Cordova+2025_NO} (squares), \cite{Cameron+2026_NO} (diamonds) and \cite{Schaerer+2026_no} (circles).}
   \label{fig:S2_trends}
\end{figure*}

Figure~\ref{fig:S2_trends} shows the S2-based MZR, compared with literature results and predictions from cosmological simulations, together with the N/O ratio as a function of oxygen abundance. Following the parametrization and fitting procedure described in Sect.~\ref{sec:mzr}, we find a slope for the relation of $\gamma = 0.35 ^{+0.13}_{-0.10}$ and a normalization $\rm Z_{m9} = 8.25 ^{+0.06}_{-0.09}$ at $\rm M_\star = 10^{9} \; M_\odot$ (see Tab.~\ref{tab:MZRfit}). The slope is consistent within the uncertainties with the N2-based one ( $\gamma = 0.42^{+0.10}_{-0.07}$), indicating that the shape of the MZR is robust against the choice of calibration. The normalization, however, is higher by $\sim0.3$~dex, reflecting systematic differences between the two strong-line calibrations in this regime. As for the N2-based relation, at the high-mass end, where our measurements provide the strongest constraints on the simulation predictions, the S2-based MZR follows the trends predicted by COLIBRE and FLARES.

In the N/O--O/H plane, the higher S2-based metallicities shift our stacks horizontally, since N/O is unchanged, bringing them into full agreement with the sequence traced by local H\textsc{ii} regions and star-forming galaxies from DESI DR2 \citep{Scholte+2026_DESIabundances}. Since N2S2 is anti-correlated with S2-based metallicities, this agreement cannot be driven by the shared line fluxes (see Sect.~\ref{sec:NO_ratio}).

\section{Hydrodynamical simulations}
\label{app:sims}

In Sect. \ref{sec:mzr}, we compare our observational determination of the mass-metallicity relation with predictions from hydrodynamical simulations. The predicted MZR depends sensitively on the implementation of baryonic physics, including the prescriptions on stellar and AGN feedback, gas inflow and outflow, metal mixing, metal depletion onto dust and star formation. As a result, different simulations predict MZRs with distinct slopes and normalizations. In the following, we briefly summarize the main characteristics of the simulations considered in this work, which are a combination of zoom-in (FLARES, THESAN-ZOOM, FIRE-2 and FirstLight) and cosmological (SPHINX$^{20}$, COLIBRE and \texttt{dustyGadget}) models. 

FLARES is a suite of zoom simulations drawn from a parent dark-matter only simulation \citep{Barnes+2017} and re-run until $z=4.67$ using full hydrodynamics configuration of the EAGLE galaxy formation model \citep{Schaye+2015_eagle, Crain+2015_eagle}. The initial gas mass particles reach a resolution of $m_g = 1.8 \cdot 10^6 \, \rm M_\odot$. The stellar masses are computed considering all the stellar particles within a 30 pkpc aperture centered on the galaxy center, defined as the location of the most bound particle (i.e. minimum gravitational potential). The gas-phase metallicity is derived from the oxygen abundance in the gas phase of FLARES galaxies. To ensure robustly resolved systems, we restrict the sample to galaxies with $\rm log(M_\star/M_\odot) > 8$ (priv. comm.). 

The THESAN-ZOOM simulations \cite{Kannan+2025_thesanzoom} are drawn from the parent THESAN run \cite{Kannan+2022-thesan} and re-simulated at higher resolution. The baryonic mass of the re-simulation is $m_b \sim 140 - 9000 \rm M_\odot$. In THESAN-ZOOM, the stellar masses are computed accounting for all the stellar particles within the virial radius and the mass-weighted metallicity are derived from gas particles within twice the stellar half-mass radius. In Fig. \ref{fig:mzr}, we report the best-fitting relation derived in \cite{McClymont+2026_mzrTHESAN} at $z = 5$, which is calculated using the median metallicity in 0.5 dex bins of stellar mass in the range $6 \leq \rm log(M_\star/M_\odot) \leq 11$. We limit the stellar mass to the maximum value available at each redshift \citep[see][]{Kannan+2025_thesanzoom}.

The FIRE-2 zoom-in simulations are drawn from a parent dark-matter only simulation run to $z=5$ \citep{HOpkins2018_FIRE2, Ma+2018_FIRE2sr}, with baryonic mass resolution $m_b \sim 100 - 7000 \rm M_\odot$ (i.e. $\rm M_\star = 8\cdot10^5 -1.5 \cdot 10^{10} \; M_\odot$). The median metallicity values are presented in \cite{Marszewski+2024_mzrFIRE2}, where the authors estimate the stellar masses and mass-weighted metallicity considering all particles within 20\% of each sub-halo. Following \cite{Marszewski+2024_mzrFIRE2}, we only consider mass bins with at least 5 galaxies. 

The FirstLight simulation is drawn from a low resolution N-body simulation, with a lower limit on baryonic mass resolution of $m_b = 100 \rm M_\odot$ \citep{Ceverino+2017_firstlight}. The galaxy gas-phase metallicity are defined as the mass-weighted average nebular metallicity including all star particles younger than 100 Myr \citep{Langan+2020_firstlight}. In Fig. \ref{fig:mzr}, we report the best-fitting line up to $\rm M_\star \sim 10^9 \; M_\odot$ derived in \cite{Langan+2020_firstlight} at $z=5$.

The SPHINX$^{20}$ hydrodynamical simulation is the largest volume in the SPHINX suite, which was run to $z=4.64$. It has a box size of 20 co-moving Mpc on a side, which limits the number of galaxies at high-mass end, but its resolution, with stellar particles having initial masses $m_s = 400 \; \rm M_\odot$, allows to probe the low-mass end. The stellar mass is computed as the sum of the initial mass of all star particles within the virial radius and the gas-phase metallicity is weighted by gas mass within each halo \citep{Katz+2023_sphinx}. In Fig. \ref{fig:mzr}, we report the median trend together with its scatter for galaxies at $z=4.64-6$ \footnote{SPHINX data available at: \url{https://github.com/HarleyKatz/SPHINX-20-data}}.

The \texttt{dustyGadget} hydrodynamical simulation has a side length of $\sim 74$ co-moving Mpc, it was run to $z=4$ and has stellar particles resolution of $m_s \sim 2\cdot 10^6 \; \rm M_\star$. The stellar mass is computed as the sum of all the star particles within the virial radius and the gas-phase metallicity is computed as the ratio of the total metal mass to the total gas mass of the gas particles within a radius R $\sim 70$ pc of stellar particles. In computing the gas-phase metallicity, \texttt{dustyGadget} accounts for the depletion of metals into dust grains. Following \cite{DiCesare+2023}, \cite{DAngelo+2026} consider galaxies with $\rm log(M_\star/M_\odot) > 8$ to ensure robustly resolved systems. In Fig. \ref{fig:mzr}, we report the median trend together with its scatter for galaxies at $z=4.495$ (priv. comm.).

The COLIBRE L200m6 hydrodynamical simulation has a side length of 200 co-moving Mpc, was run to $z = 0$ and has a baryon particle mass of $m_b = 1.84 \cdot 10^6 \; \rm M_\odot$ \citep{Schaye+2026_COLIBRE}. Among hydrodynamical simulations, COLIBRE is the only one that removes the pressure floor, allowing the gas to cool down to 10 K and thereby explicitly modeling the cold gas phase of the ISM. The stellar mass is computed considering all the stellar particles bound to the galaxy sub-halo that lie within a 50 pkpc circular aperture centered at the location of the most bound particle. The gas-phase metallicity is computed as the mass-weighted ratio of the abundances of oxygen over hydrogen in the gas, which accounts for depletion of oxygen into dust grains. Following \cite{Feldmann+2023}, COLIBRE uses a smaller aperture (3 pkpc) for the oxygen abundance (see \cite{Sharda+2026_mzrCOLIBRE} for the discussion on the impact of aperture sizes on the metallicity.). Following \cite{Sharda+2026_mzrCOLIBRE} we only consider galaxies with $\rm log(M_\star/M_\odot) > 7$, which are resolved with at least 10 stellar particles. In Fig. \ref{fig:mzr}, we report the median trend of COLIBRE L200m6 together with its scatter for galaxies at $z = 5$ \footnote{COLIBRE data available at : \url{https://github.com/psharda/colibre_mzr_sharda26/tree/main}}.

\end{appendix}
\end{document}